\documentclass[onecolumn]{aastex631}
\usepackage{amsmath, amsthm, amssymb, amsfonts}
\usepackage{multirow}
\usepackage{fontawesome}
\usepackage{tabularx}

\newcommand{\summ}{\sum_{l=0}^{L}}
\newcommand{\Fo}{F_\odot}
\newcommand{\Aint}{A_{n,\text{int}}}

\newcommand{\ftcld}{f_{\tau_\text{cld}}}
\newcommand{\ftray}{f_{\tau_\text{ray}}}
\shorttitle{Spherical Harmonics for 1D Radiative Transfer I}
\shortauthors{Rooney et al.}

\usepackage[toc,page]{appendix}
\begin{document}

\title{Spherical Harmonics for the 1D Radiative Transfer Equation I: Reflected Light}
% I. Reflected Light: Spherical Harmonics for the 1D Radiative Transfer Equation 
% Spherical Harmonics for the 1D Radiative Transfer Equation I: Reflected Light

\correspondingauthor{Natasha E. Batalha}
\email{natasha.e.batalha@nasa.gov}

\author[0000-0001-9005-2872]{Caoimhe M. Rooney}
\affiliation{NASA Ames Research Center,
Moffett Field,
CA, 94035 USA}

\author[0000-0003-1240-6844]{Natasha E. Batalha}
\affiliation{NASA Ames Research Center,
Moffett Field, 
CA, 94035 USA}

\author[0000-0002-5251-2943]{Mark S. Marley}
\affiliation{Department of Planetary Sciences, Lunar and Planetary Laboratory, University of Arizona, Tucson AZ 85721}

\begin{abstract}
A significant challenge in radiative transfer theory for atmospheres of exoplanets and brown dwarfs is the derivation of computationally efficient methods that have adequate fidelity to more precise, numerically demanding solutions.
In this work, we extend the capability of the first open-source radiative transfer model for computing the reflected light of exoplanets at any phase geometry, \texttt{PICASO}: Planetary Intensity Code for Atmospheric Spectroscopy Observations. 
Until now, \texttt{PICASO} has implemented two-stream approaches to the solving the radiative transfer equation for reflected light, in particular following the derivations of \cite{toon1989rapid} (Toon89).
In order to improve the model accuracy, we have considered higher-order approximations of the phase functions, namely, we have increased the order of approximation from 2 to 4, using spherical harmonics.
The spherical harmonics approximation decouples spatial and directional dependencies by expanding the intensity and phase function into a series of spherical harmonics, or Legendre polynomials, allowing for analytical solutions for low-order approximations to optimize computational efficiency.
We rigorously derive the spherical harmonics method for reflected light and benchmark the 4-term method (SH4) against Toon89 and two independent and higher-fidelity methods (\texttt{CDISORT} 
\& doubling-method). On average, the SH4 method provides an order of magnitude increase in accuracy, compared to Toon89. 
Lastly, we implement SH4 within \texttt{PICASO} and observe only modest increase in computational time, compared to two-stream methods (20\% increase). 
\end{abstract}



\section{Introduction}
A serious challenge in atmospheric studies is the derivation of computationally efficient methods to solve the radiative transfer equation in a scattering and absorbing medium.
Exact solutions typically do not exist; thus, we rely on approximate methods to estimate solutions \citep{stephens1984multimode, thomas2002radiative, chandrasekhar1960radiative, liou2002introduction}.
In recent decades, there has been particular focus on the derivation of simple and effective approximate methods that have sufficient fidelity to more exact, numerically intricate solutions.    
This problem is further complicated by the vast range of parameter values relevant for atmospheric models that make spectral inference for exoplanets and Brown Dwarf problems computationally expensive \citep[e.g.][]{madhu2009temp,line2012info,barstow20comp}.

The most popular approximate methods for solving the radiative transfer equation are the (1) discrete-ordinates method \citep{chandrasekhar1960radiative, stamnes1988numerically, stamnes2000disort}, (2) Monte-Carlo method \citep{modest2013radiative, iwabuchi2006efficient} and (3) spherical harmonics method \citep{modest1989modified, modest2013radiative, olfe1967modification, https://doi.org/10.48550/arxiv.2205.09713}.

The discrete-ordinates method (DOM) is arguably the most well-studied and widely used approach when both rapidity and accuracy are required.
The problem to be solved is the transfer of monochromatic radiation in a scattering, absorbing and emitting plane-parallel atmosphere with a given function for bidirectional reflectivity at the lower boundary.
The general approach is to discretize the solid angle by a finite number $N$ of directions or ``streams'', along which the radiative intensities are tracked.
DISORT \citep{stamnes1988numerically, stamnes2000disort} is an example of a discrete ordinate algorithm for radiative transfer in media that is assumed to be non-isothermal, vertically inhomogeneous, but horizontally homogeneous. 
DISORT solves the radiative transfer equation within a single layer without boundary conditions.
The boundary conditions, along with continuity conditions at interfaces between layers, are used to establish a system of linear equations, which are solved numerically to obtain the intensity. 
It is capable of simulating thermal emission, absorption, and scattering for arbitrary phase functions across the electromagnetic spectrum from UV to radio wavelengths.
However, the convergence of DOM has been shown to depreciate for optically thick media \citep{modest2013radiative, fiveland1996acceleration, lewis1984computational}. 
This is due to the strongly coupled directional equations, which make the DOM method very computationally expensive when calculating accurate solutions \citep{modest2013radiative, ravishankar2009spherical}.
However, there exist a number of acceleration schemes to improve the convergence rate of DOM \citep{fiveland1996acceleration, lewis1984computational}.

Monte-Carlo methods operate by tracking emitted photons throughout the media. 
As a renowned accurate method to model radiation, it is often used for bench-marking, however, it is computationally slow, making it unsuitable for practical applications \citep{iwabuchi2006efficient, mayer2009radiative}. 
Furthermore, due to the significant noise that arises as a consequence of the stochastic nature of Monte-Carlo, the method cannot be easily coupled with other deterministic solvers.

The spherical harmonics (SH) approximation, denoted $P_N$, decouples spatial and directional dependencies by expanding the intensity and phase function into a series of spherical harmonics, or Legendre polynomials.
This approach results in fewer equations than the discrete-ordinates method and potentially obtains higher levels of accuracy at comparable computational expense.
This method has been widely applied to study radiative transfer in planetary and stellar atmospheres \citep{https://doi.org/10.48550/arxiv.2205.09713, li1996four, zhang2013doubling}.
However, higher order $P_N$ are mathematically complex and increasingly difficult to implement as $N$ increases \citep{ge2015implementation, https://doi.org/10.48550/arxiv.2205.09713}.
This method is appropriate for scattering media, however, low-order spherical harmonics can produce inaccuracies for optically thin atmospheres, and are invalid for non-participating media\footnote{Media is described as non-participating if it does not absorb, emit or scatter radiation.}
\citep{ravishankar2009spherical}.

Both the discrete-ordinates and spherical harmonics methods allow for increasing levels of complexity by modifying the order of the approximation. 
For DOM, this involves increasing the number of streams, whereas for SH this involves increasing the order of the Legendre expansion.
The simplest and widely used choice within the discrete-ordinate method is to limit the number of ordinates, or streams $N$, to two. 
In other words, this retains only hemispherical asymmetry information.
We refer to the $N=2$ DOM method as the ``two-stream'' approximation. 

The two-stream problem is formulated as a pair of ordinary differential equations which are solved within an atmosphere partitioned into homogeneous layers.
Within these layers, the differential equations have constant coefficients and hence can be solved analytically to produce simple exponential expressions for fluxes.
There exist a number of variations of the two-stream approach, including the two-stream source function technique \citep{toon1989rapid, batalha2019exoplanet} and the analytical formulation described in \citet{heng2014analytical, heng2018analytical}. 
The equivalent, lowest order approximation in spherical harmonics is the $P_1$ or Eddington approximation \citep{meador1980two, irvine1975multiple, chandrasekhar1960radiative, mihalas2013foundations}, where we expand the intensity and phase function to only two terms.

This low-order angular variation is generally justified for problems where scattering dominates or is the principle interest; in particular, for slowly varying phase functions with angle or for large optical depths when intensity has been smoothed by multiple scattering \citep{cuzzi1982delta}.
However, there are significant limitations associated with restricting angular variation to such low-orders, and these methods are often inappropriate for use with highly asymmetric phase functions typical of particulate scattering by water and silicate clouds \citep{modest2013radiative}.
For example, \cite{wiscombe1977range} studied the accuracy of albedos and absorptivity computed using this Eddington approximation, for a homogeneous layer with varying values for optical depth, single-scattering albedo and zenith angle of the incident beam.
When compared to the doubling method, which was assumed by \cite{wiscombe1977range} to give ``exact'' solutions to the problem, the errors tended to worsen when going from the Rayleigh to the Henyey-Greenstein to the Mie scattering cases. Errors generally fell within the 10\%--33\% range for Mie scattering.
This decrease in accuracy is due to the increase in the forward scattering peak of the phase function.
Such errors in the Mie case are unacceptably large for many practical applications. 

To account for inaccuracies encountered for asymmetric phase functions, the implementation of the delta ($\delta$)-function adjustment, which does not increase computational time, was introduced \citep{joseph1976delta}.  
The $\delta-$adjustment provides a third term closure through the second moment of the phase function expansion \cite{liou1988simple}.
This adjustment results in an improvement in the accuracy of radiative flux calculations by accounting for the strong forward scattering of large particles.
\cite{wiscombe1977range} analysed the impact of the $\delta$-adjustment on the accuracy of albedos and absorptivity mentioned above, and found that the $\delta$-Eddington approximation reduced the above relative error for Mie scattering of 10\%--33\% to less than 10\%.
Such improvement implies that the forward peak of the Mie phase function can be better captured using $\delta$-adjusted approximations.

However, \cite{king1986comparative} found that relative errors of 15\%--20\% could still be produced for various two-stream approximations.
\cite{liou1988simple} conducted a similar analysis for the discrete-ordinate $\delta$-two-stream model.
The $\delta$-two-stream performed comparatively to the spherical harmonics $\delta$-Eddington approximation studied by \cite{king1986comparative}. For conservative scattering, both approximations display low accuracy for reflection values, on the order of 10\%--30\% for incident solar angle $\mu_0<0.5$ and $\mu_0>0.9$ with $\tau<1$.
For transmission, errors exceeded 10\% for $\mu_0<0.2$.

Further improvement in angular accuracy, whilst maintaining analytical capacity, can be obtained by increasing the number of streams $N$ or polynomials to four.  
Considering $N=4$ in both DOM and SH results in an improved representation of the phase function and subsequently greater accuracy of the angle-integrated parameters \citep{liou1973numerical, liou1974analytic, li1996four}.
The $\delta$-function adjustment to account for the forward diffraction peak can also be generalised to higher order techniques \citep{wiscombe1977delta, cuzzi1982delta, liou1988simple, shibata1992accuracy}, including the 4-stream method, and offers significant accuracy improvement.
\cite{liou1988simple} conducted the same relative accuracy analysis on the $\delta$-four-stream approximation (DOM) as discussed above for $\delta$-two-stream and $\delta$-Eddington.
In contrast to the $>$10\% errors observed for the $\delta$-two-stream approximations, the reflection and transmission values calculated using the $\delta$-four-stream approximation were within 5\% accuracy, except for a number of small regions within the parameter space where the relative accuracies remained below 10\%.
Generally, these higher inaccuracies occurred in optically thin regions with small solar zenith angles $\mu_0<0.3$, where the reflection and transmission values themselves were very small, thus the absolute errors were extremely small ($<1$\%).
The authors concluded that the $\delta$-four stream approximation applied to the computation of solar fluxes covering the entire solar spectrum can achieve averaged accuracy within 5\%, and thus may be sufficient for studying the flux distribution for solar radiative transfer in cloudy atmospheres \citep{liou1988simple,liou1973numerical}.

Despite the evidence in the literature suggesting that four-stream methods offer significant improvements in model accuracy, the conceptual simplicity and ease of implementation of two-stream methods is attractive for use in radiative transfer models.
A widely used model is that of \cite{toon1989rapid} who derived a general two-stream solution for the upward and downward fluxes within a single homogeneous layer.
This single-layer solution is extended to multiple homogeneous layers through the solving of a matrix system of equations, with flux continuity conditions enforced at the interfaces between layers.
Following this, the authors introduce the two-stream source function technique, which involves solving the radiative transfer equation \eqref{eq:RTE} with the source function written in terms of the two-stream intensity; thus this approximation of the source function can be inferred from the flux solution obtained in the first step.
With the source function specified, the radiative transfer equation \eqref{eq:RTE} can be easily solved to give the azimuthally averaged intensity $I(\tau,\mu)$ at the top ($\tau=0$) and bottom ($\tau=\partial\tau_n$) of a layer and thereby the top and bottom of the atmosphere.
The \cite{toon1989rapid} methodology has been implemented in Python code \texttt{PICASO} \citep{natasha_batalha_2022_6419943}. 
\texttt{PICASO}, however, does not have the ability to move to higher order streams, and there hasn't been an extension of \cite{toon1989rapid} to four terms.

Here we present a novel approach using spherical harmonics. 
Similar to the \cite{toon1989rapid} methodology, we derive and solve a system of equations for the layer-wise upwards and downwards fluxes, however we choose a $\delta$-adjusted four-term spherical harmonics ($P_3$) approximation in place of the two-stream approach.
We then extend the source-function technique for the $P_3$ approach to derive the azimuthally averaged intensity emerging from the top of a vertically inhomogeneous atmosphere.
This is then used in an example calculation to compute the full geometric albedo spectrum for some cloudy model planets.
The spherical harmonics method that we implement is similar to the analysis conducted by \cite{zhang2013doubling}, however, we extend the method to multi-layer atmospheres via the source-function technique, rather than the doubling-adding method studied by \cite{zhang2013doubling}.

We study these with spherical harmonics and compare our results to other models in the literature. Specifically the applications in atmospheric modeling often require accurate results in reflection, transmission and absorption for a wide variety of optical depths and solar zenith angles. Therefore, this work is outlined as follows: we begin by outlining the spherical harmonics method for solving the radiative transfer equation within a single layer in Section \ref{sec:SH}.
We describe both the $P_1$ and $P_3$ solutions before extending our analysis to multiple layers in Section \ref{sec:multi_layers}.
In Section \ref{sec:source_function}, we describe our extension to the \cite{toon1989rapid} source-function technique.
We then conduct a series of model comparisons in Section \ref{sec:comparison}.
The first of these analyses is the comparison of reflection and transmission values obtained via spherical harmonics to those obtained via the discrete-ordinate method (DOM) and the doubling method. 
This comparison follows the work of \cite{liou1973numerical} who produced the DOM and doubling data that we use.
Next, we compare various model outputs to that of DISORT, in particular, the azimuthally averaged intensity varying with both asymmetry and optical depth, as well as the layer-wise vertical fluxes.
Finally, we conclude our model comparisons with an analysis of the geometric albedo produced by the spherical harmonics method compared to \texttt{PICASO}'s original two stream \citet{toon1989rapid} method. 

We intend for this work to be a comprehensive explanation of the spherical harmonics technique for 1D radiative transfer of reflected light that can easily be cross-referenced with the numerical method implemented within \texttt{PICASO}.
For this reason, we step through the derivation of the model and include the key mathematical expressions needed to understand the methodology, and hence, the algorithm.
We have also included hyperlinks that can be accessed by clicking the following icon: \href{https://natashabatalha.github.io/picaso/}{\faCode}, that will redirect the reader to the relevant lines of code (stored on GitHub) corresponding to the mathematical expressions in question.

\section{Solving the radiative transfer equation using spherical harmonics}
\label{sec:SH}
We wish to use the spherical harmonics technique to solve the azimuthally-averaged, one-dimensional radiative transfer equation:
\begin{equation}
	\mu\frac{\partial I}{\partial \tau}(\tau,\mu) = I(\tau,\mu) 
		- \frac{w_0}{2}\int_{-1}^{1} I(\tau,\mu')\mathcal{P}(\mu,\mu')\mathrm{d}\mu'
		- \frac{w_0}{4 \pi}F_\odot e^{-\frac{\tau}{\mu_0}}\mathcal{P}(\mu,-\mu_0),
	\label{eq:RTE}
\end{equation}
where location within the atmosphere is specified by $\tau\in[0,\tau_N]$, (where $\tau_N$ is the cumulative optical depth), $I$ is the azimuthally averaged intensity and $w_0$ is the single scattering albedo.
The incoming solar flux is $F_\odot$, and the direction of incident solar and outgoing scattered radiation is defined by the cosine of the zenith angles, denoted $\mu_0$ and $\mu$ respectively.
Finally, $\mathcal{P}(\mu,\mu')$ is the azimuthally averaged scattering phase function. 

Both the phase function and intensity can be expanded in terms of Legendre polynomials, up to given order $L$:
\begin{align}
	\mathcal{P}(\mu,\mu') &= \sum_{l=0}^L \chi_l P_l(\mu) P_l(\mu'),\label{eq:PF}\\
	I(\tau,\mu) &= \sum_{l=0}^L (2l+1) I_l(\tau) P_l(\mu),\label{eq:I}
\end{align}
where the coefficients $\chi_l$ of the phase function expansion can be determined from the orthogonal property of Legendre polynomials \citep{liou2002introduction}:
\begin{equation}
    \chi_l = \frac{2l+1}{2}\int_{-1}^1 \mathcal{P}(\cos\Theta)P_l(\cos\Theta)\mathrm{d}\cos\Theta.
    \label{eq:chi_def}
\end{equation}

Substituting \eqref{eq:PF} and \eqref{eq:I} into \eqref{eq:RTE} and using both the orthogonality property and recursion relation of Legendre polynomials, we obtain
\begin{equation}
	\sum_{l=0}^L \left[(l+1)\frac{\mathrm{d}I_{l+1}}{\mathrm{d}\tau}
			+ l\frac{\mathrm{d}I_{l-1}}{\mathrm{d}\tau}\right]P_l(\mu)
            = \sum_{l=0}^L [a_lI_l(\tau)  
			- b_le^{-\frac{\tau}{\mu_0}}]
			P_l(\mu),
	\label{eq:RTE_expanded}
\end{equation}
where
\begin{align}
    a_l = (2l+1)-w_0\chi_l,\\
    b_l = \frac{w_0\chi_l F_\odot}{4\pi} P_l(-\mu_0),
    \label{eq:a_and_b}
\end{align}
for $l=0,\cdots,L$ \href{https://github.com/natashabatalha/picaso/blob/9d4cbd672a75c1faf5297c3f1d74074018cd7ef3/picaso/fluxes.py#L2708-L2710}{\faCode}.

\subsection{$P_1$ single-layer solution}
\label{sec:P1}
For clarity and to demonstrate the spherical harmonics methodology, we first consider an atmosphere consisting of a single horizontally homogeneous layer where the optical properties of which are characterized by its single scattering albedo $w_0$, asymmetry parameter $g_0$ and optical thickness $\tau_N$.
We will extend this analysis to multiple layers in Section \ref{sec:multi_layers}.
% Location within the layer is specified by the optical depth $\tau\in[0,\tau_N]$, whereas the direction of incident solar and scattered radiation is defined by the cosine of the zenith angles, denoted $\mu_0$ and $\mu$ respectively.
The two-stream spherical harmonics problem is denoted $P_1$, where $L=1$ represents the highest Legendre polynomial in the expansion.
% write \eqref{eq:RTE_expanded} as 
% \begin{align}
% 	\frac{\mathrm{d}I_1}{\mathrm{d}\tau} &= 
% 	a_0I_0(\tau) - b_0e^{-\frac{\tau}{\mu_0}},\\ 
% 	%(1-ww_0)I_0(\tau) - \frac{ww_0}{4\pi}\Fo e^{-\frac{\tau}{\mu_0}}P_0(-\mu_0),\\
% 	\frac{\mathrm{d}I_0}{\mathrm{d}\tau} &=  
% 	a_1I_1(\tau) - b_1e^{-\frac{\tau}{\mu_0}},\\ %(3-ww_1)I_1(\tau) - \frac{ww_1}{4\pi}\Fo e^{-\frac{\tau}{\mu_0}}P_1(-\mu_0),
% \end{align}
% which can be formulated 
We can formulate \eqref{eq:RTE_expanded}
as a matrix system given by
\begin{equation}
	\frac{\mathrm{d}}{\mathrm{d}\tau}
	\begin{pmatrix}
		I_0 \\ I_1
	\end{pmatrix} = 
	\begin{pmatrix}
		0 & a_1 \\ a_0 & 0
	\end{pmatrix}  
	\begin{pmatrix}
		I_0 \\ I_1
	\end{pmatrix} - 
	\begin{pmatrix}
		b_1 \\ b_0
	\end{pmatrix}e^{-\frac{\tau}{\mu_0}},
	\label{eq:P1_sys}
\end{equation}
which has solution 
\begin{equation}
	\begin{pmatrix}
		I_0 \\ I_1
	\end{pmatrix} = 
	\begin{pmatrix}
		e^{-\lambda\tau} & e^{\lambda\tau} \\ 
		-q e^{-\lambda\tau} & q e^{\lambda\tau} 
	\end{pmatrix}  
	\begin{pmatrix}
		X_0 \\ X_1
	\end{pmatrix} + 
	\begin{pmatrix}
		\eta_0 \\ \eta_1
	\end{pmatrix}e^{-\frac{\tau}{\mu_0}},
	\label{eq:P1_sol}
\end{equation}
where $\lambda=\sqrt{a_0 a_1}$, $q=\lambda/a_1$ and
\begin{align}
	\eta_0 &= \frac{1}{1/\mu_0^2-\lambda^2}\left(b_1/\mu_0-a_1b_0\right),\\
	\eta_1 &= \frac{1}{1/\mu_0^2-\lambda^2}\left(b_0/\mu_0-a_0b_1\right).
	\label{eq:P1_eta_reflected}
\end{align}  
Coefficients $X_0$ and $X_1$ are to be determined from boundary conditions.
The details of this solution are outlined in Appendix \ref{app:P1_sol}.

In order to solve for radiative-convective equilibrium temperature structure, the net upward and downward fluxes are required at every layer in the atmosphere. 
Since the ultimate goal of this work is to present the spherical harmonics method for both incident optical and emitted thermal radiation, we must calculate such fluxes. 
As the transformation to fluxes is non-trivial, especially for non-expert readers, we present the explicit equations.

We can write \eqref{eq:P1_sol} in terms of fluxes where $F^{\pm}(\tau) = 2\pi\int_0^{\pm 1} \mu I(\tau,\mu)\,\mathrm{d}\mu$. 
Recalling the expansion of intensity $I$ in Legendre polynomials \eqref{eq:I}, the upward and downward fluxes can be written as
\begin{align}
	F^{\pm}(\tau) &= 2\pi\int_0^{\pm 1} \mu[I_0(\tau) + 3\mu I_1(\tau)]\mathrm{d}\mu,\\
	&= 2\pi\left[\frac{1}{2}I_0(\tau) \pm I_1(\tau)\right].
\end{align}
Therefore, \eqref{eq:P1_sol} formulated in terms of fluxes is given as
\begin{equation}
	\begin{pmatrix}
		F^- \\ F^+
	\end{pmatrix} = 
	\begin{pmatrix}
		Q^+ e^{-\lambda\tau} & Q^- e^{\lambda\tau} \\ 
		Q^- e^{-\lambda\tau} & Q^+ e^{\lambda\tau} 
	\end{pmatrix}  
	\begin{pmatrix}
		X_0 \\ X_1
	\end{pmatrix} + 
	\begin{pmatrix}
		Z^- \\ Z^+
	\end{pmatrix},%e^{-\frac{\tau}{\mu_0}}
	\label{eq:P1_flx}
\end{equation}
where $Q^\pm=\pi(1\pm 2q)$ and $Z^\pm(\tau)=\pi(\eta_0\pm 2\eta_1)$.
For the single-layer problem, we enforce the following boundary conditions \citep{toon1989rapid} on \eqref{eq:P1_flx} 
\begin{align}
	F^-(0) &= 0,\\%b_\text{top},\\
	F^+(\tau_N) &= A_S[F^-(\tau_N) + \mu_0F_\odot e^{-\frac{\tau}{\mu_0}}],
\end{align}
where $A_S$ is the surface reflectivity.
These boundary conditions enforce that there is no incident diffuse flux at the top of the atmosphere, and the upward flux at the surface is the sum of the reflected downward diffuse flux and the reflection of the unattenuated portion of the direct beam.
%** Will be replacing $F^-(0)$ BC with $b_\text{top}$ here but need to figure out the corresponding intensity BC, $I(0,-\mu)$.**

\subsection{$P_3$ single-layer}
\label{sec:P3}
The same principles applied to the $P_1$ problem can be extended to study higher-order techniques.
We increase the order of approximation to $L=3$, more specifically, we consider the $P_3$ problem or the four-term spherical harmonics technique within a single layer.
In a similar method to the $P_1$ problem in Section \ref{sec:P1}, we can write \eqref{eq:RTE_expanded} as 
\begin{equation}
	\frac{\mathrm{d}}{\mathrm{d}\tau}
	\begin{pmatrix}
		I_0 \\ I_1 \\ I_2 \\ I_3
	\end{pmatrix} = 
	\begin{pmatrix}
		0 & a_1 & 0 & -\frac{2a_3}{3} \\ 
		a_0 & 0 & 0 & 0 \\
		0 & 0 & 0 & \frac{a_3}{3} \\
		-\frac{2a_0}{3} & 0 & \frac{a_2}{3} & 0 \\ 
	\end{pmatrix}  
	\begin{pmatrix}
		I_0 \\ I_1 \\ I_2 \\ I_3
	\end{pmatrix} - 
	\begin{pmatrix}
		b_1-\frac{2b_3}{3} \\ 
		b_0 \\
		\frac{b_3}{3} \\
		\frac{b_2}{3}-\frac{2b_0}{3}  
	\end{pmatrix}e^{-\frac{\tau}{\mu_0}},
	\label{eq:P3_sys}
\end{equation}
which has solution 
\begin{equation}
	\begin{pmatrix}
		I_0 \\ I_1 \\ I_2 \\ I_3
	\end{pmatrix}= 
	\begin{pmatrix}
		e^{-\lambda_1\tau} & e^{\lambda_1\tau} & 
			e^{-\lambda_2\tau} & e^{\lambda_2\tau} \\ 
		R_1 e^{-\lambda_1\tau} & -R_1 e^{\lambda_1\tau} & 
			R_2 e^{-\lambda_2\tau} & -R_2 e^{\lambda_2\tau} \\ 
		Q_1 e^{-\lambda_1\tau} & Q_1 e^{\lambda_1\tau} & 
			Q_2 e^{-\lambda_2\tau} & Q_2 e^{\lambda_2\tau} \\ 
		S_1 e^{-\lambda_1\tau} & -S_1 e^{\lambda_1\tau} & 
			S_2 e^{-\lambda_2\tau} & -S_2 e^{\lambda_2\tau} 
	\end{pmatrix}  
	\begin{pmatrix}
		X_0 \\ X_1 \\ X_2 \\ X_3
	\end{pmatrix} + 
	\begin{pmatrix}
		\eta_0 \\ \eta_1 \\ \eta_2 \\ \eta_3
	\end{pmatrix}e^{-\frac{\tau}{\mu_0}}
	\label{eq:P3_sol}
\end{equation}
where 
\begin{align}
    \lambda_{1,2} = \sqrt{\frac{1}{2}(\beta \pm \sqrt{\beta^2-4\gamma})},\qquad  \beta=a_0a_1 + \frac{1}{9}a_2a_3 + \frac{4}{9}a_0a_3,\qquad \gamma=\frac{1}{9}a_0a_1a_2a_3,
\end{align}
\begin{align}
    R_{1,2}=-\frac{a_0}{\lambda_{1,2}},\qquad
    Q_{1,2} = \frac{1}{2}\left(\frac{a_0a_1}{\lambda_{1,2}^2} - 1\right),\qquad
    S_{1,2} = -\frac{3}{2a_3}\left(\frac{a_0a_1}{\lambda_{1,2}} - \lambda_{1,2}\right),
\end{align}
and
$\eta_l=\Delta_l/\Delta$ for $\Delta = 9f(1/\mu_0)$, $f(x) = x^4 - \beta x^2 + \gamma$ and
\begin{align}
	\Delta_0 &= (a_1b_0 - b_1/\mu_0)(a_2a_3-9/\mu_0^2) + 2(a_3b_2 - 2a_3b_0-3b_3/\mu_0)/\mu_0^2,\\
	\Delta_1 &= (a_0b_1 - b_0/\mu_0)(a_2a_3-9/\mu_0^2) - 2a_0(a_3b_2 - 3b_3/\mu_0)/\mu_0,\\
	\Delta_2 &= (a_3b_2 - 3b_3/\mu_0)(a_0a_1-1/\mu_0^2) - 2a_3(a_0b_1 -b_0/\mu_0)/\mu_0,\\
	\Delta_3 &= (a_2b_3 - 3b_2/\mu_0)(a_0a_1-1/\mu_0^2) + 2(3a_0b_1 - 2a_0b_3 - 3b_0/\mu_0)/\mu_0^2.
\end{align}
% and
% \begin{align}
% 	\Delta = 9f(1/\mu_0) \qquad \text{where} \qquad f(x) = x^4 - \beta x^2 + \gamma.
% \end{align}

We wish to consider the problem in terms of fluxes, namely $F^{\pm}(\tau) = 2\pi\int_0^{\pm 1} I(\tau,\mu) P_1(\mu)\,\mathrm{d}\mu$ and $f^{\pm}(\tau) = 2\pi\int_0^{\pm 1} I(\tau,\mu)P_3(\mu)\,\mathrm{d}\mu$.
Considering the Legendre expansion of intensity \eqref{eq:I}, these fluxes are given as
\begin{align}
	F^{\pm}(\tau) &= 2\pi\int_0^{\pm 1} \mu\left[I_0(\tau) + 3\mu I_1(\tau) + \frac{5}{2}(3\mu^2-1)I_2(\tau) + \frac{7}{2}(5\mu^3-3\mu)I_3(\tau) \right]\mathrm{d}\mu,\\
	&= 2\pi\left(I_0(\tau) \pm I_1(\tau) + \frac{5}{8}I_2(\tau)\right),\\
	f^{\pm}(\tau) &= 2\pi\int_0^{\pm 1} \frac{1}{2}\left(5\mu^3-3\mu\right)\left[I_0(\tau) + 3\mu I_1(\tau) + \frac{5}{2}(3\mu^2-1)I_2(\tau) + \frac{7}{2}(5\mu^3-3\mu)I_3(\tau)\right]\mathrm{d}\mu,\\
	&= 2\pi\left(-\frac{1}{8}I_0(\tau) + \frac{5}{8}I_2(\tau) \pm I_3(\tau) \right).
\end{align}
System \eqref{eq:P3_sol} can therefore be rewritten in terms of fluxes as
\begin{equation}
	\begin{pmatrix}
		F^- \\ f^- \\ F^+ \\ f^-
	\end{pmatrix} = 
	\begin{pmatrix}
		p_1^-e^{-\lambda_1\tau} & p_1^+e^{\lambda_1\tau} & 
			p_2^-e^{-\lambda_2\tau} & p_2^+e^{\lambda_2\tau} \\ 
		q_1^-e^{-\lambda_1\tau} & q_1^+e^{\lambda_1\tau} & 
			q_2^-e^{-\lambda_2\tau} & q_2^+e^{\lambda_2\tau} \\ 
		p_1^+e^{-\lambda_1\tau} & p_1^-e^{\lambda_1\tau} & 
			p_2^+e^{-\lambda_2\tau} & p_2^-e^{\lambda_2\tau} \\ 
		q_1^+e^{-\lambda_1\tau} & q_1^-e^{\lambda_1\tau} & 
			q_2^+e^{-\lambda_2\tau} & q_2^-e^{\lambda_2\tau} \\ 
	\end{pmatrix}  
	\begin{pmatrix}
		X_0 \\ X_1 \\ X_2 \\ X_3
	\end{pmatrix} + 
	\begin{pmatrix}
		Z_1^- \\ Z_2^-\\ Z_1^+\\ Z_2^+
	\end{pmatrix}%e^{-\frac{\tau}{\mu_0}}
	\label{eq:P3_flx}
\end{equation}
where $p_{1,2}^\pm = \pi(1\pm 2R_{1,2} + \frac{5}{4}Q_{1,2})$,
$q_{1,2}^\pm = \pi(-\frac{1}{4}+ \frac{5}{4}Q_{1,2} \pm 2S_{1,2})$,
$Z_1^\pm(\tau) = \pi(\eta_0 \pm 2\eta_1 + \frac{5}{4}\eta_2)$,
$Z_2^\pm (\tau)= \pi(-\frac{1}{4}\eta_0+ \frac{5}{4}\eta_2 \pm 2\eta_3)$.

For the single-layer problem, we enforce the following boundary conditions on \eqref{eq:P3_flx},
\begin{align}
	F^-(0) &= 0,\\%b_\text{top},\\
	f^-(0) &= 0,\label{eq:f-BC_P3} \\%-\frac{b_\text{top}}{4}, \\
	F^+(\tau_N) &= A_S[F^-(\tau_N) + \mu_0F_\odot e^{-\frac{\tau}{\mu_0}}],\\
	f^+(\tau_N) &= A_S[f^-(\tau_N) -\frac{1}{4}\mu_0F_\odot e^{-\frac{\tau}{\mu_0}}], \label{eq:bottomBC_single}
\end{align}
where $A_S$ is the surface reflectivity.
The derivation of the bottom boundary condition \eqref{eq:bottomBC_single} for $f^+$ is outlined in Appendix \ref{app:f_BC}.
%** Will be replacing $F^-(0)$ BC with $b_\text{top}$ here but need to figure out the corresponding intensity BC, $I(0,-\mu)$ to be able to calculate $f^-(0,-\mu)$. Also not confident about the bottom BC.**

\section{Extension to multiple layers}
\label{sec:multi_layers}
The procedures outlined in Sections \ref{sec:P1} and \ref{sec:P3} can be extended to multiple layers by solving the flux systems \eqref{eq:P1_flx} for $P_1$ and \eqref{eq:P3_flx} for $P_3$ within each layer and enforcing continuity boundary conditions between layers along with the boundary conditions given above for the top and bottom of the atmosphere.

We divide the atmosphere into $N$ homogeneous layers of optical depth $\partial\tau_n = \tau_n-\tau_{n-1}$ for $n=1,\dots,N$.
To solve the radiative transfer equation \eqref{eq:RTE} in the $n^\text{th}$ layer we rescale the optical depth as
\begin{equation}
	\hat\tau = \tau - \tau_{n-1},\qquad \hat\tau\in[0,\partial\tau_n].
\end{equation}
Dropping the hats, we continue with the solutions within layer $n$ for $\tau\in[0,\partial\tau_n]$.

\subsection{$P_1$ multiple layers}
For $N$ layers, we must solve system \eqref{eq:P1_sys} within every homogeneous layer.
The solution at the top of layer $n$ is given by
\begin{equation}
	\begin{pmatrix}
		I_{0,n} \\ I_{1,n}
	\end{pmatrix} = 
	\begin{pmatrix}
		e^{-\lambda_n\tau} & e^{\lambda_n\tau} \\ 
		-q_n e^{-\lambda_n\tau} & q_n e^{\lambda_n\tau} 
	\end{pmatrix}  
	\begin{pmatrix}
		X_{0,n} \\ X_{1,n}
	\end{pmatrix} - 
	\begin{pmatrix}
		\eta_{0,n} \\ \eta_{1,n}
	\end{pmatrix}e^{-\frac{1}{\mu_0}(\tau+\tau_{n-1})},
	\label{eq:P1_nlayer_sol}
\end{equation}
for $\tau\in[0,\partial\tau_n]$, and where we extend the single-layer definitions for $\eta$ given by \eqref{eq:P1_eta_reflected} to obtain
\begin{align}
	\eta_{0,n} &= \frac{1}{1/\mu_0^2-\lambda_n^2}\left(b_{1,n}/\mu_0-a_{1,n}b_{0,n}\right),\\
	\eta_{1,n} &= \frac{1}{1/\mu_0^2-\lambda_n^2}\left(b_{0,n}/\mu_0-a_{0,n}b_{1,n}\right),
\end{align}  
and $a_{l,n} = (2l+1)-w_{0,n}\chi_{l,n}$, $b_{l,n} = w_{0,n}\chi_{l,n} F_\odot P_l(-\mu_0)/4\pi$, $\lambda_n=\sqrt{a_{0,n} a_{1,n}}$, and $q_n=\lambda/a_{1,n}$ \href{https://github.com/natashabatalha/picaso/blob/e7d7078b8bd93cb53295b470b96006848811c62b/picaso/fluxes.py#L3094-L3100}{\faCode}, similarly extended from  
single-layer definitions \eqref{eq:a_and_b}.
In terms of fluxes,
\begin{equation}
	\begin{pmatrix}
		F_n^- \\ F_n^+
	\end{pmatrix} = 
	\begin{pmatrix}
		Q_n^+ e^{-\lambda_n\tau} & Q_n^- e^{\lambda_n\tau} \\ 
		Q_n^- e^{-\lambda_n\tau} & Q_n^+ e^{\lambda_n\tau} 
	\end{pmatrix}  
	\begin{pmatrix}
		X_{0,n} \\ X_{1,n}
	\end{pmatrix} + 
	\begin{pmatrix}
		Z_n^- \\ Z_n^+
	\end{pmatrix}e^{-\frac{1}{\mu_0}(\tau+\tau_{n-1})},
	\label{eq:P1_nlayer_flx}
\end{equation}
where $Q_n^\pm=\pi(1\pm 2q_n)$, $Z_n^\pm(\tau)=\pi(\eta_{0,n}\pm 2\eta_{1,n})$ \href{https://github.com/natashabatalha/picaso/blob/9d4cbd672a75c1faf5297c3f1d74074018cd7ef3/picaso/fluxes.py#L3109-L3110}{\faCode}, and with boundary conditions \href{https://github.com/natashabatalha/picaso/blob/9d4cbd672a75c1faf5297c3f1d74074018cd7ef3/picaso/fluxes.py#L3130-L3133}{\faCode} 
\begin{align}
	F_1^-(0) &= 0,\\%b_\text{top},\\
	F_n^-(\partial\tau_n) &= F_{n+1}^-(0),\\
	F_n^+(\partial\tau_n) &= F_{n+1}^+(0),\\
	F_N^+(\tau_N) &= A_S[F_N^-(\tau_N) + \mu_0F_\odot e^{-\frac{\tau_N}{\mu_0}}].
\end{align}
The flux problem is formulated in \texttt{PICASO} by representing the system in terms of banded matrices, and solved using the \texttt{solve\_banded} \href{https://github.com/natashabatalha/picaso/blob/9d4cbd672a75c1faf5297c3f1d74074018cd7ef3/picaso/fluxes.py#L3476}{\faCode} functionality of \texttt{SciPy} \citep{2020SciPy-NMeth}.

\subsection{$P_3$ multiple layers}
Consider $L=3$. 
In a similar method to the $P_1$ problem, the solution to the radiative transfer equation within layer $n$ is given by
\begin{equation}
	\begin{pmatrix}
		I_{0,n} \\ I_{1,n} \\ I_{2,n} \\ I_{3,n}
	\end{pmatrix} = 
	\begin{pmatrix}
		e^{-\lambda_{1,n}\tau} & e^{\lambda_{1,n}\tau} & 
			e^{-\lambda_{2,n}\tau} & e^{\lambda_{2,n}\tau} \\ 
		R_{1,n} e^{-\lambda_{1,n}\tau} & -R_{1,n} e^{\lambda_{1,n}\tau} & 
			R_{2,n} e^{-\lambda_{2,n}\tau} & -R_{2,n} e^{\lambda_{2,n}\tau} \\ 
		Q_{1,n} e^{-\lambda_{1,n}\tau} & Q_{1,n} e^{\lambda_{1,n}\tau} & 
			Q_{2,n} e^{-\lambda_{2,n}\tau} & Q_{2,n} e^{\lambda_{2,n}\tau} \\ 
		S_{1,n} e^{-\lambda_{1,n}\tau} & -S_{1,n} e^{\lambda_{1,n}\tau} & 
			S_{2,n} e^{-\lambda_{2,n}\tau} & -S_{2,n} e^{\lambda_{2,n}\tau} 
	\end{pmatrix}  
	\begin{pmatrix}
		X_{0,n} \\ X_{1,n} \\ X_{2,n} \\ X_{3,n}
	\end{pmatrix} - 
	\begin{pmatrix}
		\eta_{0,n} \\ \eta_{1,n} \\ \eta_{2,n} \\ \eta_{3,n}
	\end{pmatrix}e^{-\frac{1}{\mu_0}(\tau+\tau_{n-1})},
	\label{eq:P3_nlayer_sol}
\end{equation}
for $\tau\in[0,\partial\tau_n]$, 
where \href{https://github.com/natashabatalha/picaso/blob/9d4cbd672a75c1faf5297c3f1d74074018cd7ef3/picaso/fluxes.py#L3238-L3241}{\faCode}
\begin{align}
    \lambda_{1,2,n} = \sqrt{\frac{1}{2}(\beta_n \pm \sqrt{\beta_n^2-4\gamma_n})},\qquad  \beta_n=a_{0,n}a_{1,n} + \frac{1}{9}a_{2,n}a_{3,n} + \frac{4}{9}a_{0,n}a_{3,n},\qquad \gamma_n=\frac{1}{9}a_{0,n}a_{1,n}a_{2,n}a_{3,n},
\end{align}
and \href{https://github.com/natashabatalha/picaso/blob/9d4cbd672a75c1faf5297c3f1d74074018cd7ef3/picaso/fluxes.py#L3273-L3275}{\faCode}
\begin{align}
    R_{1,2,n}=-\frac{a_{0,n}}{\lambda_{1,2,n}},\qquad
    Q_{1,2,n} = \frac{1}{2}\left(\frac{a_{0,n}a_{1,n}}{\lambda_{1,2,n}^2} - 1\right),\qquad
    S_{1,2,n} = -\frac{3}{2a_{3,n}}\left(\frac{a_{0,n}a_{1,n}}{\lambda_{1,2,n}} - \lambda_{1,2,n}\right).
    \label{eq:RnQnSn}
\end{align}
We also define $\eta_{i,n} = \Delta_{i,n}/\Delta$ \href{https://github.com/natashabatalha/picaso/blob/9d4cbd672a75c1faf5297c3f1d74074018cd7ef3/picaso/fluxes.py#L3248-L3261}{\faCode}, for $\Delta = 9f(1/\mu_0)$, $f(x) = x^4 - \beta_n x^2 + \gamma_n$ and
\begin{align}
	\Delta_{0,n} &= (a_{1,n}b_{0,n} - b_{1,n}/\mu_0)(a_{2,n}a_{3,n}-9/\mu_0^2) + 2(a_{3,n}b_{2,n} - 2a_{3,n}b_{0,n}-3b_{3,n}/\mu_0)/\mu_0^2,\label{eq:eta_multi_start}\\
	\Delta_{1,n} &= (a_{0,n}b_{1,n} - b_{0,n}/\mu_0)(a_{2,n}a_{3,n}-9/\mu_0^2) - 2a_{0,n}(a_{3,n}b_{2,n} - 3b_{3,n}/\mu_0)/\mu_0,\\
	\Delta_{2,n} &= (a_{3,n}b_{2,n} - 3b_{3,n}/\mu_0)(a_{0,n}a_{1,n}-1/\mu_0^2) - 2a_{3,n}(a_{0,n}b_{1,n} -b_{0,n}/\mu_0)/\mu_0,\\
	\Delta_{3,n} &= (a_{2,n}b_{3,n} - 3b_{2,n}/\mu_0)(a_{0,n}a_{1,n}-1/\mu_0^2) + 2(3a_{0,n}b_{1,n} - 2a_{0,n}b_{3,n} - 3b_{0,n}/\mu_0)/\mu_0^2.\label{eq:eta_multi_end}
\end{align}

This problem can be written in terms of fluxes as
\begin{equation}
	\begin{pmatrix}
		F_n^- \\ f_n^- \\ F_n^+ \\ f_n^-
	\end{pmatrix} = 
	\begin{pmatrix}
		p_{1,n}^-e^{-\lambda_{1,n}\tau} & p_{1,n}^+e^{\lambda_{1,n}\tau} & 
			p_{2,n}^-e^{-\lambda_{2,n}\tau} & p_{2,n}^+e^{\lambda_{2,n}\tau} \\ 
		q_{1,n}^-e^{-\lambda_{1,n}\tau} & q_{1,n}^+e^{\lambda_{1,n}\tau} & 
			q_{2,n}^-e^{-\lambda_{2,n}\tau} & q_{2,n}^+e^{\lambda_{2,n}\tau} \\ 
		p_{1,n}^+e^{-\lambda_{1,n}\tau} & p_{1,n}^-e^{\lambda_{1,n}\tau} & 
			p_{2,n}^+e^{-\lambda_{2,n}\tau} & p_{2,n}^-e^{\lambda_{2,n}\tau} \\ 
		q_{1,n}^+e^{-\lambda_{1,n}\tau} & q_{1,n}^-e^{\lambda_{1,n}\tau} & 
			q_{2,n}^+e^{-\lambda_{2,n}\tau} & q_{2,n}^-e^{\lambda_{2,n}\tau} \\ 
	\end{pmatrix}  
	\begin{pmatrix}
		X_{0,n} \\ X_{1,n} \\ X_{2,n} \\ X_{3,n}
	\end{pmatrix} + 
	\begin{pmatrix}
		Z_{1,n}^- \\ Z_{2,n}^-\\ Z_{1,n}^+\\ Z_{2,n}^+
	\end{pmatrix}e^{-\frac{1}{\mu_0}(\tau+\tau_{n-1})},
\end{equation}
where $p_{1,2,n}^\pm = \pi(1\pm 2R_{1,2,n} + \frac{5}{4}Q_{1,2,n})$,
$q_{1,2,n}^\pm = \pi(-\frac{1}{4}+ \frac{5}{4}Q_{1,2,n} \pm 2S_{1,2,n})$ \href{https://github.com/natashabatalha/picaso/blob/9d4cbd672a75c1faf5297c3f1d74074018cd7ef3/picaso/fluxes.py#L3277-L3284}{\faCode},
$Z_{1,n}^\pm = \pi(\eta_{0,n} \pm 2\eta_{1,n} + \frac{5}{4}\eta_{2,n})$,
$Z_{2^,n}\pm = \pi(-\frac{1}{4}\eta_{0,n} + \frac{5}{4}\eta_{2,n} \pm 2\eta_{3,n})$ \href{https://github.com/natashabatalha/picaso/blob/9d4cbd672a75c1faf5297c3f1d74074018cd7ef3/picaso/fluxes.py#L3263-L3266}{\faCode}, and with boundary conditions
\begin{alignat}{2}
	F_1^-(0) &= 0, &&\quad\qquad f_1^-(0) = 0, \\
	F_n^-(\partial\tau_n) &= F_{n+1}^-(0), &&\qquad f_n^-(\partial\tau_n) = f_{n+1}^-(0),\\
	F_n^+(\partial\tau_n) &= F_{n+1}^+(0), &&\qquad f_n^+(\partial\tau_n) = f_{n+1}^+(0),\\
	F_N^+(\tau_N) &= A_S[F^-_N(\tau_N) + \mu_0F_\odot e^{-\frac{\tau_N}{\mu_0}}],&&\,\,\qquad 
	f_N^+(\tau_N) = A_S[f_N^-(\tau_N) -\frac{1}{4}\mu_0F_\odot e^{-\frac{\tau}{\mu_0}}]
\end{alignat}
for $n={1,2,\cdots,N-1}$ \href{https://github.com/natashabatalha/picaso/blob/9d4cbd672a75c1faf5297c3f1d74074018cd7ef3/picaso/fluxes.py#L3320-L3344}{\faCode}. 
The derivation of the bottom boundary condition for $f_N^+(\tau_N)$ is outlined in Appendix \ref{app:f_BC}.
As for the $P_1$ case, the flux problem is formulated in \texttt{PICASO} by representing the system in terms of banded matrices, and solved using the \texttt{solve\_banded} \href{https://github.com/natashabatalha/picaso/blob/9d4cbd672a75c1faf5297c3f1d74074018cd7ef3/picaso/fluxes.py#L3476}{\faCode} functionality of \texttt{SciPy} \citep{2020SciPy-NMeth}.

\section{Source function technique}
\label{sec:source_function}
We follow the methodology outlined in \cite{toon1989rapid} and apply the source function technique to calculate the emergent intensity from the top of the atmosphere.
The radiative transfer equation \eqref{eq:RTE} can be solved to yield the azimuthally integrated intensity at angle $\mu$ at the top of the $n^\text{th}$ layer ($\tau=0$) as
\begin{equation}
	I_n(0,\mu) = I_n(\partial\tau_n,\mu)e^{-\frac{\partial\tau_n}{\mu}} + \frac{1}{\mu}\int_0^{\partial\tau_n} S_{vt}e^{-\frac{\tau}{\mu}}\mathrm{d}\tau,
	\label{eq:azimuth_av_int}
\end{equation}
for
\begin{equation}
	S_{vt} = \frac{w_{0,n}}{2}\int_{-1}^{1} I_t(\tau, \mu')\mathcal{P}(\mu,\mu')\mathrm{d}\mu' + S_n(\tau,\mu) ,
	\label{eq:Svt}
\end{equation}
where 
\begin{equation}
	S_n(\tau,\mu) = \frac{w_0}{4\pi} F_\odot e^{-\frac{\tau+\tau_{n-1}}{\mu_0}} \mathcal{P}(\mu,-\mu_0).
\end{equation}
Note that we have used $I_t$ to denote the approximated intensity \eqref{eq:I} in place of the true intensity in the source term \eqref{eq:Svt}.
Therefore we can rewrite \eqref{eq:Svt} as
\begin{equation}
	S_{vt} = w_{0,n}\summ \chi_l I_l(\tau)P_l(\mu) + S_n(\tau,\mu). 
	\label{eq:Svt_1}
\end{equation}
Let us consider the integral term in \eqref{eq:azimuth_av_int}. Using \eqref{eq:Svt_1}, this can be written as
\begin{equation}
	\int_0^{\partial\tau_n} S_{vt}e^{-\frac{\tau}{\mu}}\mathrm{d}\tau = 
		w_0\summ \chi_l P_l(\mu) \int_0^{\partial\tau_n}I_l(\tau) e^{-\frac{\tau}{\mu}}\mathrm{d}\tau 
		+ \int_0^{\partial\tau_n}  S_n(\tau,\mu) e^{-\frac{\tau}{\mu}}\mathrm{d}\tau.
		%\frac{w}{4\pi} F_\odot \mathcal{P}(\mu,-\mu_0) \int_0^{\partial\tau_n} e^{-\tau\left(\frac{\mu+\mu_0}{\mu\mu_0}\right)} \mathrm{d}\tau %.
		\label{eq:Svt_int}
\end{equation}
We can  calculate the second term on the right-hand side of  \eqref{eq:Svt_int} to be \href{https://github.com/natashabatalha/picaso/blob/9d4cbd672a75c1faf5297c3f1d74074018cd7ef3/picaso/fluxes.py#L2812-L2814}{\faCode}
\begin{equation}
	\int_0^{\partial\tau_n}  S_n(\tau,\mu) e^{-\frac{\tau}{\mu}}\mathrm{d}\tau=\frac{w_0}{4\pi} F_\odot \mathcal{P}(\mu,-\mu_0) \frac{\mu\mu_0}{\mu+\mu_0}\left(1-e^{-\partial\tau_n\frac{\mu+\mu_0}{\mu\mu_0}}\right)
	e^{-\frac{\tau_n}{\mu_0}}.
\end{equation}
Next, let us write $\Aint = \int_0^{\partial\tau_n}I_l(\tau) e^{-\frac{\tau}{\mu}}\mathrm{d}\tau $. Therefore we can calculate $\Aint$ for the $P_1$ problem \href{https://github.com/natashabatalha/picaso/blob/9d4cbd672a75c1faf5297c3f1d74074018cd7ef3/picaso/fluxes.py#L2766-L2772}{\faCode}:
\begin{equation}
	\Aint = 
	\begin{pmatrix}
		 A_{0,n} &
		A_{1,n}\\ 
		-Q A_{0,n}&
		Q A_{1,n}\\ 
	\end{pmatrix}  
	\begin{pmatrix}
		X_{0,n} \\ X_{1,n}
	\end{pmatrix} + 
	\begin{pmatrix}
        N_{0,n} \\ N_{1,n}
	\end{pmatrix} e^{-\frac{\tau_n}{\mu_0}}
	\label{eq:P1_Aint}
\end{equation}
%b_n\begin{pmatrix}
%		\eta_{0,n} \\ \eta_{1,n}
%	\end{pmatrix} 
where \href{https://github.com/natashabatalha/picaso/blob/9d4cbd672a75c1faf5297c3f1d74074018cd7ef3/picaso/fluxes.py#L2758-L2763}{\faCode}
\begin{align}
	A_{0,n} &= \frac{1}{1/\mu+\lambda_n} \left(1-e^{-\partial\tau_n\left( \frac{1}{\mu}+\lambda_n \right)}\right),\\
	A_{1,n} &= \frac{1}{1/\mu-\lambda_n} \left(1-e^{-\partial\tau_n\left( \frac{1}{\mu}-\lambda_n \right)}\right),%\\
%	b_{n} &= \frac{\mu\mu_0}{\mu+\mu_0}\left(1-e^{-\partial\tau_n \frac{\mu+\mu_0}{\mu\mu_0}}\right)
%							e^{-\frac{\tau_n}{\mu_0}}.
\end{align}
and
\begin{equation}
	N_{i,n} = \eta_{i,n}\frac{\mu\mu_0}{\mu+\mu_0}\left(1-e^{-\partial\tau_n \frac{\mu+\mu_0}{\mu\mu_0}}\right)
	\label{eq:N_reflected}
\end{equation}
where $n_{i,n}=\Delta_{i,n}/\Delta$ are defined by equations \eqref{eq:eta_multi_start}--\eqref{eq:eta_multi_end}.

Similarly for the $P_3$ problem \href{https://github.com/natashabatalha/picaso/blob/9d4cbd672a75c1faf5297c3f1d74074018cd7ef3/picaso/fluxes.py#L2790-L2802}{\faCode}:
\begin{equation}
	\Aint = 
	\begin{pmatrix}
		A_{00,n} & A_{01,n} &  A_{02,n} & A_{03,n}\\ 
 		A_{10,n} & A_{11,n} &  A_{12,n} & A_{13,n}\\ 
 		A_{20,n} & A_{21,n} &  A_{22,n} & A_{23,n}\\ 
		A_{30,n} & A_{31,n} &  A_{32,n} & A_{33,n}\\ 
	\end{pmatrix}  
	\begin{pmatrix}
		X_{0,n} \\ X_{1,n} \\ X_{2,n} \\ X_{3,n}
	\end{pmatrix} + 
	\begin{pmatrix}
		N_{0,n} \\ N_{1,n} \\ N_{2,n} \\ N_{3,n} 
	\end{pmatrix} e^{-\frac{\tau_n}{\mu_0}},
	\label{eq:P3_Aint}
\end{equation}
where \href{https://github.com/natashabatalha/picaso/blob/9d4cbd672a75c1faf5297c3f1d74074018cd7ef3/picaso/fluxes.py#L2775-L2787}{\faCode}
\begin{alignat}{4}
	A_{00,n} &= \alpha_{1,n}, \quad\qquad A_{01,n} = \beta_{1,n}, \,\,\,\quad\qquad A_{02,n} = \alpha_{2,n}, \quad\qquad A_{03,n} = \beta_{2,n},\\
	A_{10,n} &= R_{1,n}\alpha_{1,n}, \quad A_{11,n} = -R_{1,n}\beta_{1,n}, \quad A_{12,n} = R_{2,n}\alpha_{2,n}, \quad A_{13,n} = -R_{2,n}\beta_{2,n},\\
	A_{20,n} &= Q_{1,n}\alpha_{1,n}, \quad A_{21,n} = Q_{1,n}\beta_{1,n}, \,\,\,\,\quad A_{22,n} = Q_2\alpha_{2,n}, \,\,\,\,\quad A_{23,n} = Q_{2,n}\beta_{2,n},\\
	A_{30,n} &= S_{1,n}\alpha_{1,n}, \,\quad A_{31,n} = -S_{1,n}\beta_{1,n}, \,\quad A_{32,n} = S_2\alpha_{2,n}, \,\,\,\,\,\quad A_{33,n} = -S_{2,n}\beta_{2,n},\\
\end{alignat}
and
\begin{align}
	\alpha_{i,n} = \frac{1}{1/\mu+\lambda_{i,n}} \left(1-e^{-\partial\tau_n\left( \frac{1}{\mu}+\lambda_{i,n} \right)}\right), \qquad 
	\beta_{i,n} = \frac{1}{1/\mu-\lambda_{i,n}} \left(1-e^{-\partial\tau_n\left( \frac{1}{\mu}-\lambda_{i,n} \right)}\right). 
\end{align}
% \begin{align}
% 	R_i = -\frac{a_0}{\lambda_i},\qquad Q_i = \frac{1}{2}\left(\frac{a_0a_1}{\lambda_i^2}-1\right), \qquad S_i = -\frac{3}{2a_3}\left(\frac{a_0a_1}{\lambda_i}-\lambda_i\right).
% \end{align}
Coefficients $R_{i,n}, Q_{i,n}, S_{i,n}$ are defined using equations \eqref{eq:RnQnSn}, and $N_{i,n}$ using \eqref{eq:N_reflected}.

% \section{delta-M scaling}
% To adjust the strong forward peak of the phase function, we implement the $\delta$-M scaling method \cite{wiscombe1977delta}. The optical parameters are adjusted as 
% \begin{align}
% 	w^* &= \frac{w(1-f)}{1-wf},\\
% 	\tau^* &= (1-wf)\tau,\\
% 	\chi_l^* &= \frac{\chi_l - f}{1-f},
% \end{align}
% where $f = g_0^L$ for the Henyey-Greenstein phase function.

\section{Analysis}
\label{sec:comparison}

\subsection{Liou Comparison}
\label{sec:liou}
\begin{table}[t!]
\centering
\begin{tabular}{|ccc|ccc||ccccccc|ccccccc|}
\hline
\multicolumn{3}{|c|}{} & & & & \multicolumn{7}{c|}{Reflection ($w_0=1$)} & \multicolumn{7}{c|}{Transmisssion ($w_0=1$)} \\
& & & & & & \multicolumn{7}{c|}{$\mu_0$} & \multicolumn{7}{c|}{$\mu_0$} \\
& $\tau$ &  & \multicolumn{3}{c||}{Method} & & 0.1 & & 0.5 & & 0.9 & & & 0.1 & & 0.5 & & 0.9 &  \\[0.25cm]
\hline

& \multirow{4}{*}{0.25} & & & Toon89 & & & 0.41133 & & 0.07635 & & -0.01294 & & & 0.58867 & & 0.92365 & & 1.01294 &  \\
& & & & SH2 & & & 0.41742 & & 0.09176 & & 0.00423 & & & 0.58258 & & 0.90824 & & 0.99577 &  \\
& & & & SH4 & & & 0.40745 & & 0.06746 & & 0.02331 & & & 0.59255 & & 0.93254 & & 0.97669 & \\
& & & & Doubling & & & 0.41610 & & 0.07179 & & 0.02250 & & & 0.58390 & & 0.92821 & & 0.97751 & \\[0.25cm]
    
& \multirow{4}{*}{1} & & & Toon89 & & & 0.51778 & & 0.07635 & & 0.02389 & & & 0.48222 & & 0.77441 & & 0.97610 & \\
& & & & SH2 & & & 0.51577 & & 0.09176 & & 0.05904 & & & 0.48423 & & 0.75109 & & 0.94096 & \\
& & & & SH4 & & & 0.55781 & & 0.06746 & & 0.09758 & & & 0.44218 & & 0.76739 & & 0.90242 & \\
& & & & Doubling & & & 0.58148 & & 0.07179 & & 0.09672 & & & 0.41852 & & 0.75952 & & 0.90328 & \\[0.25cm]
    
& \multirow{4}{*}{4} & & & Toon89  & & & 0.68564 & & 0.07635 & & 0.31611 & & & 0.31436 & & 0.50001 & & 0.68388 & \\
& & & & SH2      & & & 0.67143 & & 0.09176 & & 0.32974 & & & 0.32857 & & 0.50002 & & 0.67025 & \\
& & & & SH4      & & & 0.72698 & & 0.06746 & & 0.34904 & & & 0.27301 & & 0.48146 & & 0.65095 & \\
& & & & Doubling & & & 0.73254 & & 0.07179 & & 0.34823 & & & 0.26746 & & 0.48069 & & 0.65178 & \\[0.25cm]
    
& \multirow{4}{*}{16} & & & Toon89  & & & 0.86859 & & 0.07635 & & 0.71338 & & & 0.13140 & & 0.20899 & & 0.28659 & \\
& & & & SH2      & & & 0.85624 & & 0.09176 & & 0.70623 & & & 0.14374 & & 0.21874 & & 0.29373 & \\
& & & & SH4      & & & 0.87891 & & 0.06746 & & 0.70853 & & & 0.12107 & & 0.21351 & & 0.29143 & \\
& & & & Doubling & & & 0.88103 & & 0.07179 & & 0.70722 & & &  0.11897 & & 0.21342 & & 0.29279 & \\[0.5cm]
\hline

\multicolumn{3}{|c|}{} & & & & \multicolumn{7}{c|}{Reflection ($w_0=0.8$)} & \multicolumn{7}{c|}{Transmisssion ($w_0=0.8$)} \\
& & & & & & \multicolumn{7}{c|}{$\mu_0$} & \multicolumn{7}{c|}{$\mu_0$} \\
& $\tau$ &  & \multicolumn{3}{c||}{Method} & & 0.1 & & 0.5 & & 0.9 & & & 0.1 & & 0.5 & & 0.9 &  \\[0.25cm]
\hline

& \multirow{4}{*}{0.25} & & & Toon89  & & & 0.31802 & & 0.05739 & & -0.01125 & & & 0.46566 & & 0.84979 & & 0.95403 & \\
    & & & & SH2      & & & 0.31987 & & 0.06802 & & 0.00111 & & & 0.45934 & & 0.83725 & & 0.94049 & \\
    & & & & SH4      & & & 0.30146 & & 0.04662 & & 0.01676 & & & 0.45385 & & 0.85230 & & 0.92677 & \\
    & & & & Doubling & & & 0.28961 & & 0.04855 & & 0.01547 & & & 0.43017 & & 0.84756 & & 0.92669 & \\[0.25cm]
    
    & \multirow{4}{*}{1} & & & Toon89 & & & 0.37519 & & 0.05739 & & -0.00064 & & & 0.29023 & & 0.55267 & & 0.76333 & \\
    & & & & SH2      & & & 0.36348 & & 0.06802 & & 0.01710 & & & 0.28732 & & 0.53231 & & 0.73607 & \\
    & & & & SH4      & & & 0.36158 & & 0.04662 & & 0.05275 & & & 0.22624 & & 0.52033 & & 0.72010 & \\
    & & & & Doubling & & & 0.35487 & & 0.04855 & & 0.04929 & & & 0.20556 & & 0.51606 & & 0.71772 & \\[0.25cm]
    
    & \multirow{4}{*}{4} & & & Toon89  & & & 0.40411 & & 0.05739 & & 0.05152 & & & 0.06605 & & 0.12281 & & 0.20828 & \\
    & & & & SH2    & & & 0.38034 & & 0.06802 & & 0.05024 & & & 0.06584 & & 0.11729 & & 0.19690 & \\
    & & & & SH4      & & & 0.38125 & & 0.04662 & & 0.09327 & & & 0.04441 & & 0.10603 & & 0.22080 & \\
    & & & & Doubling & & & 0.37148 & & 0.04855 & & 0.08925 & & & 0.04539 & & 0.10718 & & 0.21953 &  \\[0.25cm]
    
    & \multirow{4}{*}{16} & & & Toon89 & & & 0.40571 & & 0.05739 & & 0.05636 & & & 0.00018 & & 0.00034 & & 0.00060 & \\
    & & & & SH2      & & & 0.38128 & & 0.06802 & & 0.05297 & & & 0.00018 & & 0.00033 & & 0.00057 & \\
    & & & & SH4      & & & 0.38204 & & 0.04662 & & 0.09700 & & & 0.00027 & & 0.00063 & & 0.00138 & \\
    & & & & Doubling & & & 0.37229 & & 0.04855 & & 0.09297 & & & 0.00027 & & 0.00062 & & 0.00139 & \\[0.25cm]
\hline
\end{tabular}
    \caption{Comparison of reflection and transmission as computed by Toon89, 2-term spherical harmonics (SH2), 4-term spherical harmonics (SH4) and by the doubling method for conservative and non-conservative scattering. The doubling method results are taken from \cite{liou1973numerical}.}
    \label{tab:liou}
\end{table}

\begin{figure*}[t!]
\centering
    \gridline{\fig{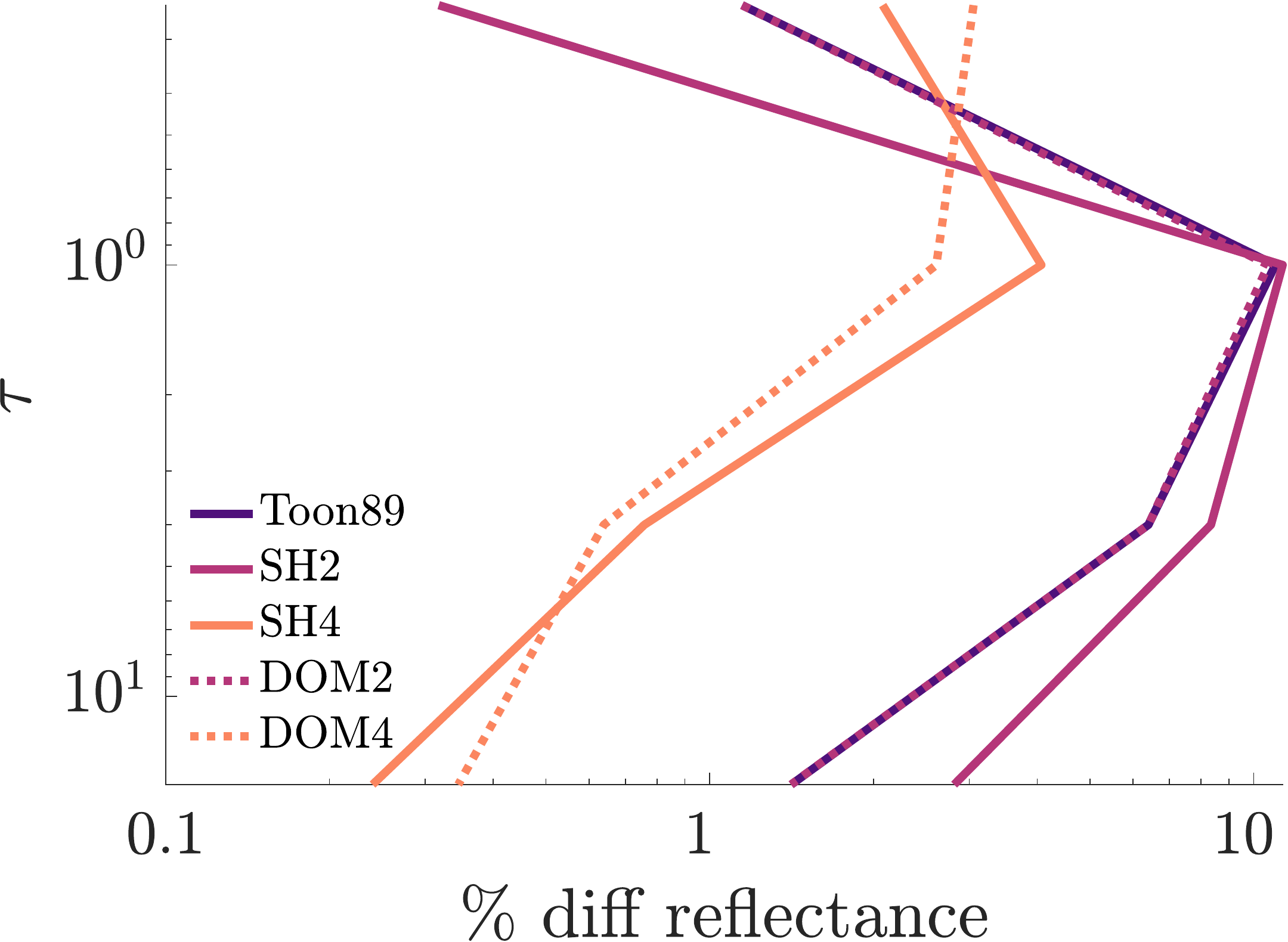}{.3\textwidth}{(a) Reflectivity for $w_0=1$, $\mu_0=0.1$.}
	\fig{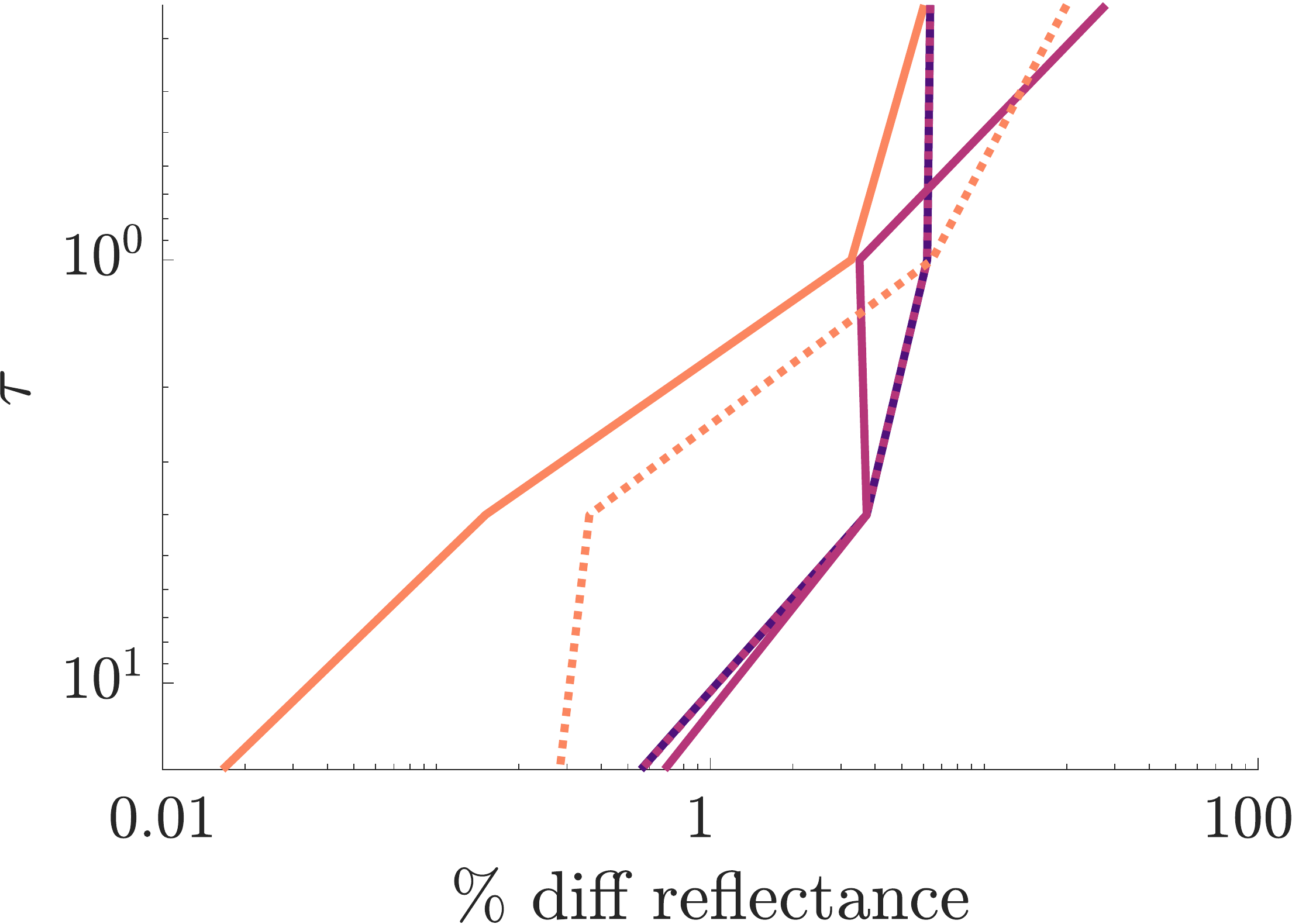}{.3\textwidth}{(b) Reflectivity for $w_0=1$, $\mu_0=0.5$.}
	\fig{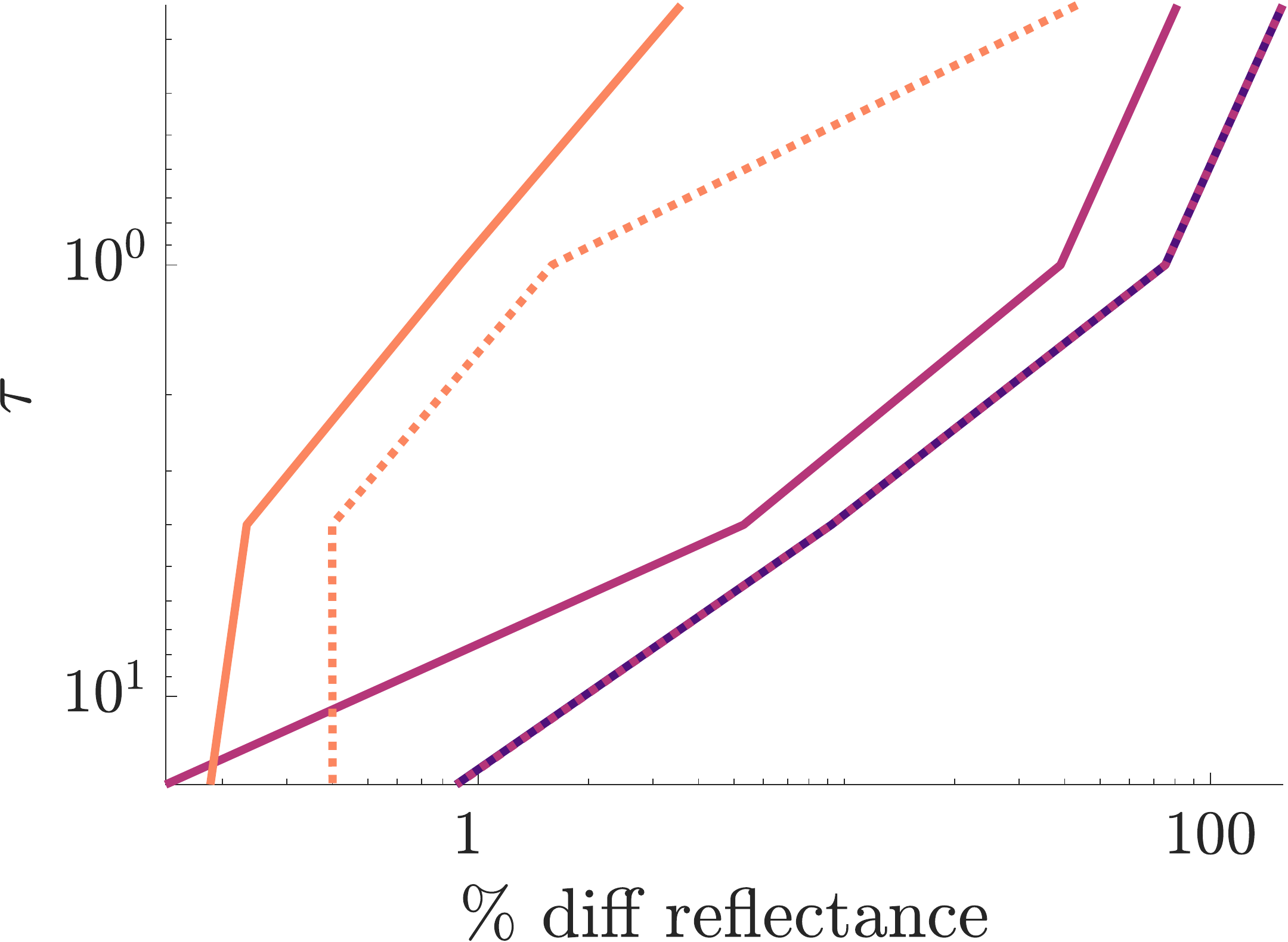}{.3\textwidth}{(c) Reflectivity for $w_0=1$, $\mu_0=0.9$.}}
	
	\gridline{\fig{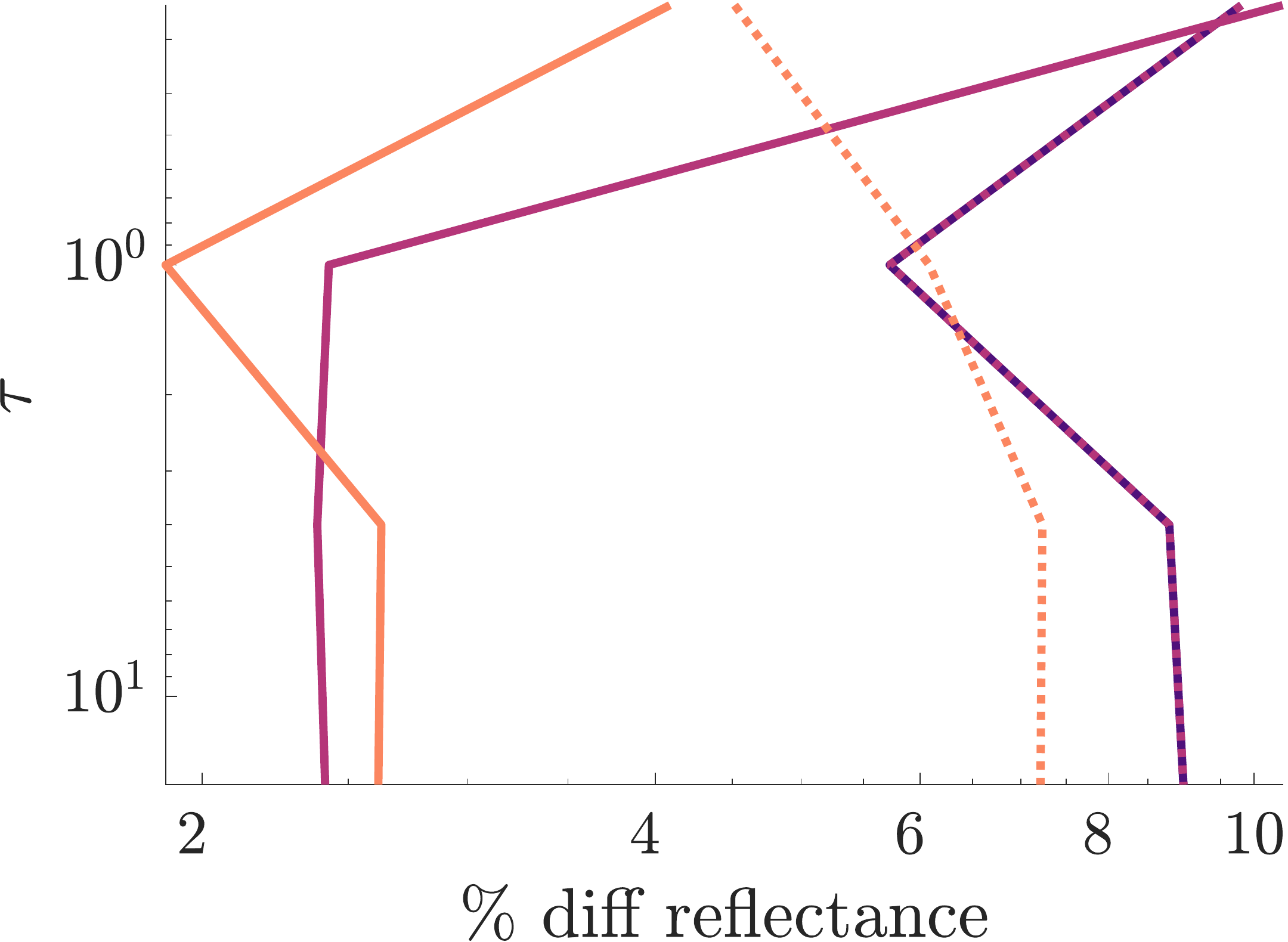}{.3\textwidth}{(d) Reflectivity for $w_0=0.8$, $\mu_0=0.1$.}
	\fig{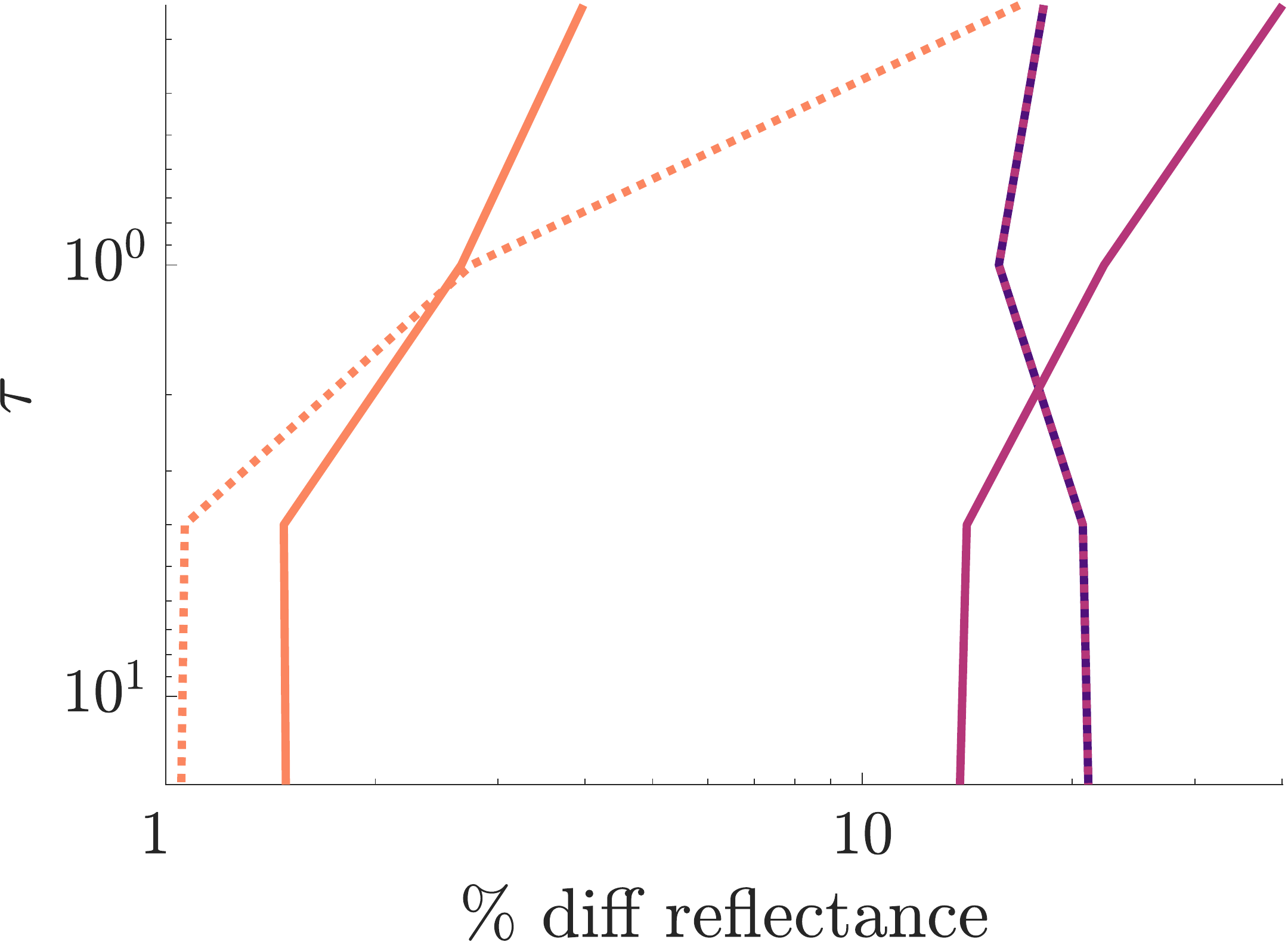}{.3\textwidth}{(e) Reflectivity for $w_0=0.8$, $\mu_0=0.5$.}
	\fig{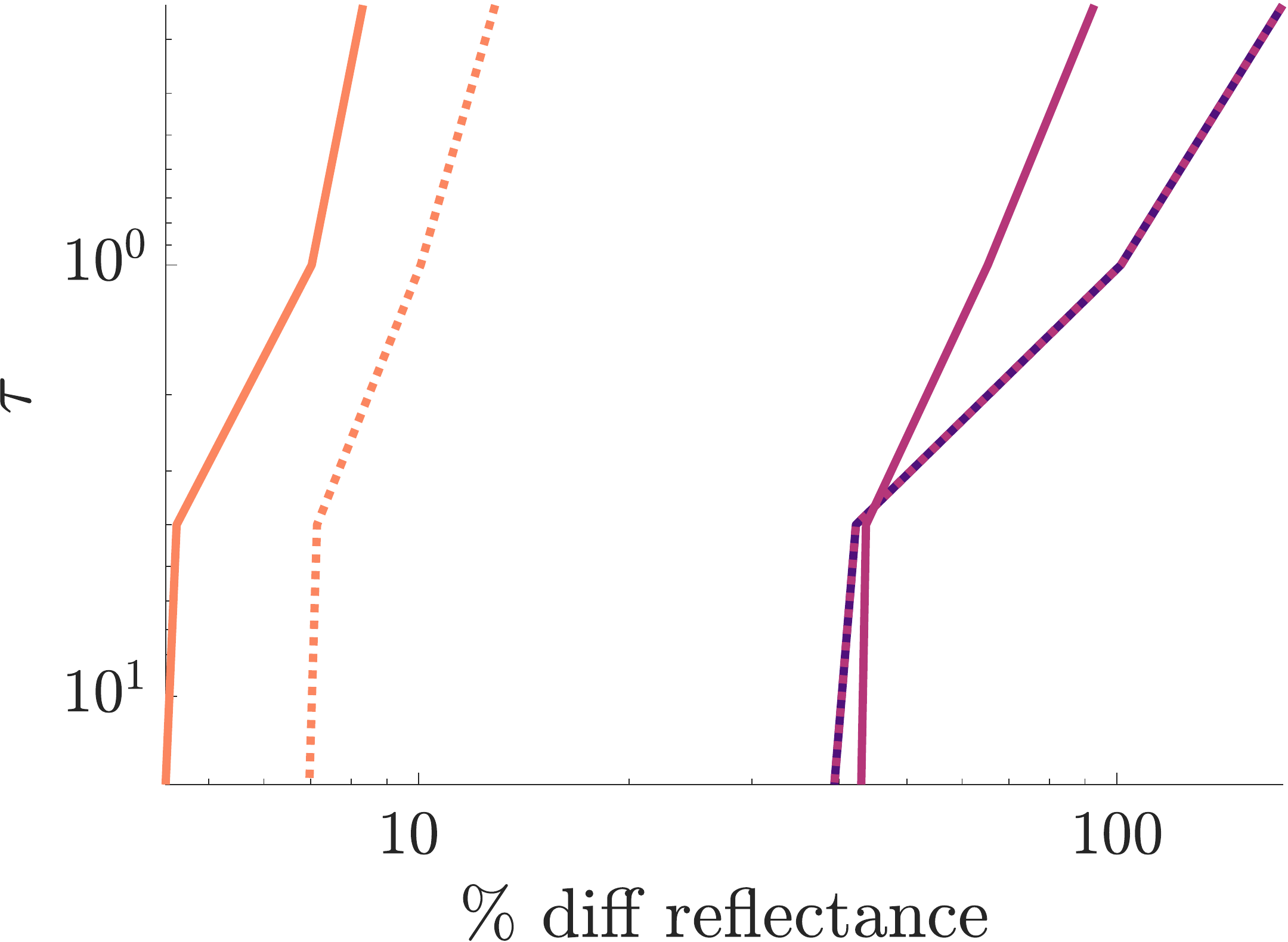}{.3\textwidth}{(f) Reflectivity for $w_0=0.8$, $\mu_0=0.9$.}}
	
	\gridline{\fig{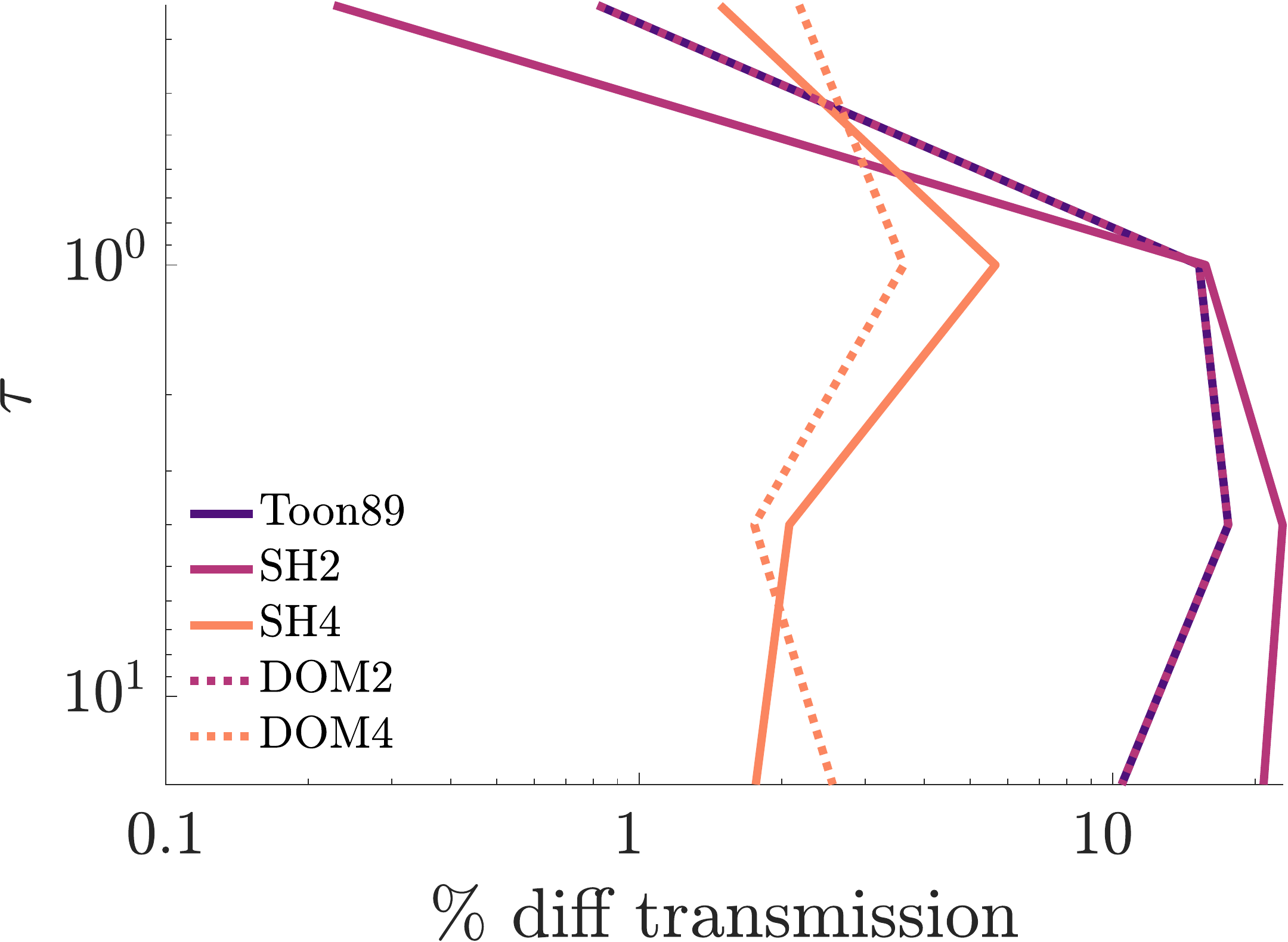}{.3\textwidth}{(g) Transmission for $w_0=1$, $\mu_0=0.1$.}
	\fig{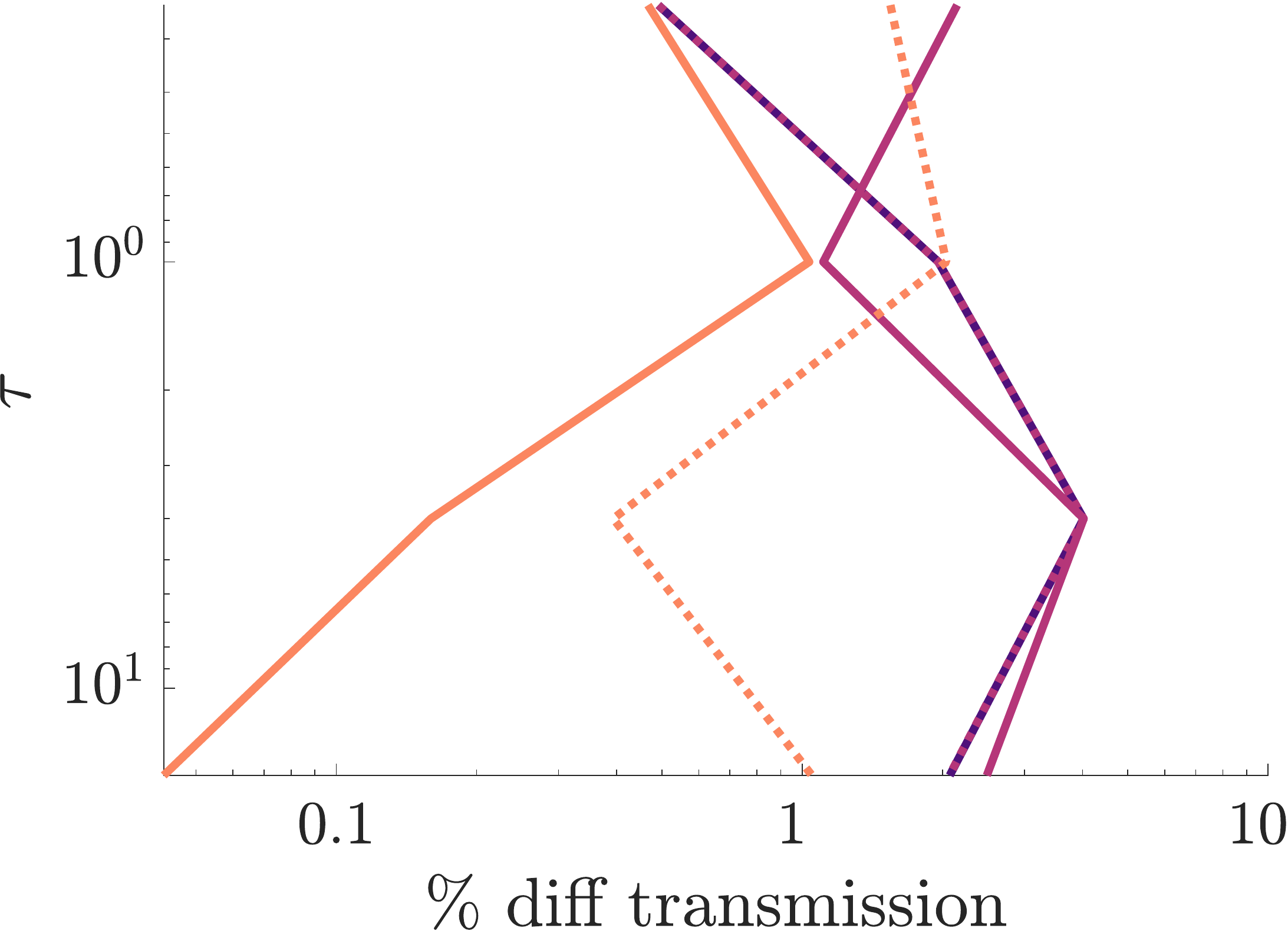}{.3\textwidth}{(h) Transmission for $w_0=1$, $\mu_0=0.5$.}
	\fig{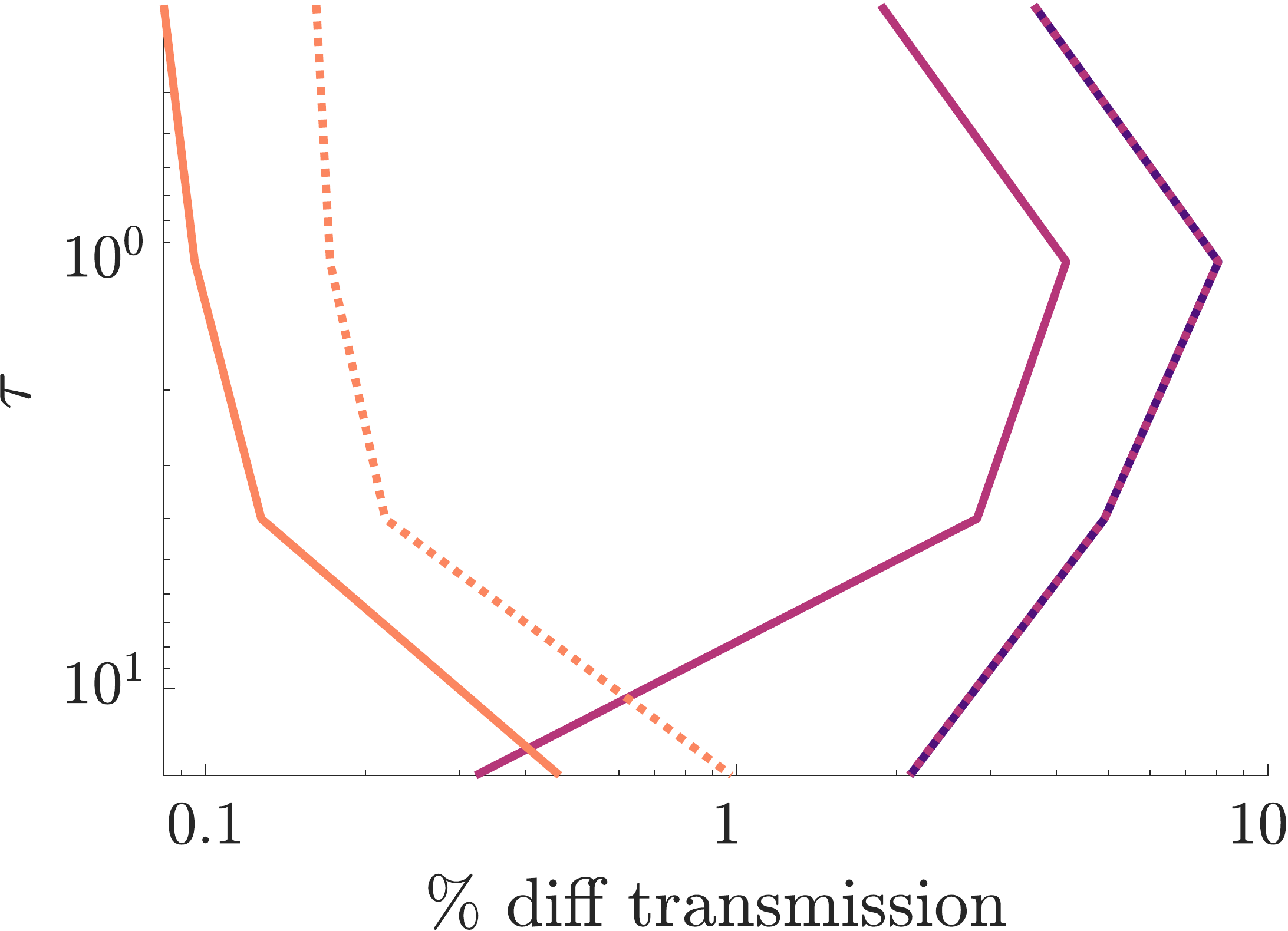}{.3\textwidth}{(i) Transmission for $w_0=1$, $\mu_0=0.9$.}}
	
	\gridline{\fig{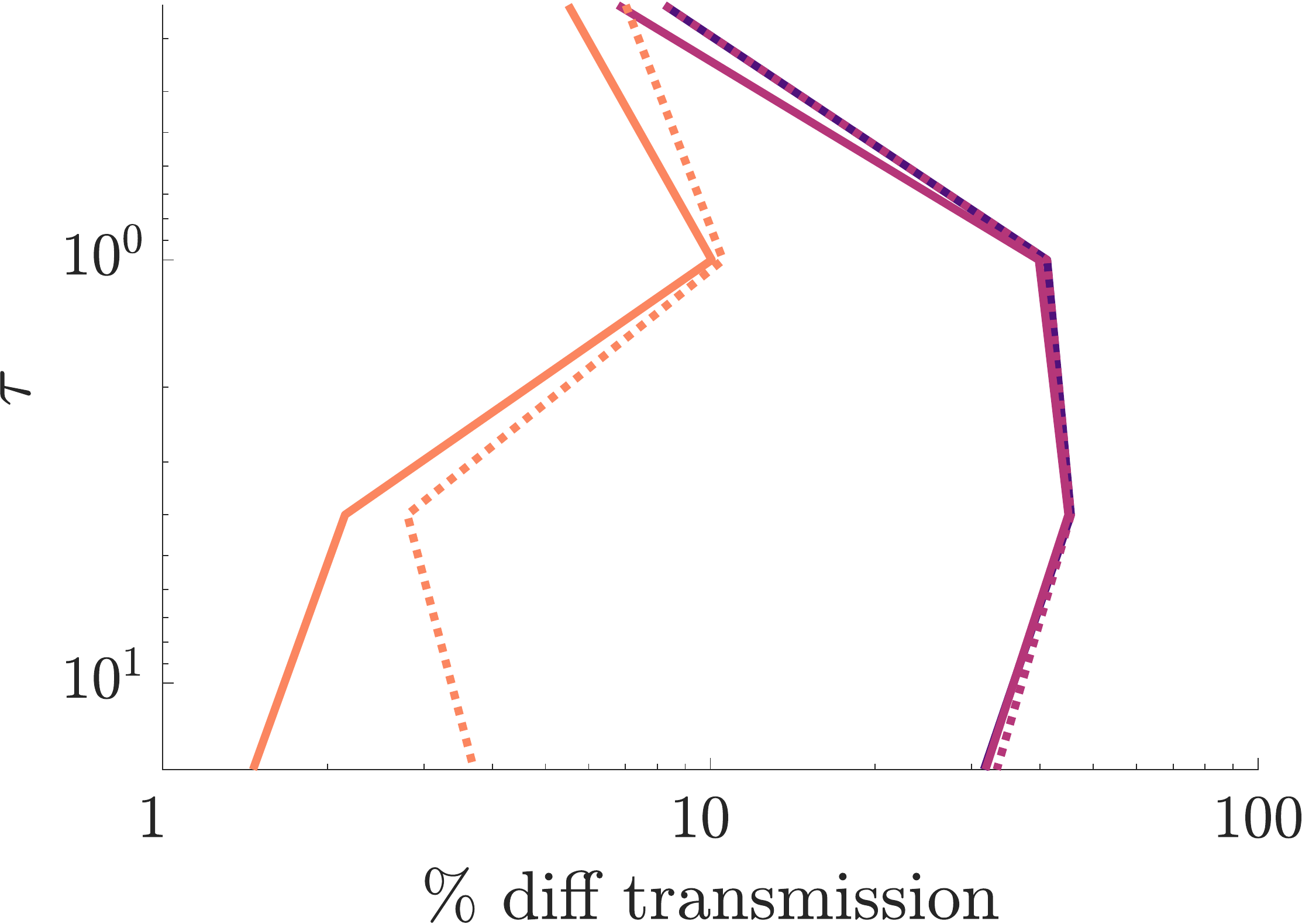}{.3\textwidth}{(j) Transmission for $w_0=0.8$, $\mu_0=0.1$.}
	\fig{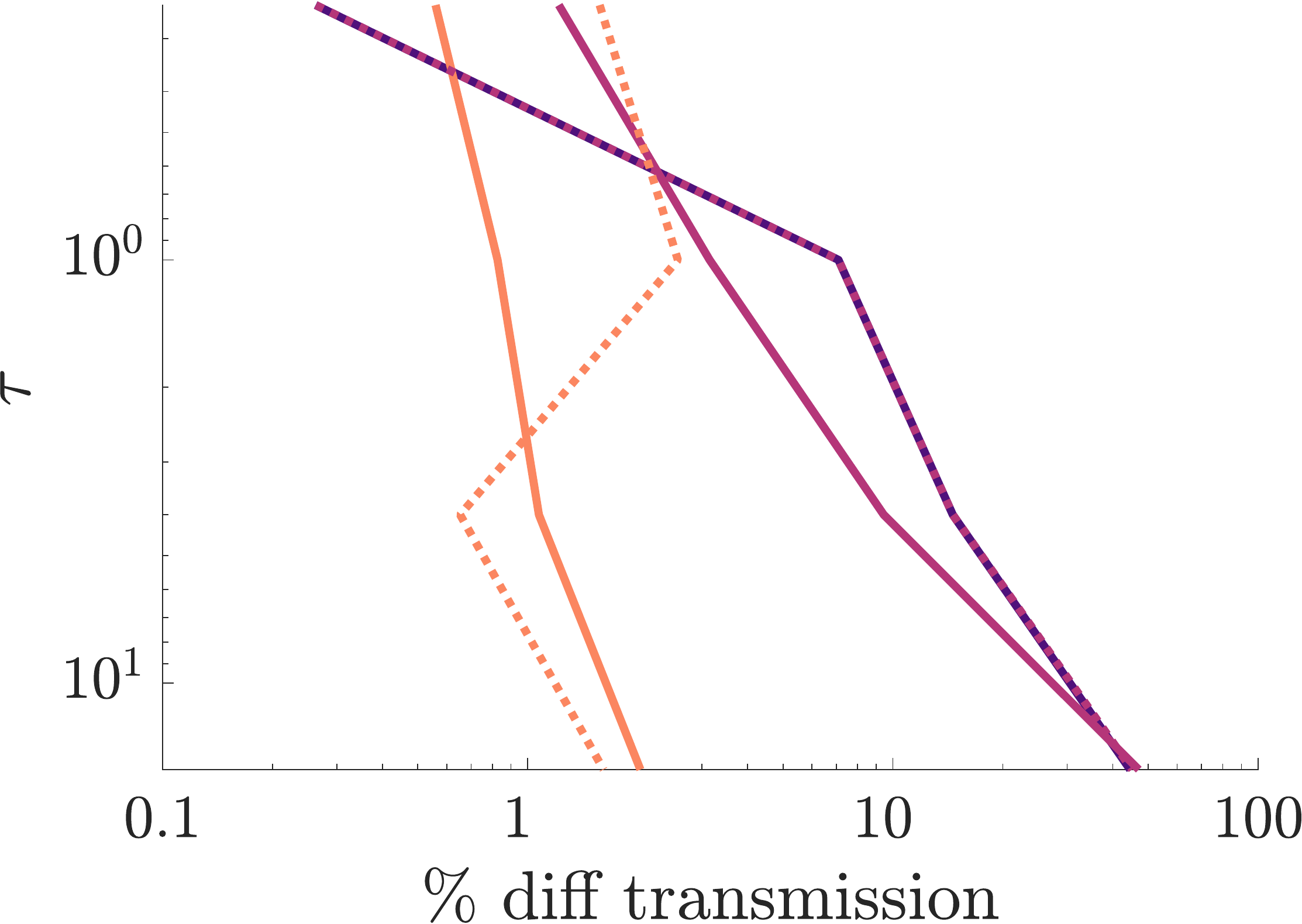}{.3\textwidth}{(k) Transmission for $w_0=0.8$, $\mu_0=0.5$.}
	\fig{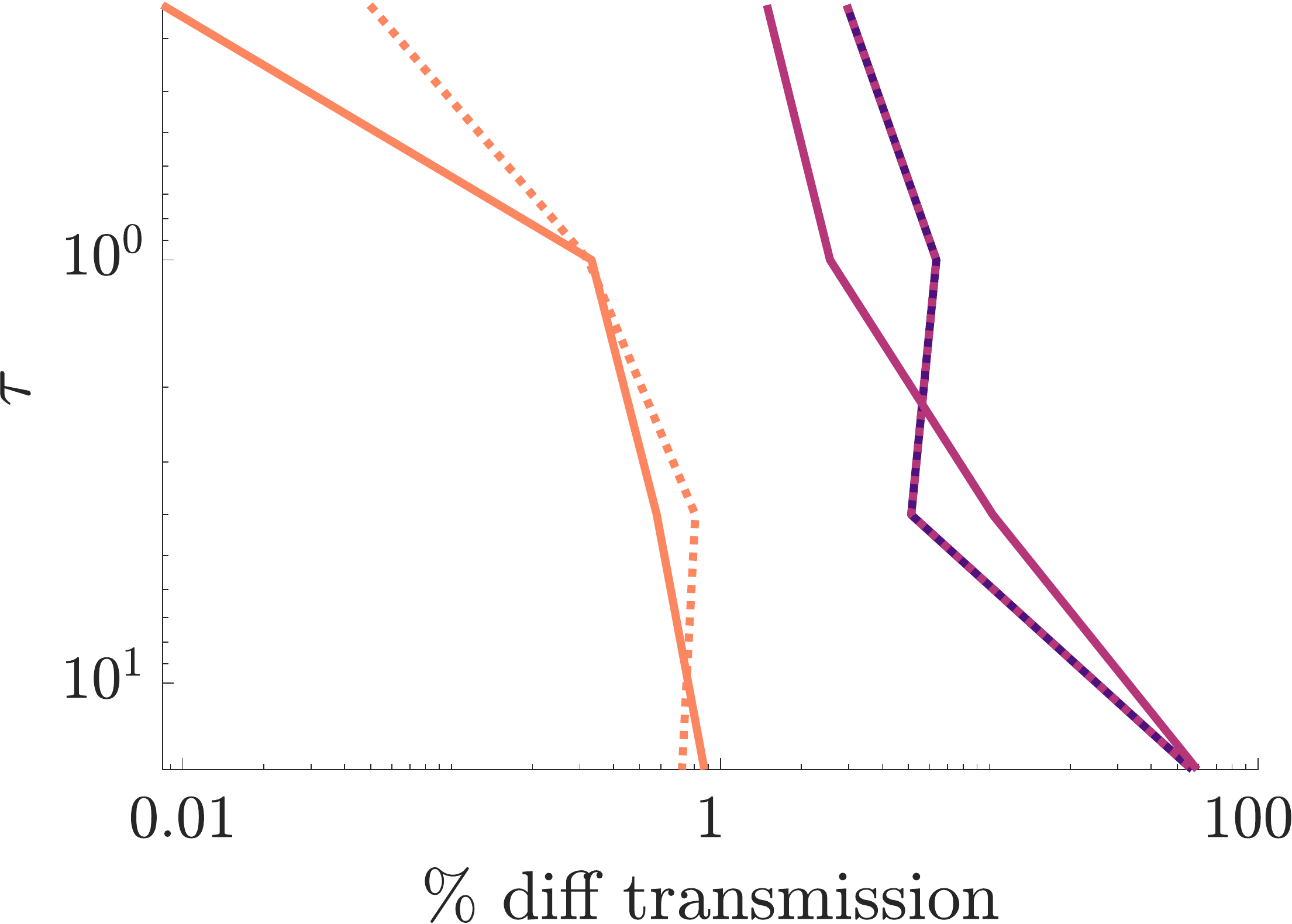}{.3\textwidth}{(l) Transmission for $w_0=0.8$, $\mu_0=0.9$.}}
	
	\caption{To illustrate model performance, we calculated the percentage difference between each model (Toon89, 2 and 4-term spherical harmonics [SH2, SH4], 2 and 4-stream discrete ordinates [DOM2, DOM4]) and the doubling method for reflection and transmission. The raw data for Toon89 and spherical harmonics is given in Table \ref{tab:liou} and the DOM and doubling method results are taken from \cite{liou1973numerical}, where the results for the doubling method are considered to be the ``true'' solution.}
	\label{fig:liou_comparison}
\end{figure*}

To quantitatively justify the effective performance of the spherical harmonics method for radiative transfer, we compare it with established methods in the literature.
The physical quantities of interest for radiative balance studies in planetary atmospheres are the vertical distributions of fluxes.
To study the capabilities of the spherical harmonics method in flux calculation for solar radiation, we repeat the analysis of \cite{liou1973numerical}, who numerically studied the discrete-ordinate method (DOM) for radiative transfer by comparing the reflection and transmission values obtained via DOM with those obtained using the doubling method.

The doubling method, first introduced by \cite{van1963new}, is widely considered one of the most accurate tools for multiple-scattering calculations \citep{liu2006advanced, zhang2013doubling, chou1992solar, li1996four,ayash2008implementing}.
In this approach, it is assumed that the reflection and transmission properties of a single, very thin, homogeneous layer are known.
The reflection and transmission properties of a layer twice as thick is then found by considering two adjacent layers with properties described above and summing the contributions from each layer \citep{van1980multiple}.
The reflection and transmission values for an arbitrarily thick layer is thus calculated by repeatedly doubling until arriving at the desired thickness.
Derived from the work of \cite{stokes1862iv}, the method computes the radiative properties of the medium instead of solving for the radiance field explicitly \citep{evans1991new}.
This allows for the easy computation of radiance exiting the atmosphere for many boundary conditions after the solution has been found. Although a simple concept compared to the discrete ordinates and spherical harmonics methods, the doubling method is very computational expensive due to the large iteration loop over the layers, therefore it is not optimal for retrievals or data assimilation. 
Instead, the doubling method is often used as an accurate and detailed benchmark for comparison with fast, approximate radiative transfer models.

We calculated the reflection and transmission values obtained by using the Toon89 implementation in \texttt{PICASO} \citep{natasha_batalha_2022_6419943} and 2-term and 4-term spherical harmonics (SH2 \& SH4). Following the analysis of \cite{liou1973numerical}, we consider single-layer atmospheres of optical thickness $\tau_N=0.25, 1, 4$ and $16$, where the scattering phase function is given by the Henyey-Greenstein phase function \eqref{eq:HG} with asymmetry factor $g_0=0.75$.
For single scattering albedos we choose values that represent the potential values for visible and near infrared of the solar spectrum for clouds and aerosols,  $w_0=1$ and $w_0=0.8$. Additionally, we choose to benchmark a range of cosine solar zenith angles $\mu_0=0.1,0.5$ and 0.9.  
We note that for comparison with the DOM and doubling methods discussed by \cite{liou1973numerical}, we are restricted to the same parameter values considered by the authors.
The reflection $r$, diffuse transmission $t_\text{dif}$ and direct transmission $t_\text{dir}$ for solar radiation are given by
\begin{align}
    r &= \frac{F^+(0)}{\pi\mu_0 F_0},\\
    t_\text{dif} &= \frac{F^-(\tau_N)}{\pi\mu_0 F_0},\\
    t_\text{dir} &= \exp(\tau_N/\mu_0),
\end{align}
The total transmission $t$ is equal to $t_\text{dif}+t_\text{dir}$.
Table \ref{tab:liou} compares the reflection and total transmission values for each model to those obtained using the doubling method which are taken from Table 1 in \cite{liou1973numerical}.

In Figure \ref{fig:liou_comparison} we plot the percentage difference between each model (Toon89, 2 and 4-term spherical harmonics [SH2, SH4], 2 and 4-stream discrete ordinates [DOM2, DOM4]) and the doubling method. The values for DOM2, DOM4 and the doubling method are taken from \cite{liou1973numerical}, where the results for the doubling method are considered to be the ``true'' solution.
We use orange lines for 4-stream models (SH4 and DOM4), pink for 2-stream models (SH2 and DOM2) with the exception of Toon89, which is plotted in purple.
The solid lines represent the results produced by the authors of this work (SH and Toon89) whereas the dashed lines represent those taken from \citep{liou1973numerical} (DOM).

Figure \ref{fig:liou_comparison} shows that the results produced by Toon89 and DOM2 are visibly identical ($<0.01$\%).
This is not surprising since Toon89 is built upon the \cite{toon1989rapid} 2-stream methodology which utilizes the quadrature approximation when deriving the solution coefficients.
The quadrature approximation is analogous to the 2-stream discrete-ordinates approach to solving the radiative transfer equation. 

In most cases considered, SH4 agrees more closely with the accurate doubling method than SH2, DOM2 and Toon89.
This is expected due to the higher-order approximation of the phase function.
The 4-term advantage is most evident for the $\mu_0=0.9$ tests (Figures \ref{fig:liou_comparison}(c), \ref{fig:liou_comparison}(f), \ref{fig:liou_comparison}(i) and \ref{fig:liou_comparison}(l)), where the percentage difference between the 2-stream methods and the doubling method is approximately two orders of magnitude greater than that of the 4-stream methods.

However, in a few cases, particularly for small cosine zenith angle $\mu_0=0.1$, the 2-stream methods incur a lower percentage difference with the doubling method than their 4-stream counterparts for optically thin layers. 
For example, both the reflection and transmission values produced by SH2 in the $w_0=1,\mu_0=0.1$ case (Figures \ref{fig:liou_comparison}(a)\&(g)), agree more closely than SH4 with the doubling benchmark for optical depth less than around 0.6.
It is important to note, however, that in this case, the 4-stream DOM method exhibits similar behaviour to SH4 and also suffers from worse agreement with the doubling method than its 2-stream counterpart, indicating that this behaviour in the optical thin region is not due to the choice of modelling method (SH versus DOM) but rather that 4-stream techniques are still limited in the improvement that can be achieved over 2-stream techniques in particular cases. 

In general, in the regions where SH2 exceeds the accuracy of the SH4, such as in the case discussed above, both models remain within a 5\% difference with the doubling method.
Therefore, even though SH4 might not be the ``optimal'' model in these regions, it is still exhibiting excellent agreement with the benchmark model.
Notably, when considering the full range of parameters considered in this analysis, the SH4 values for reflectivity and transmission are consistently within 10\% of the doubling method, whereas SH2 can reach differences of close to 100\%.
This can be clearly realised in Figure \ref{fig:liou_comparison}(f). 

Furthermore, SH4 performs comparatively to, if not better than, DOM4.
We notice that in cases such as Figures \ref{fig:liou_comparison}(c), for $w_0=1,\mu_0=0.9$ the maximum percentage difference of both SH4 and DOM4 with the doubling method occurs at the optically thin region at the top of the atmosphere.
However, SH4 observes a percentage difference of approximately 4\% versus a 40\% difference for DOM4.
Similar discrepancies are observed for other cases throughout the analysis.
\cite{liou1973numerical} concluded that the 4-stream DOM method may be of adequate accuracy for studies of the flux distribution in the transfer of solar irradiance through cloudy atmospheres.
We thus extend this conclusion to the 4-term spherical harmonics method.

\subsection{\texttt{CDISORT} Comparison}
\label{sec:cdisort}
The discrete ordinate solver, DISORT \citep{stamnes1988numerically,stamnes2000disort} is a versatile, well-tested and one of the most widely used one-dimensional radiative transfer solvers. 
The numerical capabilities of DISORT extend to $N$-stream discrete ordinates approximations, where $N$ is arbitrary and considerably greater than 4 (typically DISORT is run for 32 streams).
DISORT was originally written in FORTRAN \citep{stamnes2000disort}.
T. Dowling rewrote the code in C, which we will refer to as \texttt{CDISORT} \citep{mayer2005libradtran, buras2011new}.

To further study the efficacy of the spherical harmonics method of solving the azimuthally-averaged radiative transfer equation \eqref{eq:RTE}, we benchmark our   model outputs against that of \texttt{CDISORT}.

We consider a test atmosphere that is comprised of 30 layers with  cumulative optical depth $\tau=10^{-2}-10^2$, with constant values for the single scattering albedo $w_0$ and asymmetry parameter $g_0$, and use the Henyey-Greenstein phase function \eqref{eq:HG} to describe scattering.
We consider incident solar radiation $\Fo=1$ with a cosine solar zenith angle of $\mu_0$.

% where the scattering angle $\Theta$ is defined as
% \begin{align}
% 	\cos\Theta &= \mu\mu' - \sqrt{1-\mu^2}\sqrt{1-\mu'^2}\cos(\phi-\phi').
% 	\label{eq:costheta}
% \end{align}
% The azimuthally-averaged Henyey-Greenstein function is given by 
% \begin{equation}
% 	\mathcal{P}(\mu,\mu') = \sum_{l=0}^N \chi_l P_l(\mu)P_l(\mu'),
% 	\label{eq:leg_azimuthally_averaged}
% \end{equation}
% for moments \citep{henyey1941diffuse}
% \begin{equation}
%     \chi_l = (2l+1)g_0^l.
% \end{equation}

\subsubsection*{Azimuthally-averaged intensity with asymmetry}
\begin{figure*}[t!]
\centering
    \gridline{\fig{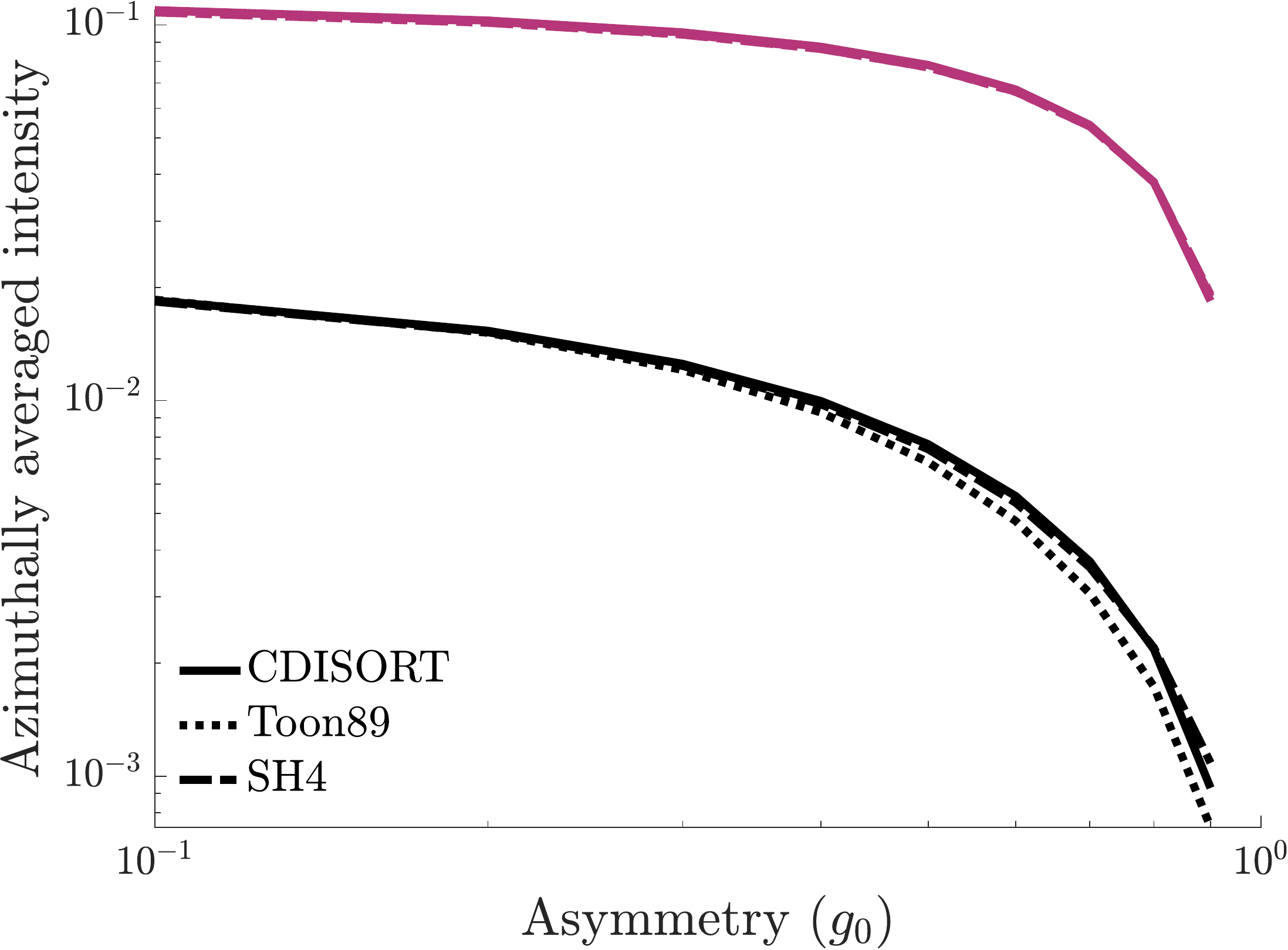}{.45\textwidth}{(a) $\mu=\mu_0=0.9$.}
	\fig{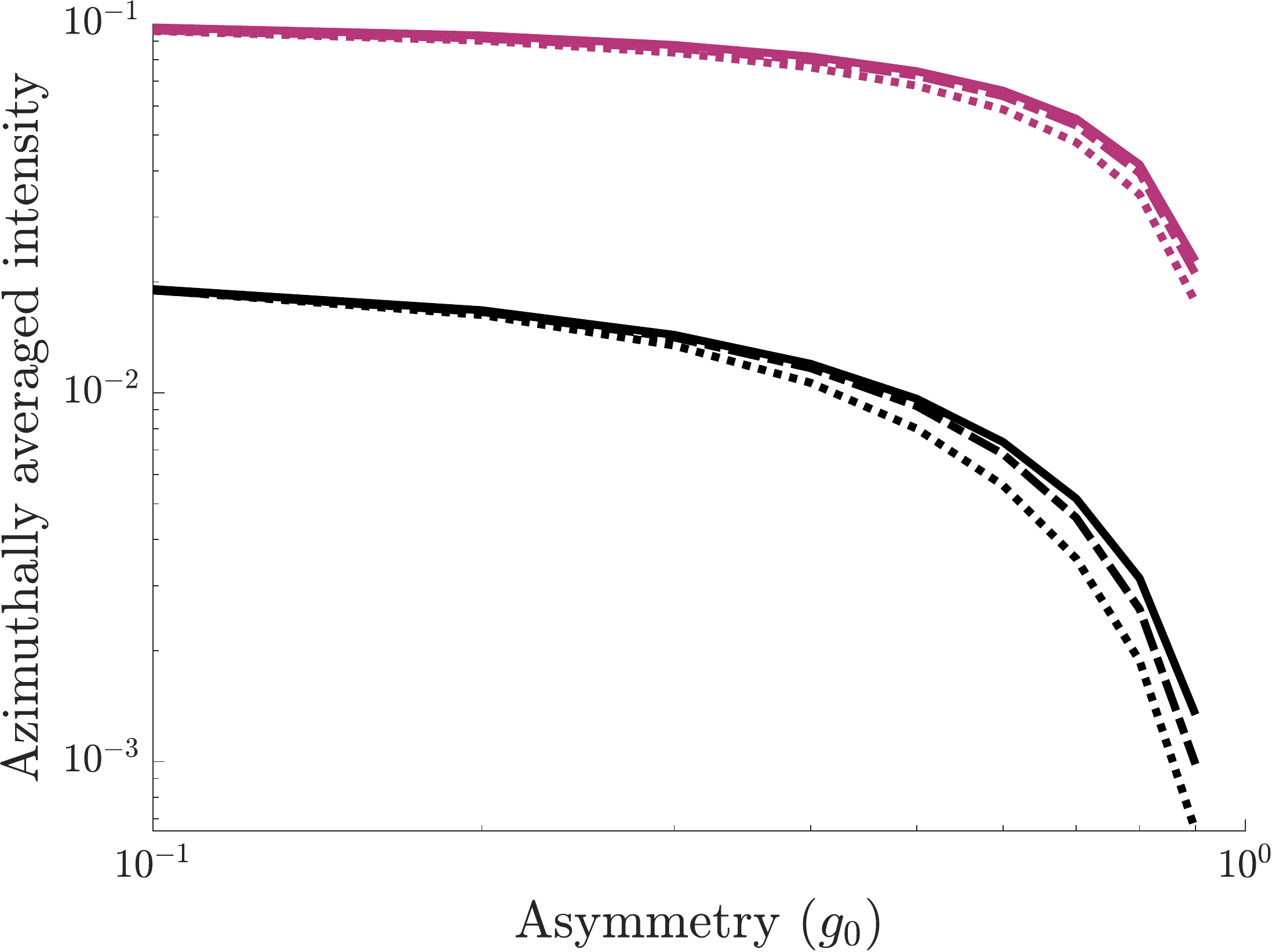}{.45\textwidth}{(b) $\mu=\mu_0=0.7$.}}
	
	\gridline{\fig{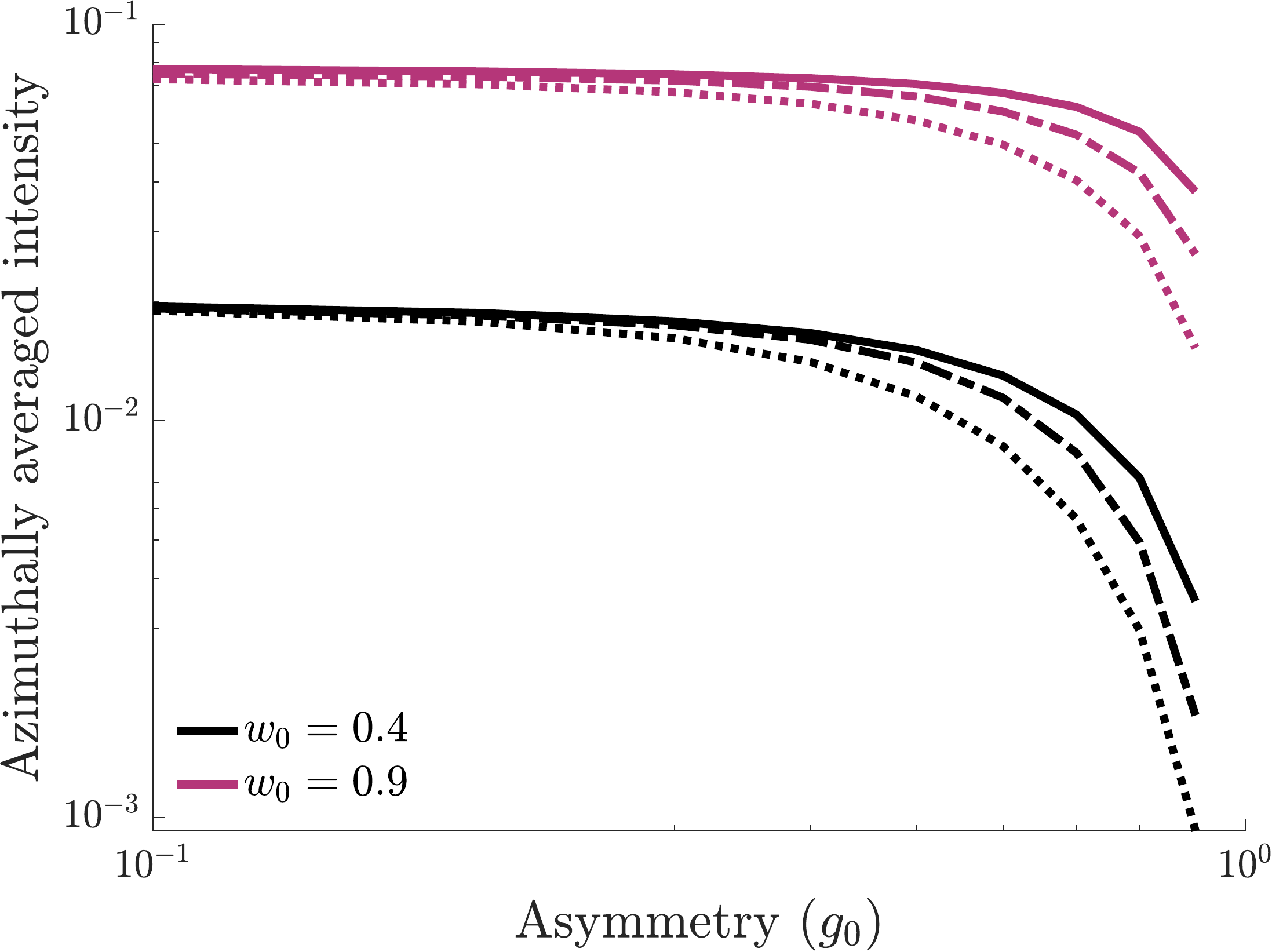}{.45\textwidth}{(c) $\mu=\mu_0=0.4$.}
	\fig{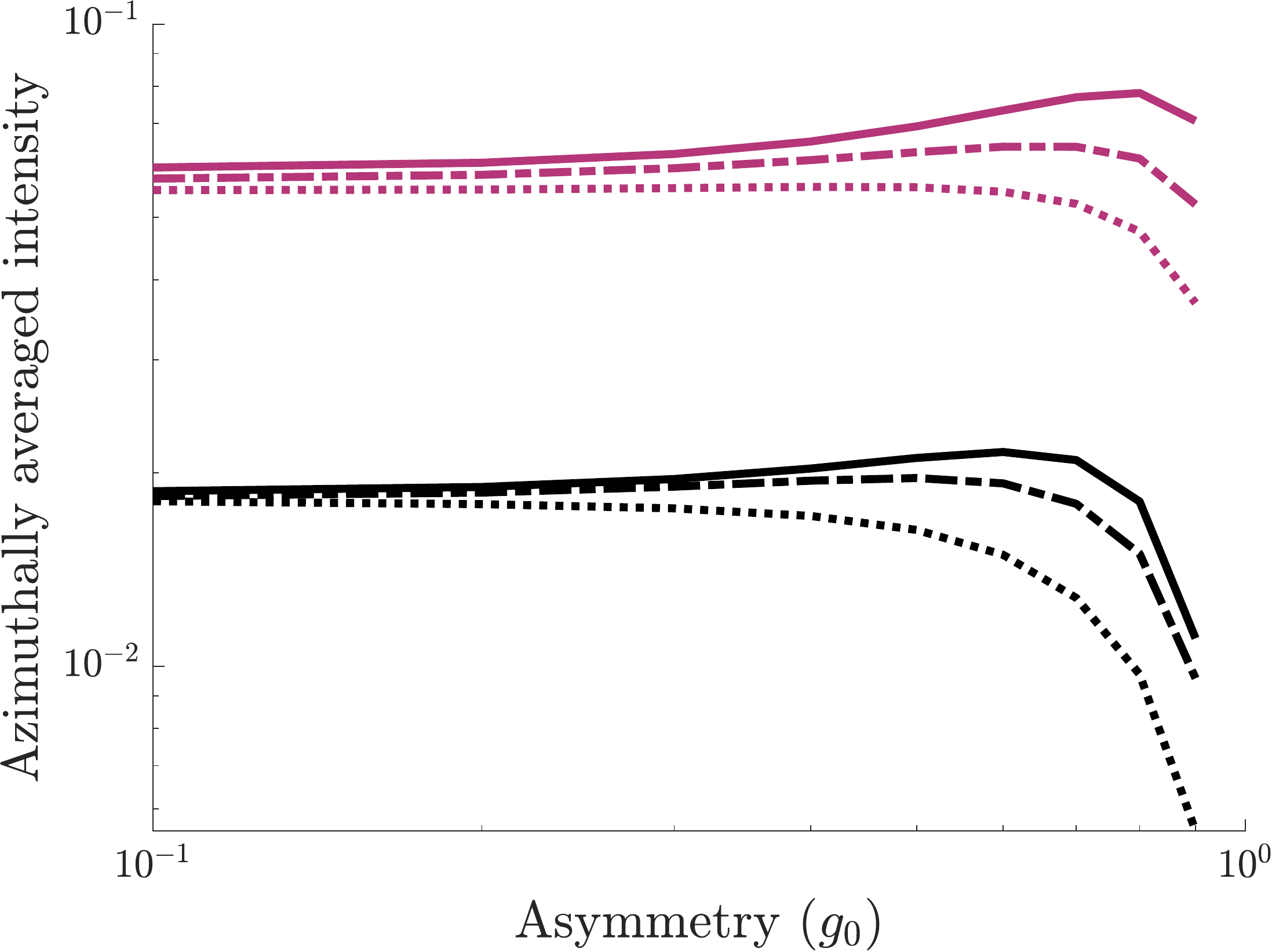}{.45\textwidth}{(d) $\mu=\mu_0=0.2$.}}
	
	\caption{Azimuthally-averaged intensity emergent from the top of the atmosphere with asymmetry $g_0$ for a variety of zenith angles $\mu,\mu_0$ and single scattering albedo $w_0=0.4,0.9$. We compare \texttt{CDISORT}, Toon89 and SH4 and notice that the agreement worsens for decreasing cosine zenith angle.}
	\label{fig:xint_with_g0}
\end{figure*}
In Figure \ref{fig:xint_with_g0} we plot the azimuthally-averaged intensity emergent from the top of the atmosphere $I(0,\mu)$ against the asymmetry parameter $g_0$ for 32-stream \texttt{CDISORT}, Toon89 and 4-term spherical harmonics (SH4).
This quantity is important as it is used to calculate the geometric albedo of the planet.
Each panel of Figure \ref{fig:xint_with_g0} represents a different cosine zenith angle, namely $\mu=0.9,0.7,0.4,0.2$.
For each case, we have chosen to set $\mu$ to be equal to the cosine solar zenith angle $\mu_0$.
We also consider two single scattering albedos $w_0=0.4$ and $w_0=0.9$ in each plot, similar to our doubling method benchmarking exercise against \citet{liou1973numerical}.
In each plot, we notice that SH4 more closely matches 32-stream \texttt{CDISORT} than Toon89 does, especially towards high asymmetry. However, overall the agreement between all models worsens as $g_0$ increases. 
This is as expected because low-order approximations are inadequate representations of highly asymmetric phase functions.
We also notice that the digression of the solutions for high asymmetry is more pronounced for smaller values of $\mu$ and $\mu_0$.
This is similar to the behaviour observed in the \cite{liou1973numerical} analysis in Section \ref{sec:liou}.

\subsubsection*{Azimuthally-averaged intensity with optical depth}

\begin{figure*}[t!]
\centering
    \gridline{\fig{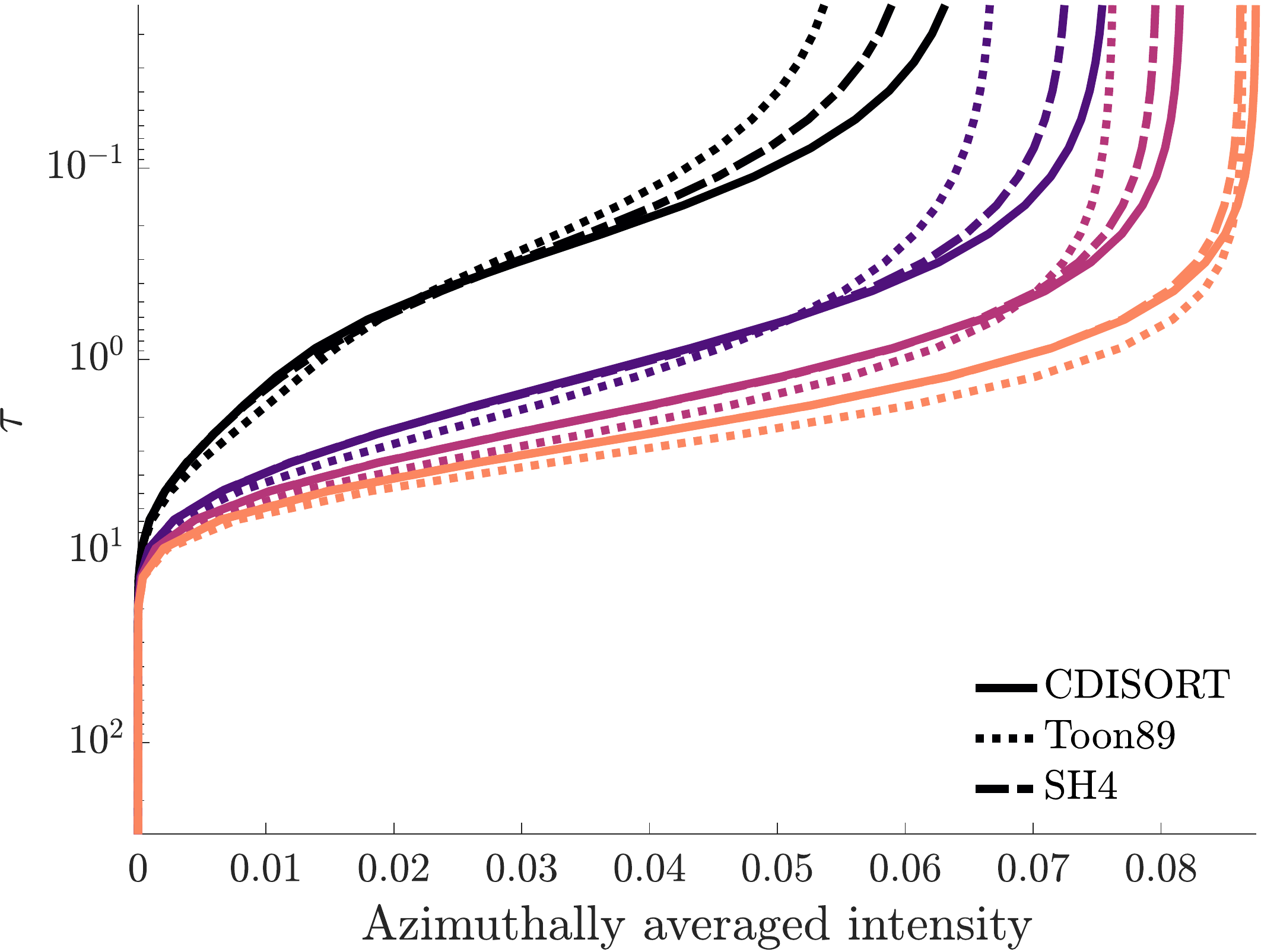}{.45\textwidth}{(a) $g_0=0.4$.}
	\fig{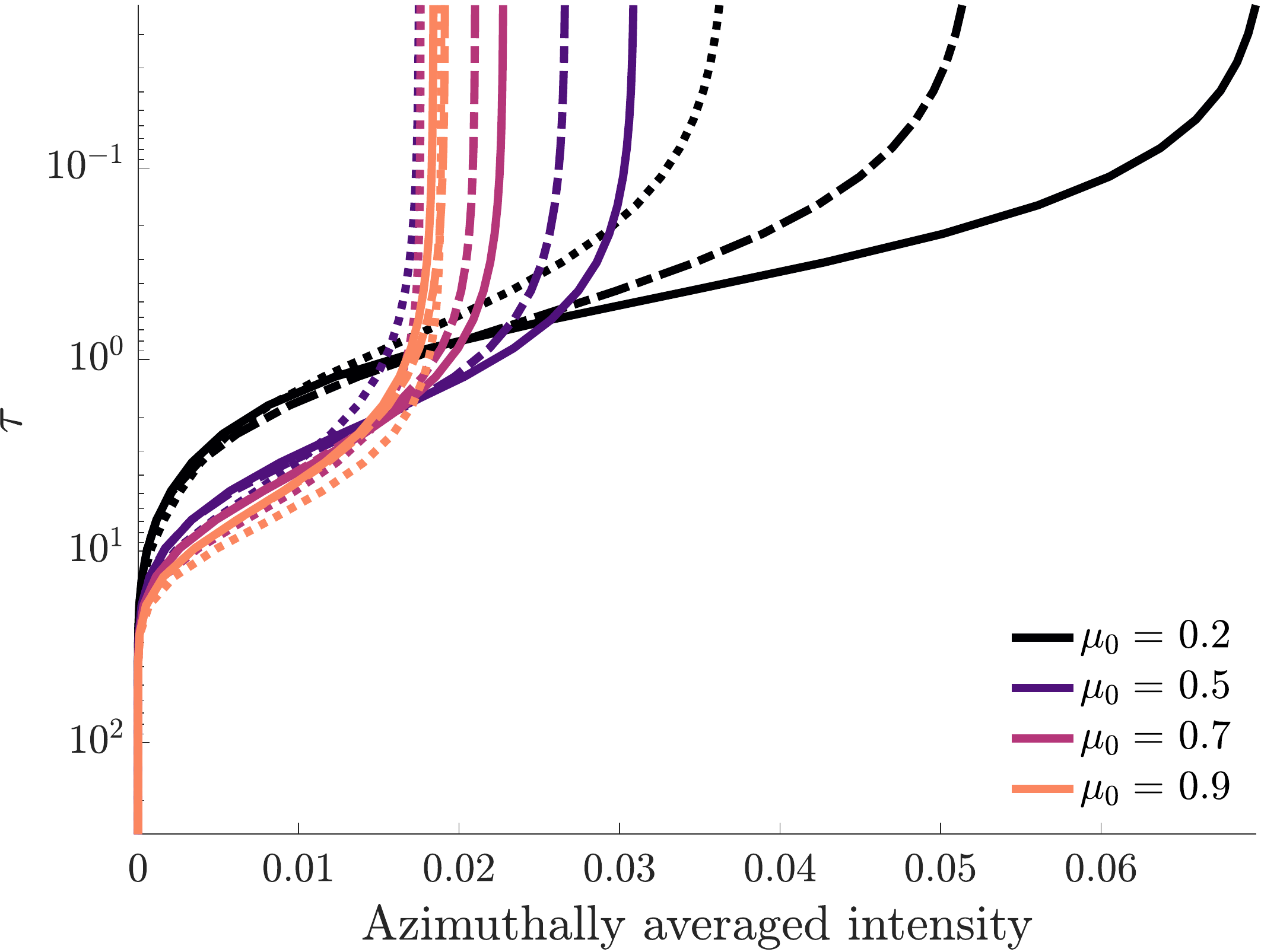}{.45\textwidth}{(b) $g_0=0.9$.}}
	
	\caption{Azimuthally-averaged intensity emergent from the top of the atmosphere with optical depth $\tau$ for a variety of zenith angles $\mu=\mu_0=0.2,0.4,0.7,0.9$ and asymmetry $g_0=0.4,0.9$. The single scattering albedo is $w_0=0.9$. We compare \texttt{CDISORT}, Toon89 and SH4 and notice that the agreement worsens for decreasing zenith angle and increasing $g_0$.}
	\label{fig:xint_with_opd}
\end{figure*}

In Figure \ref{fig:xint_with_opd} we study how the azimuthally-averaged intensity $I(\tau,\mu)$ varies with optical depth $\tau$ for each of the models.
For this analysis, we set $w_0=0.9$ and consider two values of asymmetry, namely $g_0=0.4$ and $g_0=0.9$.
We notice that the agreement between all three models is greater for low asymmetry ($g_0=0.4$) -- the models are more disparate for the high asymmetry value of $g_0=0.9$.
Similar to the analysis of Figure \ref{fig:xint_with_g0},  the agreement between the solutions worsens as $\mu=\mu_0$ decreases. This effect is consistent with what was observed in the Liou analysis in Section \ref{sec:liou}.
However, despite its limitations in high asymmetry and low cosine zenith angles, SH4 does offer a significant improvement in agreement with \texttt{CDISORT} over Toon89 for all cases.

\subsubsection*{Layer-wise vertical fluxes}

\begin{figure*}[h!]
\centering
    \gridline{\fig{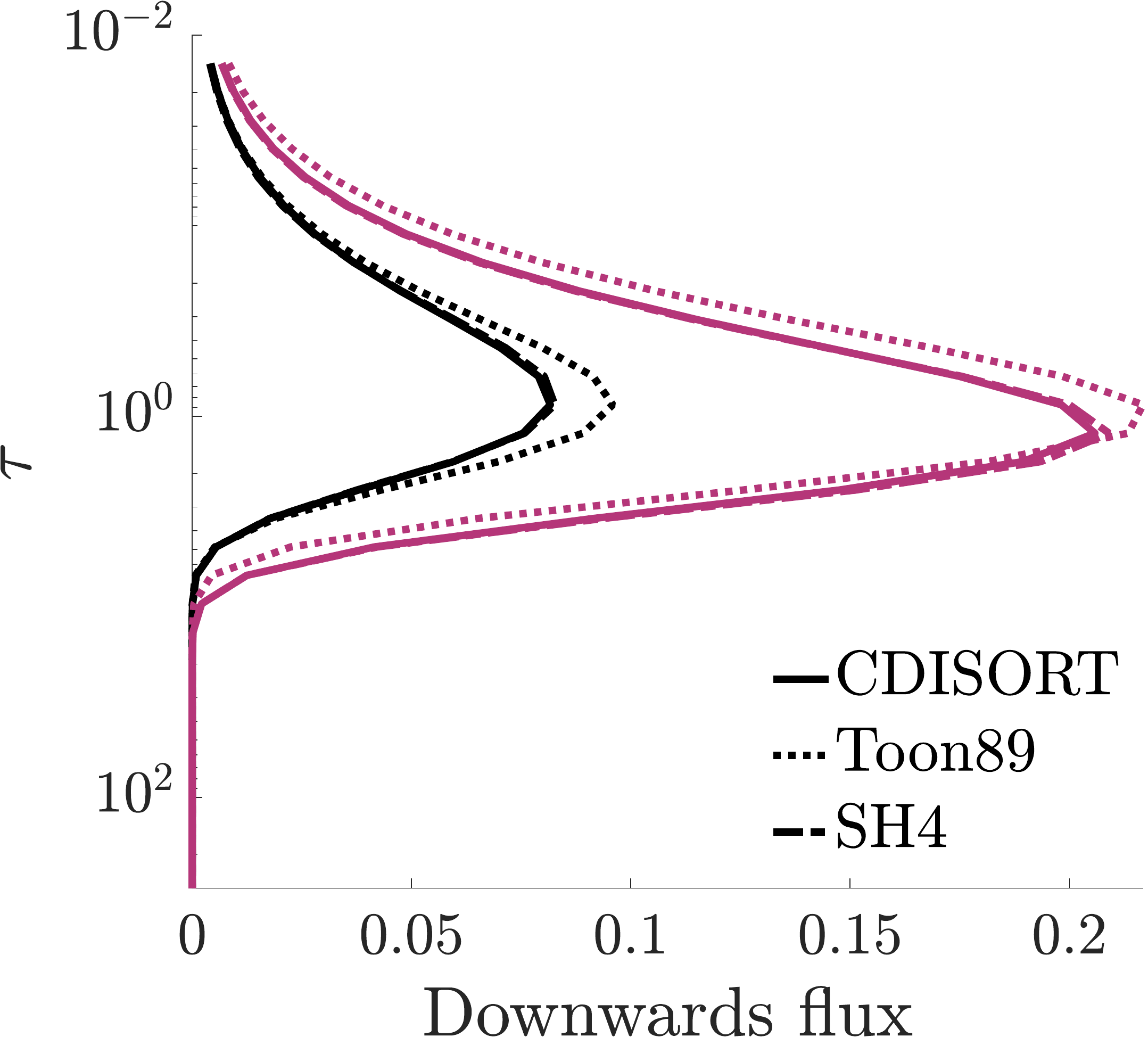}{.31\textwidth}{}
	\fig{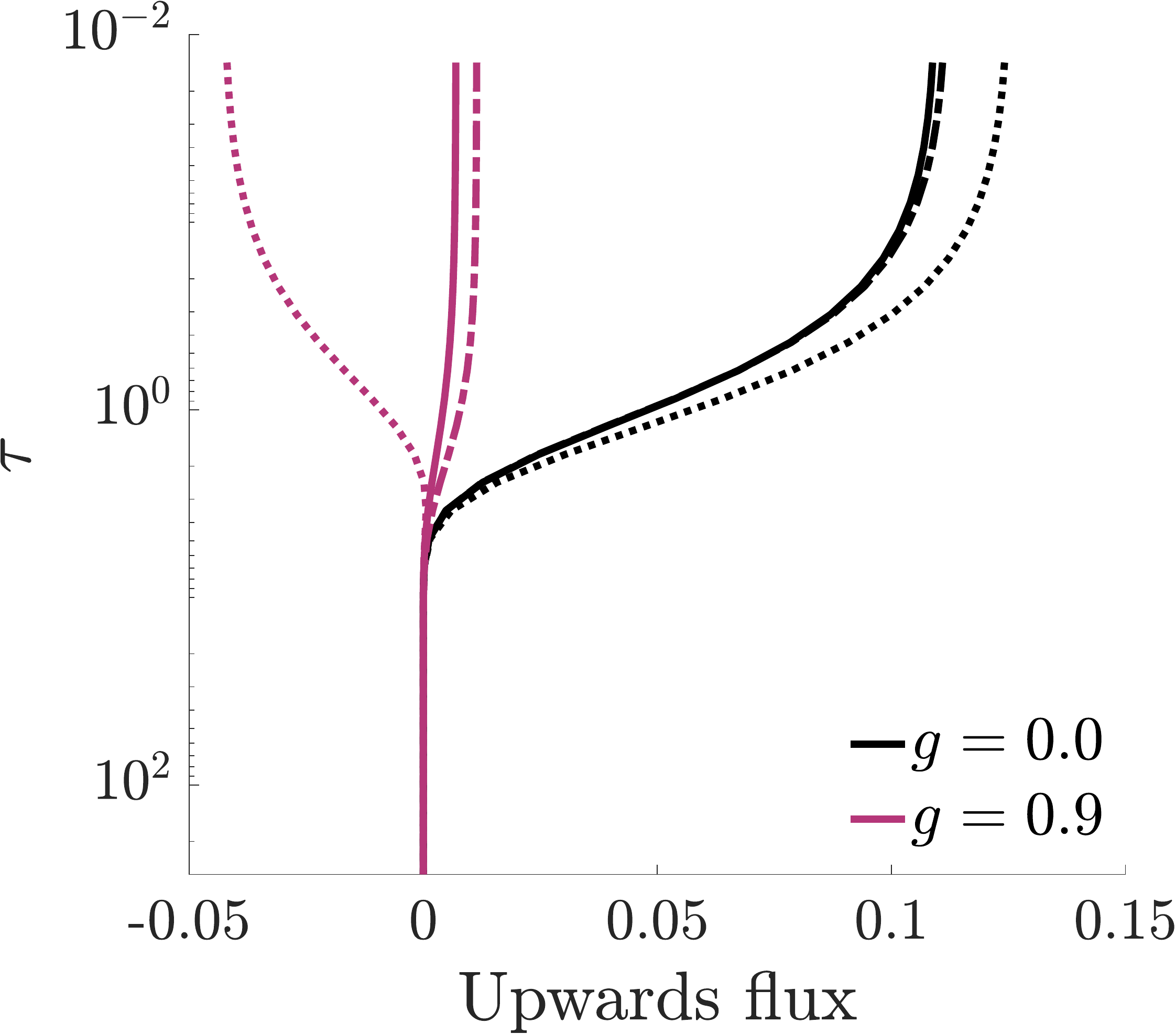}{.31\textwidth}{(a) $\mu_0=0.9$.}
	\fig{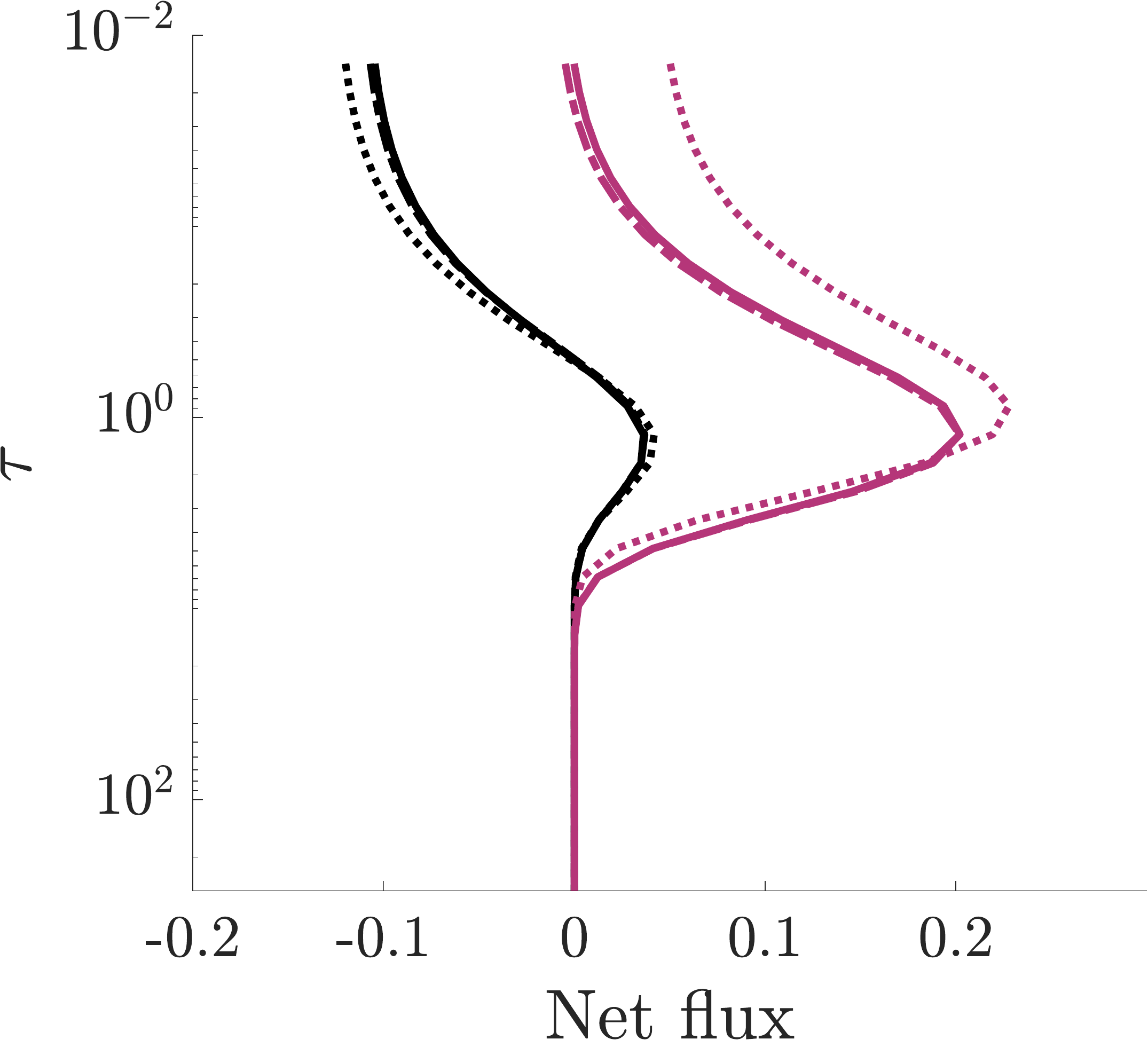}{.31\textwidth}{}}
	
	\gridline{\fig{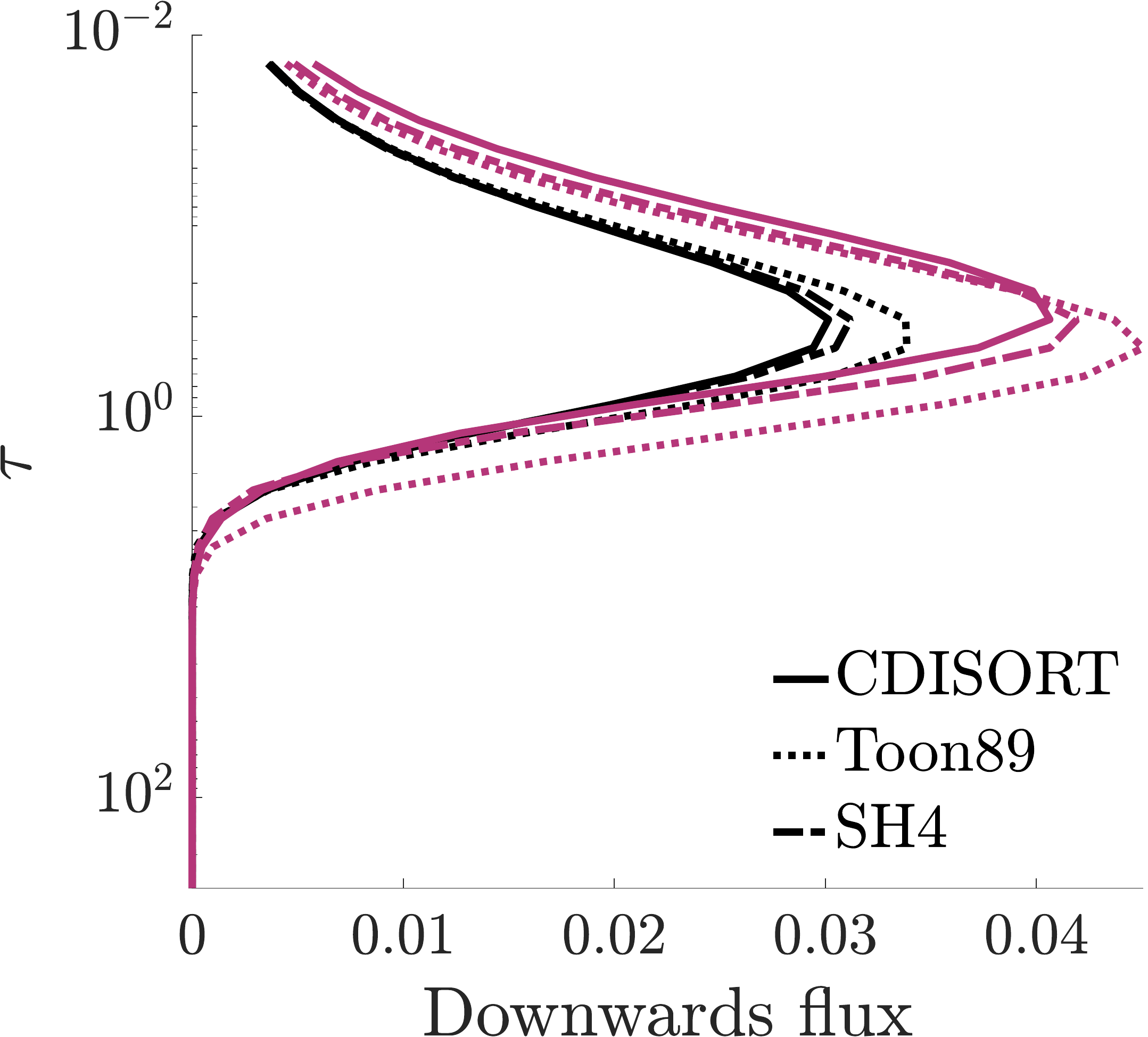}{.31\textwidth}{}
	\fig{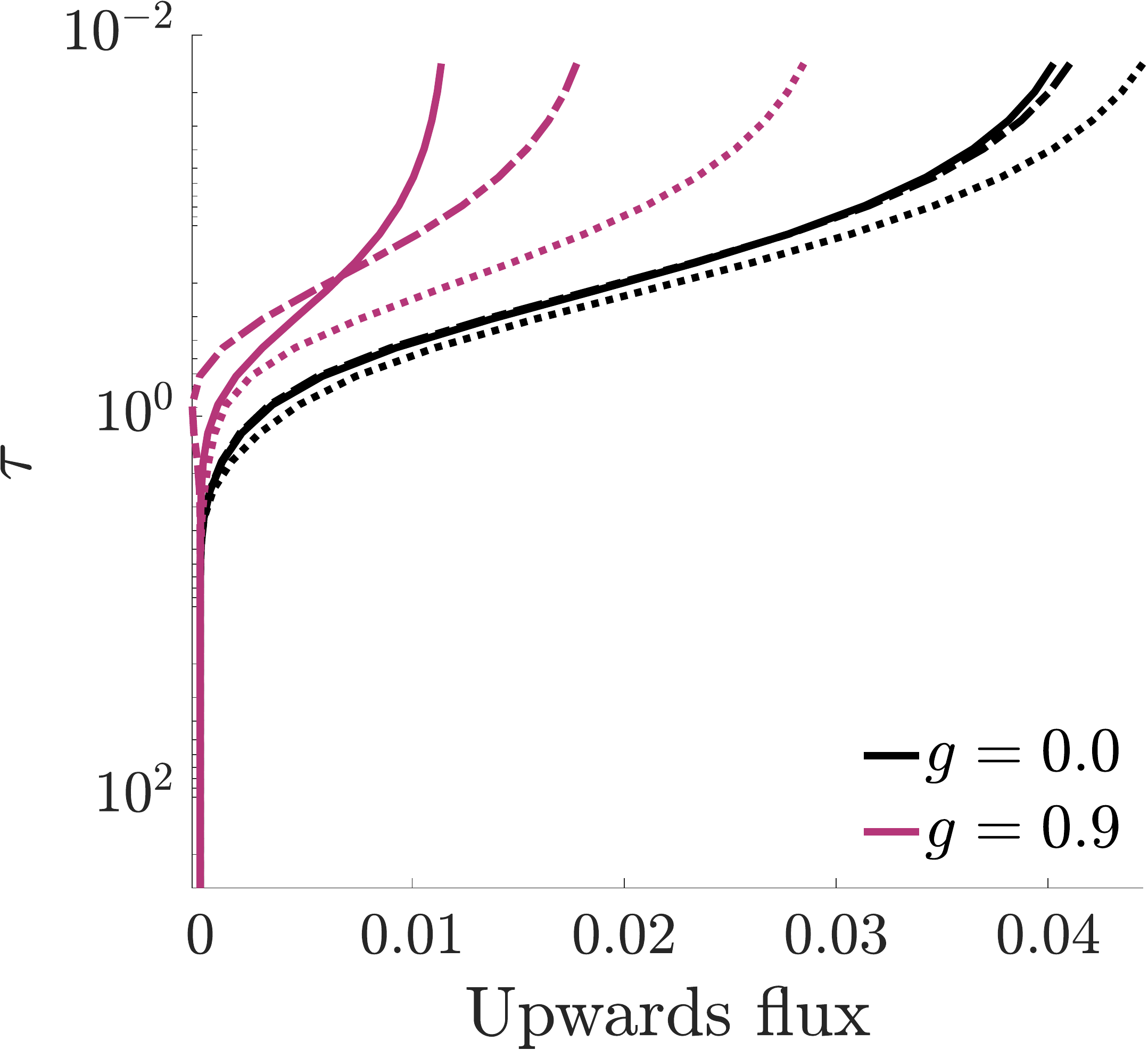}{.31\textwidth}{(b) $\mu_0=0.2$.}
	\fig{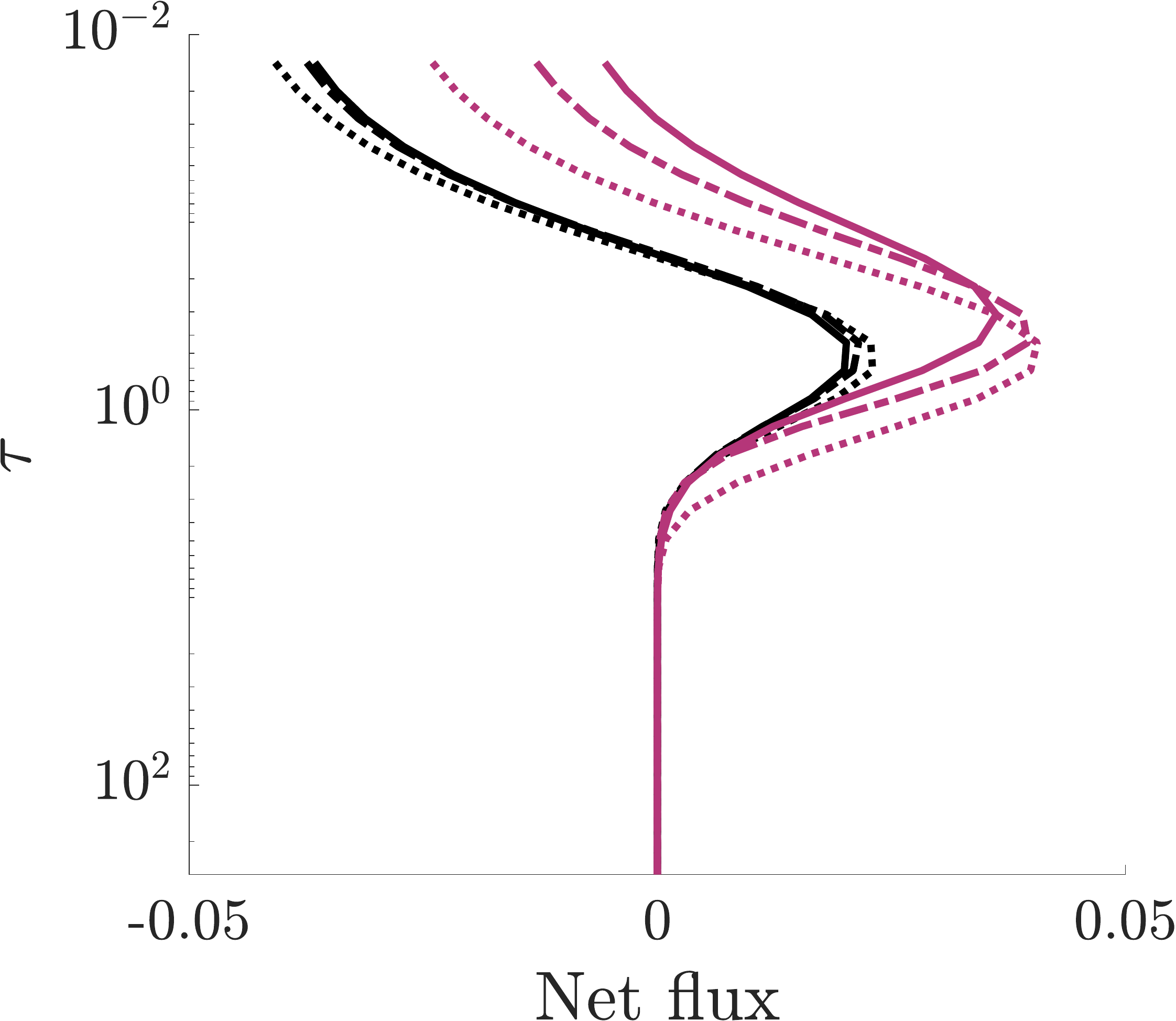}{.31\textwidth}{}}
	
	\caption{Fluxes with optical depth $\tau$ for cosine solar zenith angle (a) $\mu_0=0.9$ and (b) $\mu_0=0.2$, single scattering albedo $w_0=0.5$ and asymmetry $g_0=0.0, 0.9$. We compare \texttt{CDISORT} with Toon89 and SH4 and notice that SH4 agrees marginally better with \texttt{CDISORT} than Toon89.}
	\label{fig:flux}
\end{figure*}

Figure \ref{fig:flux} depicts the layer-wise downward, upward and net fluxes calculated by each model, where Figure \ref{fig:flux}(a) is for incoming angle $\mu_0=0.9$ and \ref{fig:flux}(b) is for $\mu_0=0.2$.
We consider a constant single scattering albedo of $w_0=0.5$ and include results for two extreme asymmetry values in each plot, namely $g_0=0.0$ and $g_0=0.9$.
We again notice an improvement in the agreement with \texttt{CDISORT} of SH4 from Toon89.
In particular, we observe negative upward flux values for Toon89 in Figure \ref{fig:flux}(a).
Negative fluxes have been reported for delta-Eddington calculations in the literature and are discussed in detail by \citep{wiscombe1977delta}.
The authors explain that, although unphysical, a negative flux does is not particularly more problematic than an inaccurate positive flux; the importance lies in how great the error is.
SH4 does not exhibit this negative-flux behaviour.

For each of the 30-layer, single-wavelength simulations studied in this section, the Toon89 framework in \texttt{PICASO} took on average 4e-5 seconds to run, while SH2 and SH4 took 2.5e-4 seconds and 3e-4 seconds respectively, using a 2021 Macbook Pro with the Apple M1 chip. All of the \texttt{PICASO} packages  leverage numba's CPU just-in-time compiling. \citep{numba}. 
We note that the Toon89 framework has been extensively used and sufficiently optimized, whereas the SH framework within \texttt{PICASO} is still in its infancy. 
With adequate computational optimization and code refinement, we expect SH2 to operate at similar speeds to Toon89, and for SH4 to exhibit a comparable increase in speed.
The equivalent 30-layer, single-wavelength simulations ran by 32-stream \texttt{CDISORT} took on average 2.4e-1 seconds -- three orders of magnitude slower than the current  implementation of SH4.
Therefore, even without the appropriate speed optimization and code refinement, SH4 offers significantly faster solutions than the higher-order simulations of \texttt{CDISORT}, whilst allowing notable improvement in model agreement over the two-stream Toon89. 

\subsection{Geometric Albedo Spectrum}
\label{sec:albedo}
\begin{figure*}[b!]
\centering
    \gridline{\fig{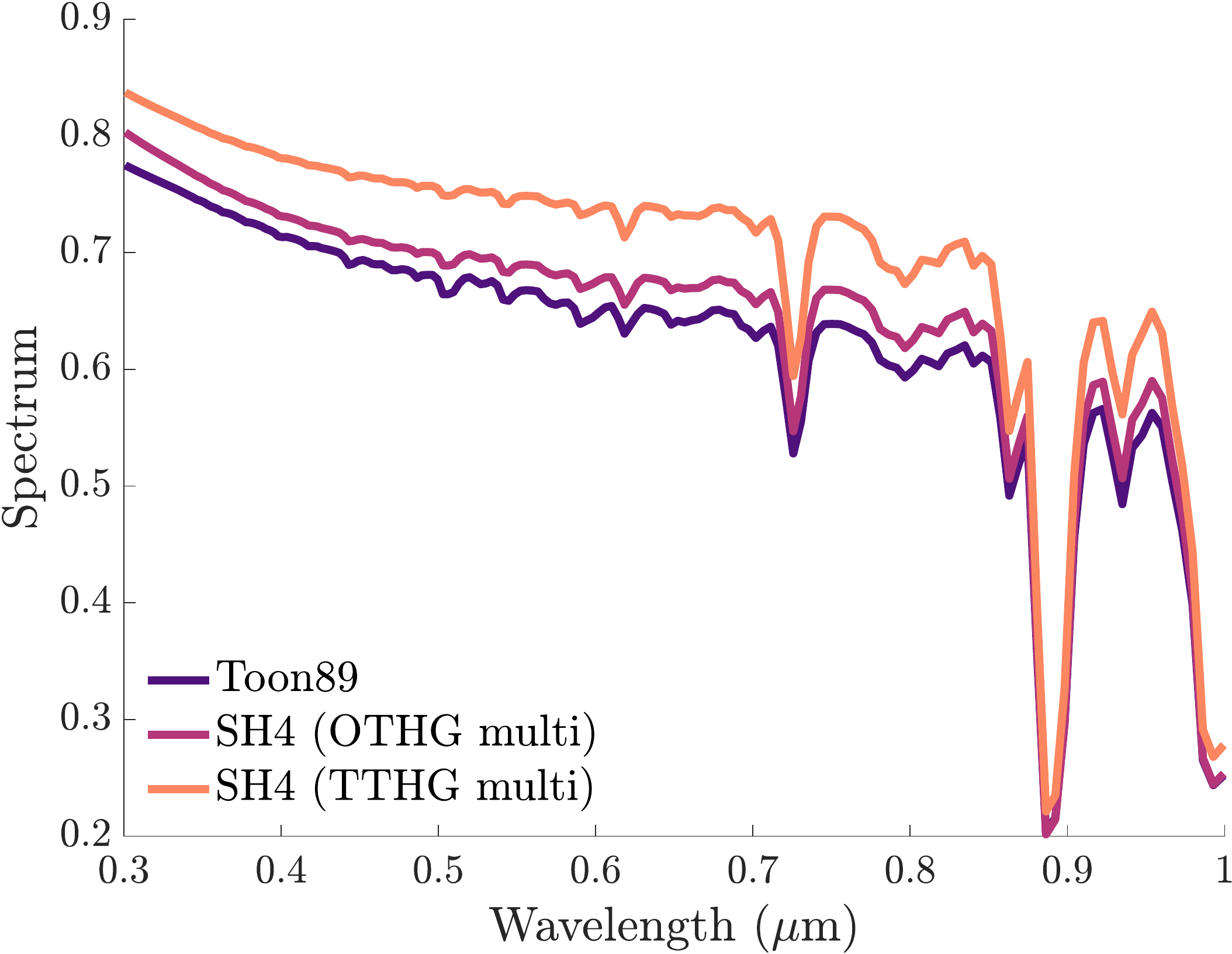}{.45\textwidth}{(a) Jupiter cloud profile.}
	\fig{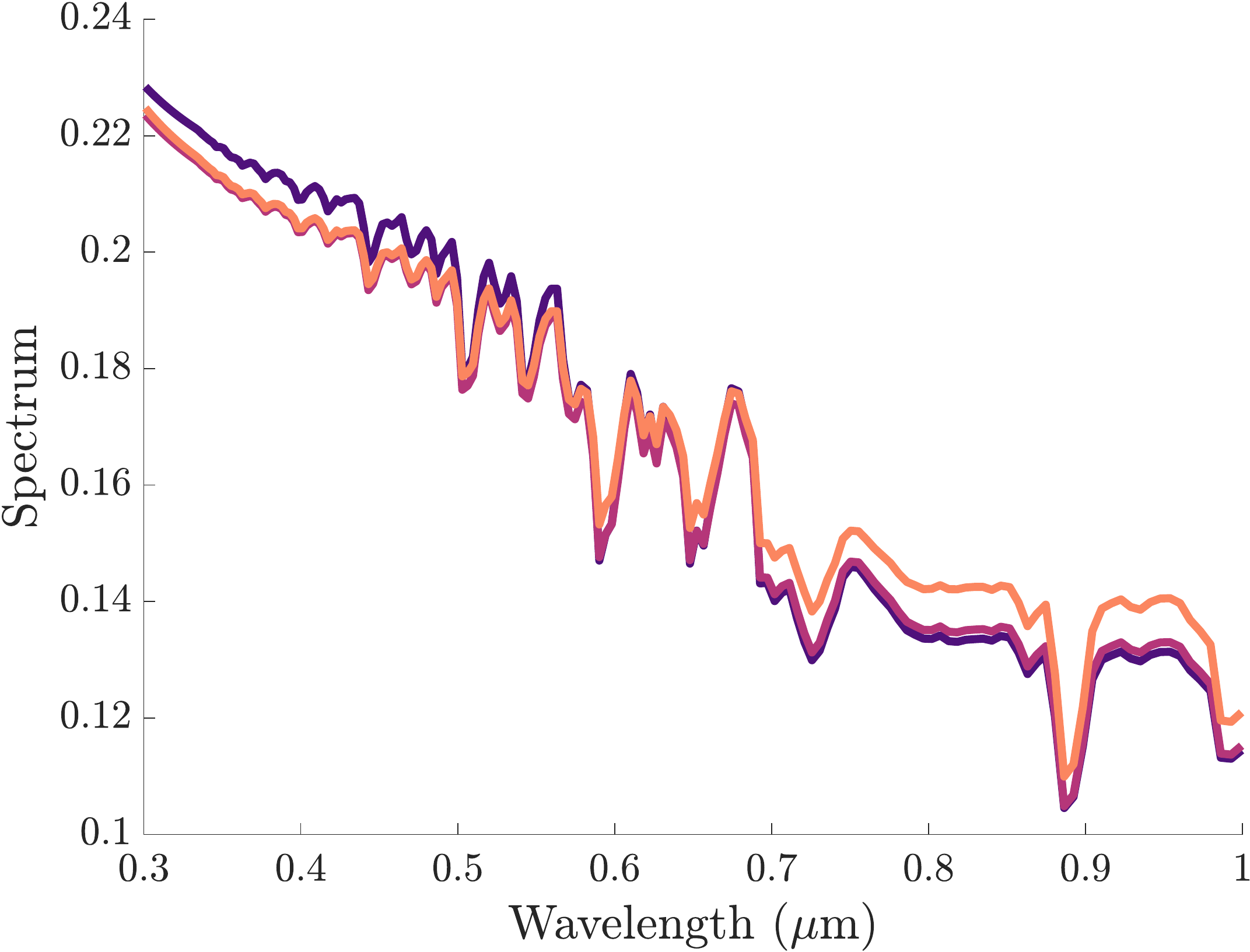}{.45\textwidth}{(b) Single parameter box cloud profile.}}
	
	\caption{Geometric albedo for a Jupiter atmosphere with (a) Jupiter clouds and (b) single parameter box cloud profile. We compare the spectra produced by Toon89 and SH4, where we consider two different scattering options for the latter; one for which the single scattering phase function is taken to be two-term Henyey-Greenstein (TTHG) while multi-scattering is one-term Henyey Greenstein, and the other for which all scattering is assumed to be TTHG. The former of these two SH4 options is most similar to the Toon89 analysis. \href{https://github.com/natashabatalha/picaso/blob/9d4cbd672a75c1faf5297c3f1d74074018cd7ef3/docs/notebooks/10b_AnalyzingApproximationsReflectedLightSH.ipynb}{\faCode}}
	\label{fig:geom_albedo}
\end{figure*}

Finally, we compare the ultimate geometric albedo produced by Toon89 and SH4 for a Jupiter-like system (1$\times$Solar metallicty, 25 ms$^{-2}$ gravity, 5 AU from a sun-like star) with two different cloud parameterizations: 1) Jupiter-like water clouds with a sedimentation efficiency \citep{ackerman2001precipitating} of f$_{\rm sed}=1$, and 2) a single-parameter box cloud profile described by a single asymmetry and single scattering albedo. The former case was also used in  \citet{batalha2019exoplanet} as the benchmark reflected light cloudy case.
For the latter, we specify a cloud layer located between $P=1$ and $P=0.1$ bars with asymmetry $g_0=0.9$, single scattering $w_0=0.8$ and layer optical depth $\tau=0.5$. 
These cases allow us to sample two different scattering regimes. The Jupiter-like water cloud represents the scenario where the cloud is dominant source of scattering throughout the 0.3-1$\mu$m. The single-parameter box cloud represents the scenario where Rayleigh is the dominant source of scattering at 0.5$<\mu$m, and clouds are the dominant source 0.5$>\mu$m. This is particularly important to test because of the different methodologies for including Rayleigh scattering.

To model these two cases, we test two different scattering approaches for SH4 against the default scattering approach used in Toon89. 
%-- that is, whether multiple-scattering is taken to be OTHG or TTHG (with Rayleigh).
In the Toon89 framework of \texttt{PICASO}, multiple-scattering for clouds is always modelled using a one-term Henyey-Greenstein (OTHG) phase function \eqref{eq:HG}, whereas single-scattering is represented using a weighted sum of a two-term Henyey-Greenstein (TTHG) phase function \eqref{eq:TTHG} for clouds, and Rayleigh, as described in Appendix \ref{app:rayleigh}.
SH4 has more flexibility: namely, we can prescribe either OTHG or TTHG for both single and multiple scattering for clouds, and we can include Rayleigh scattering in either scattering regimes.
For this comparison, we replicate the Toon89 framework for single scattering, but consider two cases for multiple scattering: (a) OTHG (with Rayleigh) and (b) TTHG (with Rayleigh). 
The geometric albedo for each parameterization is plotted in Figure \ref{fig:geom_albedo}.
The different modelling approaches produce evident differences in the geometric albedo.

For the Jupiter-like-cloud case [Figure \ref{fig:geom_albedo}(a)], the SH4 model with an OTHG multiple-scattering function agrees more closely with Toon89 than the TTHG case, where the TTHG case is significantly brighter than the other two models. The purpose of the TTHG framework is to better capture the back scattering peak, which contributes to the overall brightness \citep{batalha2019exoplanet}. Therefore, one major drawback of the standard \texttt{PICASO} methodology is its inability to account for the back scattering peak for multiply scattered photons. Figure \ref{fig:geom_albedo}(a) shows how SH4 can better account for this.

For the single parameter box cloud profile [Figure \ref{fig:geom_albedo}(b)], both SH4 cases produce different spectra than Toon89.
In the regime dominated by clouds ($>0.58\mu$m), the SH4 TTHG is still the brightest case, similar to the case shown in Figure \ref{fig:geom_albedo}(a). This suggests that, similar to the Jupiter-like case, the differences in this regime are due to the additional brightness from the inclusion of the TTHG treatment. In the region dominated by Rayleigh ($<0.58\mu$m) the behaviour flips and Toon89 is brighter than both SH4 OTHG and TTHG. Because the SH4 OTHG and TTHG spectra are similar in this regime, the spectral differences between SH4 and Toon89 are due to the differences in how the Rayleigh scattering is incorporated. \texttt{PICASO} post-processes Rayleigh into the multiple- and single-scattering phase function after the Toon89 methodology. SH2/4 incorporate Rayleigh directly in the phase functions expansion (see Appendix \ref{app:rayleigh}).

\section{Conclusion}
In this work, we presented a spherical harmonics approach to solving the radiative transfer equation for reflected light, which has been implemented in modelling software \texttt{PICASO} \cite{natasha_batalha_2022_6419943}.
This new methodology allows us to approximate scattering phase functions to higher orders than the original, two-stream implementation in \texttt{PICASO}, which we denoted Toon89 to reflect its heritage from \cite{toon1989rapid}.
The primary purpose of this work was to rigorously outline the derivation of the model, explaining the steps of the spherical harmonics analysis and explicitly stating the matrix systems being solved by the model.
To demonstrate the efficacy of the higher-order model, we compared our results to two independent models from \cite{liou1973numerical} and \cite{stamnes2000disort}.
These analyses illustrated the expected superiority of higher-order approximations in radiative transfer calculations, whilst also elucidating the extent of the improvement when moving from 2 to 4 term expansions. 
We studied the vertical distribution of fluxes by comparing the reflection and transmission values obtained via 2 and 4-term discrete ordinates methods (DOM), as well as the more accurate doubling method with those produced by 2 and 4-term spherical harmonics in Section \ref{sec:liou}.
This investigation highlighted that the choice of model (DOM or SH) is negligible compared to the order of approximation (2 or 4), where the 4-term approximations significantly out-performed the 2-term approximations.
However, in the majority of the cases studied in this analysis, the SH approach was marginally better than the DOM approach.
Extending the conclusion of \cite{liou1973numerical} regarding the comparison of low-order approximations with the doubling method, we conclude that the 4-term spherical harmonics method may be of adequate accuracy for studied of the flux distribution in the transfer of solar irradiance through cloudy atmospheres.
When comparing the 2-stream Toon89 and four-term SH4 to 32-stream \texttt{CDISORT} in Section \ref{sec:cdisort}, we noticed a significant enhancement of model accuracy when moving from 2 to 4 term approximations; we achieve notable improvement with minimal cost to computational expense.

Finally, we considered the impact of SH4 and the different scattering capabilities on the geometric albedo for a Jupiter atmosphere with two types of clouds: (a) Jupiter clouds and (b) single parameter box cloud profile.
This analysis illustrated that the choice of multiple-scattering function can significantly affect the reflected light spectra, thus the user-specification of scattering behaviour renders the SH4 model useful for retrievals and fine-tuning of atmospheric models.
A major drawback of the standard \texttt{PICASO} methodology is its inability to account for the back scattering peak for multiply scattered photons, a phenomenon that SH4 can better capture.
This analysis further highlighted the impact on spectra depending on how the Rayleigh scattering is incorporated.

The spherical harmonics technique outlined in this paper is also appropriate for modelling thermal emission.
This will be the focus of a future publication.

\begin{acknowledgements}
C.R.’s research was supported by an appointment to the NASA Postdoctoral Program at the NASA Ames Research Center, administered by Universities Space Research Association under contract with NASA. N.B. \& C.R. both acknowledge support from the NASA Astrophysics Division. Additionally, N.B. acknowledges support from NASA’S Interdisciplinary Consortia for Astrobiology Research (NNH19ZDA001N-ICAR) under award number 19-ICAR19\_2-0041. We thank Jeff Cuzzi and Sanford Davis for enlightening discussions about some of the finer points of radiative transfer in higher order approximations. Lastly we thank Arve Kylling for helpful discussions regarding \texttt{CDISORT}'s radiative transfer methodology. 
\end{acknowledgements}

\software{numba \citep{numba}, pandas \citep{mckinney2010data}, bokeh \citep{bokeh}, NumPy \citep{2020SciPy-NMeth}, \citep{walt2011numpy}, IPython \citep{perez2007ipython}, Jupyter, \citep{kluyver2016jupyter}, VIRGA \citep{natasha_batalha_2020_3759888,rooney2022new}, \texttt{PICASO} \citep{natasha_batalha_2022_6419943}, MATLAB \citep{MATLAB:2010}, A version of \text{PICASO} corresponding to these hyperlinks and the software used in this work is archived on Zenodo as v3.1 with DOI: 10.5281/zenodo.7765171}

\clearpage

\begin{appendices}
\section{Spherical harmonics methodology}
	The azimuthally-averaged radiative transfer equation for reflected light is given by \eqref{eq:RTE}. 
	By substituting \eqref{eq:PF}--\eqref{eq:I} into \eqref{eq:RTE}, we obtain
	\begin{equation}
		\begin{split}
			\summ (2l+1)\mu\frac{\mathrm{d}I_l}{\mathrm{d}\tau}P_l(\mu) &= \summ(2l+1)I_l(\tau)P_l(\mu) \\
			&- \frac{w_0}{2}\int_{-1}^1\summ (2l+1)I_l(\tau)P_l(\mu')\sum_{k=0}^L \chi_kP_k(\mu)P_k(\mu')\mathrm{d}\mu'
			-\sum_{l=0}^L b_le^{-\frac{\tau}{\mu_0}}P_l(\mu),%\frac{w \Fo}{4\pi}e^{-\frac{\tau}{\mu_0}}\chi_lP_l(\mu)P_l(-\mu_0).
		\end{split}
	\end{equation}
	where 
	\begin{equation}
	b_l=
		\frac{w_0\chi_l \Fo}{4\pi}\chi_lP_l(-\mu_0), 
	\end{equation}
	Using the recurrence relation and orthogonality principle of Legendre polynomials, i.e.
	\begin{align}
		(2l+1)\mu P_l(\mu) &= lP_{l-1}(\mu) + (l+1)P_{l+1}(\mu),\\
		\int_{-1}^1 P_l(\mu)P_k(\mu)\mathrm{d}\mu &= \frac{2}{2l+1}\delta_{l,k},
	\end{align}
	we obtain
	\begin{equation}
		\sum_{l=0}^L \left[lP_{l-1}(\mu) + (l+1)P_{l+1}(\mu)\right]
			\frac{\mathrm{d}I_l}{\mathrm{d}\tau}
			= \sum_{l=0}^L [(2l+1 - w_0\chi_l)I_l(\tau) - b_le^{-\frac{\tau}{\mu_0}} ]P_l(\mu), 
			%\frac{w\chi_l}{4\pi} F_\odot e^{-\frac{\tau}{\mu_0}} P_l(-\mu_0)] P_l(\mu),
	\end{equation}
	which can be rewritten as
	\begin{equation}
		\sum_{l=0}^L \left[(l+1)\frac{\mathrm{d}I_{l+1}}{\mathrm{d}\tau}
			+ l\frac{\mathrm{d}I_{l-1}}{\mathrm{d}\tau}\right]P_l(\mu)
			= \sum_{l=0}^L [(2l+1 - w_0\chi_l)I_l(\tau) - b_le^{-\frac{\tau}{\mu_0}} ]P_l(\mu), 
			%\frac{w\chi_l}{4\pi} F_\odot e^{-\frac{\tau}{\mu_0}} P_l(-\mu_0)] P_l(\mu). 
	\end{equation}

\section{$P_1$ single-layer solution}
	\label{app:P1_sol}
		The $P_1$ single-layer problem \eqref{eq:P1_sys} is linear in intensity, so is traditionally solved by combining the ``homogeneous'' and ``particular'' solutions, where the former must satisfy \eqref{eq:P1_sys} with no source term whereas the latter is a solution to \eqref{eq:P1_sys} with the source term included.
		We first consider the homogeneous problem, namely
		\begin{equation}
			\frac{\mathrm{d}}{\mathrm{d}\tau}
			\begin{pmatrix}
				I_0 \\ I_1
			\end{pmatrix} = 
			\begin{pmatrix}
				0 & a_1 \\ a_0 & 0
			\end{pmatrix}  
			\begin{pmatrix}
				I_0 \\ I_1
			\end{pmatrix}.
			\label{eq:P1_hom}
	\end{equation}
	We assume that this homogeneous solution is of the form $I_l = G_l e^{\lambda\tau}$ for $l=0,1$.
	Substituting this general form into \eqref{eq:P1_hom} yields
	\begin{equation}
			\begin{pmatrix}
				-\lambda & a_1 \\ a_0 & -\lambda
			\end{pmatrix}  
			\begin{pmatrix}
				G_0 \\ G_1
			\end{pmatrix} = 0.
			\label{eq:P1_hom1}
	\end{equation}
	To ensure that \eqref{eq:P1_hom1} has a solution, the value of the determinant must be zero, namely
	\begin{align}
		\lambda^2 - a_0a_1 = 0 \Rightarrow \lambda = \pm\sqrt{a_0a_1}.
	\end{align}
	The coefficients of $G_l$ are not independent, rather one can define
	\begin{equation}
		G_1 = \frac{\lambda}{a_1}G_0.
	\end{equation}
	
% 	\subsection{Reflected light}
	For the particular solution, we assume the form $I_l = \eta_l e^{-\tau/\mu_0}$ for $l=0,1$.
% 	$\eta_l = \zeta_l e^{-\tau/\mu_0}$ for $l=0,1$.
	Substituting the particular solution into \eqref{eq:P1_sys}, we can obtain the coefficients of $\zeta_l$ by solving 
	\begin{equation}
			\begin{pmatrix}
				1/\mu_0 & a_1 \\ a_0 & 1\mu_0
			\end{pmatrix}  
			\begin{pmatrix}
				% \zeta_0 \\ \zeta_1
				\eta_0 \\ \eta_1
			\end{pmatrix} = 
			\begin{pmatrix}
				b_1 \\ b_0
			\end{pmatrix}.
			\label{eq:P1_part}
	\end{equation}
	By inverting the matrix in \eqref{eq:P1_part}, we obtain the particular solution
	\begin{equation}
	\begin{aligned}
		\eta_0 = \frac{1}{1/\mu_0^2-a_0a_1}\left(\frac{b_1}{\mu_0} - a_1b_0\right),\\%e^{-\tau/\mu_0},\\
		\eta_1 = \frac{1}{1/\mu_0^2-a_0a_1}\left(\frac{b_0}{\mu_0} - a_0b_1\right).%e^{-\tau/\mu_0}.
	\end{aligned}
	\end{equation}
	
	Thus, we can write the full solution to \eqref{eq:P1_sys}:
	\begin{equation}
		\begin{pmatrix}
				I_0 \\ I_1
			\end{pmatrix} = 
			\begin{pmatrix}
				e^{-\lambda\tau} & e^{\lambda\tau} \\ -qe^{-\lambda\tau} & qe^{\lambda\tau}
			\end{pmatrix}  
			\begin{pmatrix}
				X_0 \\ X_1 
			\end{pmatrix} + 
			\begin{pmatrix}
				\eta_0 \\ \eta_1
			\end{pmatrix}e^{-\frac{\tau}{\mu_0}},
	\end{equation}
	where $q=\lambda/a_1$ and $\lambda = \sqrt{a_0a_1}$. 
	
\section{$P_3$ single-layer solution}
	\label{app:P3_sol}
	To solve the $P_3$ single-layer problem \eqref{eq:P3_sys}, we first consider the homogeneous problem, namely
	\begin{equation}
		\frac{\mathrm{d}}{\mathrm{d}\tau}
		\begin{pmatrix}
			I_0 \\ I_1 \\ I_2 \\ I_3
		\end{pmatrix} = 
		\begin{pmatrix}
			0 & a_1 & 0 & -\frac{2a_3}{3} \\ 
			a_0 & 0 & 0 & 0 \\
			0 & 0 & 0 & \frac{a_3}{3} \\
			-\frac{2a_0}{3} & 0 & \frac{a_2}{3} & 0 \\ 
		\end{pmatrix}  
		\begin{pmatrix}
			I_0 \\ I_1 \\ I_2 \\ I_3
		\end{pmatrix},
		\label{eq:P3_hom}
	\end{equation}
	As with the $P_1$ case, we assume the solution takes the form $I_l=G_le^{\lambda\tau}$ for $l=0,1,2,3$.
	Substituting this general form into \eqref{eq:P3_hom} yields
	\begin{equation}
			\begin{pmatrix}
			-\lambda & a_1 & 0 & -\frac{2a_3}{3} \\ 
			a_0 & -\lambda & 0 & 0 \\
			0 & 0 & -\lambda & \frac{a_3}{3} \\
			-\frac{2a_0}{3} & 0 & \frac{a_2}{3} & -\lambda 
		\end{pmatrix}  
			\begin{pmatrix}
				G_0 \\ G_1 \\ G_2 \\ G_3
			\end{pmatrix} = 0.
			\label{eq:P3_hom1}
	\end{equation}
	To ensure that \eqref{eq:P3_hom1} has a solution, the value of the determinant must be zero, namely
	\begin{align}
		\lambda^4 - \beta\lambda^2 + \gamma = 0 \Rightarrow \lambda_{1,2} = \sqrt{\frac{\beta\pm\sqrt{\beta^2-4\gamma}}{2}},
	\end{align}
	where
	\begin{align}
		\beta = a_0a_1 + \frac{1}{9}a_2a_3 + \frac{4}{9}a_0a_3,\qquad \gamma = \frac{1}{9}a_0a_1a_2a_3.
	\end{align}
	The coefficients of $G_l$ are not independent, rather by considering \eqref{eq:P3_hom} they are related by
	\begin{equation}
	\begin{aligned}
		\lambda G_0 &= a_1 G_1 - \frac{2a_3}{3}G_3,\\ 
		\lambda G_1 &= a_0 G_0 ,\\ 
		\lambda G_2 &= \frac{a_3}{3} G_3 ,\\
		\lambda G_3 &= -\frac{2a_0}{3} G_0 + \frac{a_2}{3}G_2,
	\end{aligned}
	\end{equation}
	and $G_l$ for $l=1,2,3$ can be written in terms of $G_0$:
	\begin{equation}
		\begin{aligned}
		G_1 &= \frac{a_0}{\lambda} G_0 ,\\ 
		G_2 &= \frac{1}{2}\left(\frac{a_0a_1}{\lambda^2} - 1\right) G_0 ,\\
		G_3 &= \frac{3}{2a_3}\left(\frac{a_0a_1}{\lambda} - \lambda\right) G_0 .
	\end{aligned}
	\end{equation}
	
% 	\subsection{Reflected light}
	For the particular solution, we assume the form $I_l = \eta_l e^{-\tau/\mu_0}$ for $l=0,1,2,3$.
% 	$\eta_l = \zeta_l e^{-\tau/\mu_0}$ for $l=0,1,2,3$.
	Substituting the particular solution into \eqref{eq:P1_sys}, we can obtain the coefficients of $\eta_l$ by solving 
	\begin{equation}
			\begin{pmatrix}
				1/\mu_0 & a_1 & 0 & -\frac{2a_3}{3} \\ 
			a_0 & 1/\mu_0 & 0 & 0 \\
			0 & 0 & 1/\mu_0 & \frac{a_3}{3} \\
			-\frac{2a_0}{3} & 0 & \frac{a_2}{3} & 1/\mu_0 
			\end{pmatrix}  
			\begin{pmatrix}
				% \zeta_0 \\ \zeta_1 \\ \zeta_2 \\ \zeta_3
				\eta_0 \\ \eta_1 \\ \eta_2 \\ \eta_3
			\end{pmatrix} = 
			\begin{pmatrix}
				-\frac{2b_3}{3} +b_1 \\ 
				b_0 \\
				\frac{b_3}{3} \\
				-\frac{2b_0}{3} +\frac{b_2}{3} 
			\end{pmatrix}.
			\label{eq:P3_part}
	\end{equation}
	By inverting the matrix in \eqref{eq:P3_part}, we obtain $\eta_l=\Delta_l/\Delta $ where
	\begin{equation}
	\begin{aligned}
		\Delta_0 &= (a_1b_0 - b_1/\mu_0)(a_2a_3-9/\mu_0^2) + 2(a_3b_2 - 2a_3b_0-3b_3/\mu_0)/\mu_0^2,\\
		\Delta_1 &= (a_0b_1 - b_0/\mu_0)(a_2a_3-9/\mu_0^2) - 2a_0(a_3b_2 - 3b_3/\mu_0)/\mu_0,\\
		\Delta_2 &= (a_3b_2 - 3b_3/\mu_0)(a_0a_1-1/\mu_0^2) - 2a_3(a_0b_1 -b_0/\mu_0)/\mu_0,\\
		\Delta_3 &= (a_2b_3 - 3b_2/\mu_0)(a_0a_1-1/\mu_0^2) + 2(3a_0b_1 - 2a_0b_3 - 3b_0/\mu_0)/\mu_0^2,
	\end{aligned}
	\end{equation}
	and
	\begin{align}
		\Delta = 9f(1/\mu_0) \qquad \text{where} \qquad f(x) = x^4 - \beta x^2 + \gamma.
	\end{align}
	
	Thus, we can write the full solution to \eqref{eq:P3_sys}:
	\begin{equation}
		\begin{pmatrix}
				I_0 \\ I_1 \\ I_2 \\ I_3
			\end{pmatrix} = 
			\begin{pmatrix}
				e^{-\lambda_1\tau} & e^{\lambda_1\tau} & e^{-\lambda_2\tau} & e^{\lambda_2\tau} \\ 
				R_1 e^{-\lambda_1\tau} & -R_1 e^{\lambda_1\tau} & R_2 e^{-\lambda_2\tau} & -R_2 e^{\lambda_2\tau} \\ 
				Q_1 e^{-\lambda_1\tau} & Q_1 e^{\lambda_1\tau} & Q_2 e^{-\lambda_2\tau} & S_2 e^{\lambda_2\tau} \\ 
				S_1 e^{-\lambda_1\tau} & -S_1 e^{\lambda_1\tau} & S_2 e^{-\lambda_2\tau} & -S_2 e^{\lambda_2\tau} 
			\end{pmatrix}  
			\begin{pmatrix}
				X_0 \\ X_1 \\ X_2 \\ X_3
			\end{pmatrix} + 
			\begin{pmatrix}
				\eta_0 \\ \eta_1 \\ \eta_2 \\ \eta_3
			\end{pmatrix} e^{-\frac{\tau}{\mu_0}},
	\end{equation}
	where 
	\begin{equation}
	\begin{aligned}
		R_{1,2} &= -\frac{a_0}{\lambda_{1,2}}, \qquad Q_{1,2} = \frac{1}{2}\left(\frac{a_0a_1}{\lambda_{1,2}^2} - 1\right), 
		\qquad S_{1,2} = -\frac{3}{2a_3}\left(\frac{a_0a_1}{\lambda_{1,2}} - \lambda_{1,2}\right).
	\end{aligned}
	\end{equation}
	
\section{Henyey-Greeinstein phase function}
Many of the analyses discussed in this work use the Henyey-Greenstein phase function \citep{henyey1941diffuse}:
\begin{equation}
	\mathcal{P}_\text{HG}(\cos\Theta) = \frac{1-g_0^2}{(1+g_0^2-2g_0\cos\Theta)^{3/2}},
	\label{eq:HG}
\end{equation}
where the scattering angle $\Theta$ is defined as
\begin{align}
	\cos\Theta &= \mu\mu' - \sqrt{1-\mu^2}\sqrt{1-\mu'^2}\cos(\phi-\phi').
	\label{eq:costheta}
\end{align}
The direction of incident and outgoing scattered radiation is defined by the cosine of the zenith angles, denoted $\mu'$ and $\mu$ respectively, and the azimuth angles, $\phi'$ and $\phi$.
The asymmetry parameter is denoted $g_0$.
The azimuthally-averaged Henyey-Greenstein function can be expressed in terms of Legendre polynomials \citep{liou2002introduction}: 
\begin{equation}
	\mathcal{P}(\mu,\mu') = \sum_{l=0}^N \chi_l P_l(\mu)P_l(\mu'),
	\label{eq:leg_azimuthally_averaged}
\end{equation}
for moments 
\begin{equation}
    \chi_l = (2l+1)g_0^l.
\end{equation}

The \texttt{PICASO} \citep{natasha_batalha_2022_6419943} methodology also utilises a two-term Henyey-Greenstein phase function in an attempt to capture back-scattering.
The two-term Henyey-Greenstein (TTHG) phase function is given by
\begin{equation}
	\mathcal{P}_\text{TTHG}(\cos\Theta) = \alpha\frac{1-g_1^2}{(1+g_1^2-2g_1\cos\Theta)^{3/2}}
	    + (1-\alpha)\frac{1-g_2^2}{(1+g_2^2-2g_2\cos\Theta)^{3/2}},
	\label{eq:TTHG}
\end{equation}
where we have two asymmetry factors, $g_1$ for forward scattering and $g_2$ for backward scattering, and new parameter $\alpha$ to determine the fraction of forward to back scattering:
\begin{equation}
    \alpha = C_1 + C_2g_2^{C_3}.
    \label{eq:alpha}
\end{equation}
Users can specify $C_1$, $C_2$, and $C_3$, however, \texttt{PICASO} defaults these to $C_1=1$, $C_2=-1$, and $C_3=2$.
The moments for TTHG are given by
\begin{equation}
    \chi_l = (2l+1)\left[\alpha g_1^l + (1-\alpha)g_2^l\right].
\end{equation}

\section{$\delta-$M approximation}
\label{app:deltaM}
Low order Legendre expansions are not capable of accurately representing highly forward scattering phase functions.
High scattering asymmetries are produced by Mie scattering particles with sizes larger than typical optical wavelengths, thus we must consider this limitation of the Legendre methodology.
To combat this, we apply the $\delta-$M technique \citep{joseph1976delta,wiscombe1977delta}, which involves approximating the phase function by a Dirac delta function forward scatter peak and an M-term expansion of the phase function.
Following the derivations of  \cite{joseph1976delta, wiscombe1977delta, cuzzi1982delta}, we outline the $\delta-$M methodology for our problem.

Consider the radiative transfer equation \eqref{eq:RTE} without the source function:
\begin{equation}
	\mu\frac{\partial I}{\partial \tau}(\tau,\mu) = I(\tau,\mu) 
		- \frac{w_0}{2}\int_{-1}^{1} I(\tau,\mu')\mathcal{P}(\mu,\mu')\mathrm{d}\mu'.
	\label{eq:RTE_nosource}
\end{equation}
The azimuthally-averaged phase function $\mathcal{P}(\mu,\mu')$ can be expanded in Legendre polynomials $P_l$ \citep{chandrasekhar1960radiative}
\begin{align}
	\mathcal{P}(\mu,\mu') &= \mathcal{P}(\cos\Theta) = \sum_{l=0}^{M-1} \chi_l P_l(\mu)P_l(\mu'),\\
	\chi_l &= \frac{2l+1}{2}\int_{-1}^1 \mathcal{P}(\cos\Theta)P_l(\cos\Theta)\,\mathrm{d}\cos\Theta,
\end{align}
Approximating the phase function by a Dirac delta function forward scatter peak and an M-term expansion, we obtain
\begin{align}
	\mathcal{P}_{\delta}(\mu,\mu')  &=2f\delta(\mu-\mu') + (1-f)\sum_{l=0}^{M-1} \chi_l^*P_l(\mu)P_l(\mu'),\\
	&=2f\delta(\mu-\mu') + (1-f)\mathcal{P}^*(\mu,\mu'),
	\label{eq:deltaM_PF}
\end{align}
where $f$ is the fractional scattering into the forward peak.
If we substitute $\mathcal{P}_{\delta}(\mu,\mu')$ \eqref{eq:deltaM_PF} in place of the azimuthally-averaged phase function $\mathcal{P}(\mu,\mu')$ in \eqref{eq:RTE_nosource} we obtain
\begin{equation}
	\mu\frac{\partial I}{\partial \tau}(\tau,\mu) = (1-w_0f)I(\tau,\mu) 
		- \frac{(1-f)w_0}{2}\int_{-1}^{1} I(\tau,\mu')\mathcal{P}^*(\mu,\mu')\mathrm{d}\mu',
\end{equation}
which we rewrite as
\begin{equation}
	\mu\frac{\partial I}{\partial \tau^*}(\tau^*,\mu) = I(\tau^*,\mu) 
		- \frac{w_0^*}{2}\int_{-1}^{1} I(\tau^*,\mu')\mathcal{P}^*(\mu,\mu')\mathrm{d}\mu',
		\label{eq:RTE_afterdelta}
\end{equation}
where \href{https://github.com/natashabatalha/picaso/blob/9d4cbd672a75c1faf5297c3f1d74074018cd7ef3/picaso/optics.py#L387-L397}{\faCode}
\begin{equation}
	\tau^* = (1-w_0f)\tau, \qquad w_0^* = \frac{(1-f)w_0}{1-w_0f}.
\end{equation}
%Note that we require $\mathcal{P}_{\delta}(\mu,\mu')$ to have the same asymmetry factor $g_0$ as the original phase function $\mathcal{P}(\mu,\mu)$, thus
%\begin{equation}
%	g_0 = \int_{4\pi}\mathcal{P}_{\delta}(\cos\Theta)\frac{\mathrm{d}\cos\Theta}{4\pi} = f + (1-f)g_0^*,
%\end{equation}
%from which $g_0^*$, the asymmetry of $\mathcal{P}^*(\mu,\mu')$, is determined to be
%\begin{equation}
%	g_0^* = \frac{g_0-f}{1-f}.
%\end{equation}
Note that we require the moments of $\mathcal{P}_\delta(\mu,\mu')$ be identical to that of the original phase function $\mathcal{P}(\mu,\mu')$ which allows us to determine the coefficients $\chi_l^*$ of $\mathcal{P}^*(\mu,\mu')$ to be
\begin{align}
	\chi_l^* = \frac{\chi_l-(2l+1)f}{1-f}.
	\label{eq:deltaM_coeff}
\end{align}
The fractional scattering coefficient $f$ is evaluated by ensuring that the $M^\text{th}$-order coefficient $\chi^*_M$ in the new phase function $\mathcal{P}^*(\mu,\mu)$ is equal to zero, thus
\begin{equation}
	f = \frac{\chi_M}{2M+1}.
	\label{eq:frac_scat}
\end{equation}
In \texttt{PICASO} \citep{natasha_batalha_2022_6419943}, we primarily use either one-term Henyey-Greenstein (OTHG) phase functions or two-term Henyey-Greenstein (TTHG) phase function.
For OTHG given by \eqref{eq:HG}, the fractional scattering coefficient is 
\begin{equation}
	f = g_0^M,
\end{equation}
for asymmetry $g_0$ \href{https://github.com/natashabatalha/picaso/blob/e7d7078b8bd93cb53295b470b96006848811c62b/picaso/optics.py#L389}{\faCode}.
% For TTHG, given by \eqref{eq:TTHG}, the fractional scattering coefficient is \href{https://github.com/natashabatalha/picaso/blob/891343fcc41faa345f8b85aaa8d50c4939c421a3/picaso/fluxes.py#L2644-L2649}{\faCode}
% \begin{equation}
% 	f = \alpha g_1^M + (1-\alpha)g_2^M.
% \end{equation}
% for forward and backward asymmetry parameters $g_1$ and $g_2$ respectively, and coefficient $\alpha$ for determining the fraction of forward to back scattering \eqref{eq:alpha}.

\section{Scattering capabilities}
The spherical harmonics method implemented within the \texttt{PICASO} framework \citep{natasha_batalha_2022_6419943} has user-specific scattering functionality; namely, the user can specify the type of phase function for single and multiple scattering separately.
The three types of phase functions currently offered by SH4 are one-term Henyey-Greenstein (OTHG), two-term Henyey-Greenstein (TTHG) and Rayleigh.
The choice of phase function impacts the coefficients $\chi_l$ in the Legendre approximation, which we recall is given by
\begin{equation}
    \mathcal{P}(\mu,\mu') = \sum_{l=0}^L \chi_l P_l(\mu) P_l(\mu').
\end{equation}
For cloud-scattering captured by a OTHG phase function \eqref{eq:HG} with asymmetry $g_0$, the coefficients are given by 
\begin{equation}
    \chi_l^\text{cld} = (2l+1)g_0^l,
    \label{eq:coeff_OTHG}
\end{equation}
whereas for a TTHG phase function \eqref{eq:TTHG}, we have
\begin{equation}
    \chi_l^\text{cld} = (2l+1)\left[\alpha g_1^l + (1-\alpha)g_2^l\right].
    \label{eq:coeff_TTHG}
\end{equation}
We note that for OTHG and TTHG we apply the $\delta-$M approximation described in Appendix \ref{app:deltaM}, thus the coefficients \eqref{eq:coeff_OTHG} and \eqref{eq:coeff_TTHG} must be adjusted using \eqref{eq:deltaM_coeff} and \eqref{eq:frac_scat}.

\subsection{Rayleigh scattering}
\label{app:rayleigh}
We include Rayleigh scattering by constructing a phase function that is comprised of a weighted sum of the cloud-scattering and Rayleigh properties \citep{batalha2019exoplanet}
\begin{equation}
	\mathcal{P}(\mu,\mu') = \frac{\tau_\text{cld}}{\tau_\text{scat}}\mathcal{P}_\text{cld}(\mu,\mu') +  \frac{\tau_\text{ray}}{\tau_\text{scat}}\mathcal{P}_\text{ray}(\mu,\mu'),
	\label{eq:weighted_PF}
\end{equation} 
where $\tau_\text{cld}$ and $\tau_\text{ray}$ denote the opacity contributed from the clouds and Rayleigh scattering respectively, and $\tau_\text{scat}$ is the total scattering opacity.
The Rayleigh scattering phase function is taken to be
\begin{equation}
	\mathcal{P}_\text{ray}(\cos\Theta) = \frac{3}{4}(1+\cos^2\Theta),
	\label{eq:rayleigh}
\end{equation}
and the cloud-scattering phase function $\mathcal{P}_\text{cld}(\mu,\mu')$ will either be the OTHG \eqref{eq:HG} or TTHG phase function \eqref{eq:TTHG}.
We note that before constructing this weighted sum, the cloud-scattering phase function has already been adjusted using the $\delta-$M approximation discussed in Appendix \ref{app:deltaM}.
Let us denote 
\begin{equation}
    \ftcld = \frac{\tau_\text{cld}}{\tau_\text{scat}}, \qquad \ftray = \frac{\tau_\text{ray}}{\tau_\text{scat}}.
\end{equation}
Thus, to conduct the spherical harmonics analysis outlined in this paper, we must be able to approximate the phase function in terms of Legendre polynomials; that is, we must be able to write the phase function in the form
\begin{equation}
    \mathcal{P}(\mu,\mu') = \sum_{l=0}^L \chi_l P_l(\mu) P_l(\mu').
\end{equation}
The coefficients $\chi_l$ for the weighted phase function \eqref{eq:weighted_PF} are \href{https://github.com/natashabatalha/picaso/blob/9d4cbd672a75c1faf5297c3f1d74074018cd7ef3/picaso/fluxes.py#L2683-L2690}{\faCode}
\begin{equation}
    \chi_l = \ftcld \chi_l^\text{cld} + \ftray \chi_l^\text{ray}.
\end{equation}
Since the cloud-scattering phase function has already been adjusted using the $\delta-$M approximation, the Legendre coefficients $\chi_l^\text{cld}$ are given by \eqref{eq:deltaM_coeff}.
The Rayleigh phase function \eqref{eq:rayleigh} can be written in terms of Legendre polynomials as
\begin{equation}
    \mathcal{P}_\text{ray}(\cos\Theta) = 1+\frac{1}{2}P_2(\mu)P_2(\mu'),
\end{equation}
where $P_2$ is the third Legendre polynomial. 
Therefore, the Legendre coefficients $\chi_l^\text{ray}$ for the Rayleigh phase function are
\begin{equation}
    \chi_l^\text{ray} = \begin{cases}
				1, &\quad l=0,\\
				0.5, &\quad l=2,\\
				0, &\quad \text{otherwise}.
				\end{cases}
\end{equation}
We notice that for 2-term expansions the Rayleigh contribution to the scattering function is neglected, as it only appears in the third term ($l=2$) of the phase function expansion.
This is an issue in the Toon89 framework within \texttt{PICASO}, but is managed by including the Rayleigh contribution by forcing a third moment equal to $0.5\ftray$, so that when Rayleigh dominates the total opacity, the correct scattering moments are included. 
This is discussed in more detail in \cite{batalha2019exoplanet}.
No such alternative measures are needed for 4-term spherical harmonics as the critical $l=2$ contribution appears naturally in the phase function expansion.

\section{Derivation of boundary condition}
\label{app:f_BC}
To derive the boundary condition for upward flux $F^+(\tau_N)$ at a Lambertian surface, we must determine the total amount of flux being reflected off of the surface.
This will be the sum of the reflected diffuse downward flux $F^-(\tau_N)$ and the reflected direct flux from the solar beam.
The intensity of the solar beam at the surface is 
\begin{equation}
    I_{direct}(\tau_N,\mu) = \frac{\Fo}{\pi}\mu_0e^{\frac{-\tau_N}{\mu_0}},
\end{equation}
from which we can derive the direct flux to be 
\begin{equation}
	F^-_{direct}(\tau_N) = 2\pi\int_0^{-1} I_{direct}(\tau_N,\mu)\mu\mathrm{d}\mu = \Fo\mu_0e^{\frac{-\tau_N}{\mu_0}}.
\end{equation}
Thus, for a Lambertian surface with surface reflectivity $A_S$, the upward flux $F^+(\tau_N)$ is given by the reflected portion of the downward diffuse and direct fluxes:
\begin{equation}
    F^+(\tau_N) = A_S[F^-(\tau_N) + \mu_0F_\odot e^{-\frac{\tau}{\mu_0}}].
\end{equation}
We can derive the equivalent boundary condition for $f^+(\tau_N)$ by first calculating $f^-_{direct}(\tau_N)$ using 
\begin{equation}
	f^-_{direct}(\tau_N) = 2\pi\int_0^{-1} I_{direct}(\tau_N,\mu)\frac{1}{2}(5\mu^3-3\mu)\mathrm{d}\mu = -\frac{1}{4}\Fo\mu_0e^{\frac{-\tau_N}{\mu_0}}.
\end{equation}
Similarly, for a Lambertian surface with surface reflectivity $A_S$, the upward $f^+(\tau_N)$ is given by 
\begin{equation}
    f^+(\tau_N) = A_S[f^-(\tau_N) - \frac{1}{4} \mu_0F_\odot e^{-\frac{\tau}{\mu_0}}].
\end{equation}

\section{Modelling recommendations}
In Section \ref{sec:albedo} we investigated the geometric albedo produced by the Toon89 and SH4 frameworks within \texttt{PICASO}.
Here, we aggregate our modelling recommendations in a single table.
In the current version of \texttt{PICASO}, these represent the radiative transfer defaults. 
All of these toggles are controlled through the \texttt{PICASO} \texttt{approx}  \href{https://github.com/natashabatalha/picaso/blob/9d4cbd672a75c1faf5297c3f1d74074018cd7ef3/picaso/justdoit.py#L3405-L3411}{picaso.justdoit.iputs.approx} function. 
A tutorial on the use of this function is available as \href{https://github.com/natashabatalha/picaso/blob/9d4cbd672a75c1faf5297c3f1d74074018cd7ef3/docs/notebooks/10a_AnalyzingApproximationsReflectedLightToon.ipynb}{Jupyter notebook}.

\begin{itemize}
    \item \textbf{Single Scattering}
    \begin{itemize}
        \item Use default specification for direct scattering:
        \begin{itemize}
        \item Toon89: \texttt{single\_phase=`TTHG\_ray'}.
        \item SH4: \texttt{psingle\_form=`TTHG'} with \texttt{psingle\_rayleigh=`on'}, and \texttt{w\_single\_form=`TTHG'} with \texttt{w\_single\_rayleigh=`on'}.
        \end{itemize}
        \item Specify the functional form of the fraction, \texttt{f}, of forward and backward scattering according to the problem being addressed.
    \end{itemize}
    \item \textbf{Multiple Scattering}
    \begin{itemize}
        \item For planet cases with some degree of asymmetric clouds scatterers:
        \begin{itemize}
            \item Toon89: always use N=2 Legendre polynomial expansion with the $
        \delta-$Eddington approximation.
            \item SH4: choose either \texttt{w\_multi\_form=`TTHG'} if you expect the back-scattering fraction to be non-negligible, or \texttt{`OTHG'} if you wish to primarily capture forward scattering. Both phase functions should be adjusted with the $\delta-$Eddington approximation.
        \end{itemize}
        \item Additionally for SH4, include Rayleigh effects in multiple scattering by setting \texttt{w\_multi\_rayleigh=`on'}.
    \end{itemize}
\end{itemize}
\end{appendices}

\clearpage
\bibliography{spherical_harmonics}

\begin{thebibliography}{}
\expandafter\ifx\csname natexlab\endcsname\relax\def\natexlab#1{#1}\fi
\providecommand{\url}[1]{\href{#1}{#1}}
\providecommand{\dodoi}[1]{doi:~\href{http://doi.org/#1}{\nolinkurl{#1}}}
\providecommand{\doeprint}[1]{\href{http://ascl.net/#1}{\nolinkurl{http://ascl.net/#1}}}
\providecommand{\doarXiv}[1]{\href{https://arxiv.org/abs/#1}{\nolinkurl{https://arxiv.org/abs/#1}}}

\bibitem[{Ackerman \& Marley(2001)}]{ackerman2001precipitating}
Ackerman, A.~S., \& Marley, M.~S. 2001, The Astrophysical Journal, 556, 872

\bibitem[{Ayash {et~al.}(2008)Ayash, Gong, \& Jia}]{ayash2008implementing}
Ayash, T., Gong, S., \& Jia, C.~Q. 2008, Journal of the atmospheric sciences,
  65, 2448

\bibitem[{{Barstow} {et~al.}(2020){Barstow}, {Changeat}, {Garland}, {Line},
  {Rocchetto}, \& {Waldmann}}]{barstow20comp}
{Barstow}, J.~K., {Changeat}, Q., {Garland}, R., {et~al.} 2020, \mnras, 493,
  4884, \dodoi{10.1093/mnras/staa548}

\bibitem[{Batalha {et~al.}(2020)Batalha, caoimherooney11, \&
  sagnickm}]{natasha_batalha_2020_3759888}
Batalha, N., caoimherooney11, \& sagnickm. 2020, natashabatalha/virga: Initial
  Release, v0.0,  Zenodo, \dodoi{10.5281/zenodo.3759888}

\bibitem[{Batalha {et~al.}(2022)Batalha, Rooney, Blanch, \&
  MacDonald}]{natasha_batalha_2022_6419943}
Batalha, N., Rooney, C., Blanch, N.~R., \& MacDonald, R. 2022,
  natashabatalha/picaso: Release 2.3, v2.3.0,  Zenodo,
  \dodoi{10.5281/zenodo.6419943}

\bibitem[{Batalha {et~al.}(2019)Batalha, Marley, Lewis, \&
  Fortney}]{batalha2019exoplanet}
Batalha, N.~E., Marley, M.~S., Lewis, N.~K., \& Fortney, J.~J. 2019, The
  Astrophysical Journal, 878, 70

\bibitem[{{Bokeh Development Team}(2014)}]{bokeh}
{Bokeh Development Team}. 2014, Bokeh: Python library for interactive
  visualization.
\newblock \url{http://www.bokeh.pydata.org}

\bibitem[{Buras {et~al.}(2011)Buras, Dowling, \& Emde}]{buras2011new}
Buras, R., Dowling, T., \& Emde, C. 2011, Journal of Quantitative Spectroscopy
  and Radiative Transfer, 112, 2028

\bibitem[{Chandrasekhar(1960)}]{chandrasekhar1960radiative}
Chandrasekhar, S. 1960, Chandrasekhar, S, 20, 2

\bibitem[{Chou(1992)}]{chou1992solar}
Chou, M.-D. 1992, Journal of Atmospheric Sciences, 49, 762

\bibitem[{Cuzzi {et~al.}(1982)Cuzzi, Ackerman, \& Helmle}]{cuzzi1982delta}
Cuzzi, J.~N., Ackerman, T.~P., \& Helmle, L.~C. 1982, Journal of the
  Atmospheric Sciences, 39, 917

\bibitem[{Evans \& Stephens(1991)}]{evans1991new}
Evans, K.~F., \& Stephens, G. 1991, Journal of Quantitative Spectroscopy and
  Radiative Transfer, 46, 413

\bibitem[{Fiveland \& Jessee(1996)}]{fiveland1996acceleration}
Fiveland, V., \& Jessee, J. 1996, Journal of thermophysics and heat transfer,
  10, 445

\bibitem[{Ge {et~al.}(2015)Ge, Marquez, Modest, \& Roy}]{ge2015implementation}
Ge, W., Marquez, R., Modest, M.~F., \& Roy, S.~P. 2015, Journal of Heat
  Transfer, 137

\bibitem[{Heng {et~al.}(2018)Heng, Malik, \& Kitzmann}]{heng2018analytical}
Heng, K., Malik, M., \& Kitzmann, D. 2018, The Astrophysical Journal Supplement
  Series, 237, 29

\bibitem[{Heng {et~al.}(2014)Heng, Mendon{\c{c}}a, \& Lee}]{heng2014analytical}
Heng, K., Mendon{\c{c}}a, J.~M., \& Lee, J.-M. 2014, The Astrophysical Journal
  Supplement Series, 215, 4

\bibitem[{Henyey \& Greenstein(1941)}]{henyey1941diffuse}
Henyey, L.~G., \& Greenstein, J.~L. 1941, The Astrophysical Journal, 93, 70

\bibitem[{Irvine(1975)}]{irvine1975multiple}
Irvine, W.~M. 1975, Icarus, 25, 175

\bibitem[{Iwabuchi(2006)}]{iwabuchi2006efficient}
Iwabuchi, H. 2006, Journal of the atmospheric sciences, 63, 2324

\bibitem[{Joseph {et~al.}(1976)Joseph, Wiscombe, \& Weinman}]{joseph1976delta}
Joseph, J.~H., Wiscombe, W., \& Weinman, J. 1976, Journal of Atmospheric
  Sciences, 33, 2452

\bibitem[{King(1986)}]{king1986comparative}
King, M.~D. 1986, Journal of Atmospheric Sciences, 43, 784

\bibitem[{Kluyver {et~al.}(2016)Kluyver, Ragan-Kelley, P{\'e}rez, Granger,
  Bussonnier, Frederic, Kelley, Hamrick, Grout, Corlay,
  {et~al.}}]{kluyver2016jupyter}
Kluyver, T., Ragan-Kelley, B., P{\'e}rez, F., {et~al.} 2016, in ELPUB, 87--90

\bibitem[{Lam {et~al.}(2015)Lam, Pitrou, \& Seibert}]{numba}
Lam, S.~K., Pitrou, A., \& Seibert, S. 2015, in Proceedings of the Second
  Workshop on the LLVM Compiler Infrastructure in HPC, LLVM '15 (New York, NY,
  USA: ACM), 7:1--7:6, \dodoi{10.1145/2833157.2833162}

\bibitem[{Lewis \& Miller(1984)}]{lewis1984computational}
Lewis, E.~E., \& Miller, W.~F. 1984

\bibitem[{Li \& Ramaswamy(1996)}]{li1996four}
Li, J., \& Ramaswamy, V. 1996, Journal of the atmospheric sciences, 53, 1174

\bibitem[{{Line} {et~al.}(2012){Line}, {Zhang}, {Vasisht}, {Natraj}, {Chen}, \&
  {Yung}}]{line2012info}
{Line}, M.~R., {Zhang}, X., {Vasisht}, G., {et~al.} 2012, \apj, 749, 93,
  \dodoi{10.1088/0004-637X/749/1/93}

\bibitem[{Liou(1973)}]{liou1973numerical}
Liou, K.-N. 1973, Journal of Atmospheric Sciences, 30, 1303

\bibitem[{Liou(1974)}]{liou1974analytic}
Liou, K.-n. 1974, J. Atmos. Sci, 31, 1473

\bibitem[{Liou(2002)}]{liou2002introduction}
Liou, K.-N. 2002, An introduction to atmospheric radiation (Elsevier)

\bibitem[{Liou {et~al.}(1988)Liou, Fu, \& Ackerman}]{liou1988simple}
Liou, K.-N., Fu, Q., \& Ackerman, T.~P. 1988, Journal of Atmospheric Sciences,
  45, 1940

\bibitem[{Liu \& Weng(2006)}]{liu2006advanced}
Liu, Q., \& Weng, F. 2006, Journal of the Atmospheric Sciences, 63, 3459

\bibitem[{{Madhusudhan} \& {Seager}(2009)}]{madhu2009temp}
{Madhusudhan}, N., \& {Seager}, S. 2009, \apj, 707, 24,
  \dodoi{10.1088/0004-637X/707/1/24}

\bibitem[{MATLAB(2010)}]{MATLAB:2010}
MATLAB. 2010, version 7.10.0 (R2010a) (Natick, Massachusetts: The MathWorks
  Inc.)

\bibitem[{Mayer(2009)}]{mayer2009radiative}
Mayer, B. 2009, in EPJ web of Conferences, Vol.~1, EDP Sciences, 75--99

\bibitem[{Mayer \& Kylling(2005)}]{mayer2005libradtran}
Mayer, B., \& Kylling, A. 2005, Atmospheric Chemistry and Physics, 5, 1855

\bibitem[{McKinney(2010)}]{mckinney2010data}
McKinney, W. 2010, in Proceedings of the 9th Python in Science Conference, ed.
  S.~van~der Walt \& J.~Millman, 51 -- 56

\bibitem[{Meador \& Weaver(1980)}]{meador1980two}
Meador, W., \& Weaver, W. 1980, Journal of Atmospheric Sciences, 37, 630

\bibitem[{Mihalas \& Mihalas(2013)}]{mihalas2013foundations}
Mihalas, D., \& Mihalas, B.~W. 2013, Foundations of radiation hydrodynamics
  (Courier Corporation)

\bibitem[{Modest(1989)}]{modest1989modified}
Modest, M.~F. 1989, Journal of Thermophysics and Heat Transfer, 3, 283

\bibitem[{Modest(2013)}]{modest2013radiative}
---. 2013, Radiative Heat Transfer (Academic Press)

\bibitem[{Olfe(1967)}]{olfe1967modification}
Olfe, D. 1967, AIAA Journal, 5, 638

\bibitem[{P{\'e}rez \& Granger(2007)}]{perez2007ipython}
P{\'e}rez, F., \& Granger, B.~E. 2007, Computing in Science \& Engineering, 9

\bibitem[{Ravishankar(2009)}]{ravishankar2009spherical}
Ravishankar, M. 2009, PhD thesis, The Ohio State University

\bibitem[{Rooney {et~al.}(2022)Rooney, Batalha, Gao, \& Marley}]{rooney2022new}
Rooney, C.~M., Batalha, N.~E., Gao, P., \& Marley, M.~S. 2022, The
  Astrophysical Journal, 925, 33

\bibitem[{Shibata \& Uchiyama(1992)}]{shibata1992accuracy}
Shibata, K., \& Uchiyama, A. 1992, Journal of the Meteorological Society of
  Japan. Ser. II, 70, 1097

\bibitem[{Stamnes {et~al.}(1988)Stamnes, Tsay, Wiscombe, \&
  Jayaweera}]{stamnes1988numerically}
Stamnes, K., Tsay, S.-C., Wiscombe, W., \& Jayaweera, K. 1988, Applied optics,
  27, 2502

\bibitem[{Stamnes {et~al.}(2000)Stamnes, Tsay, Wiscombe, \&
  Laszlo}]{stamnes2000disort}
Stamnes, K., Tsay, S.-C., Wiscombe, W., \& Laszlo, I. 2000

\bibitem[{Stephens \& Preisendorfer(1984)}]{stephens1984multimode}
Stephens, G.~L., \& Preisendorfer, R.~W. 1984, Journal of the atmospheric
  sciences, 41, 725

\bibitem[{Stokes(1862)}]{stokes1862iv}
Stokes, G.~G. 1862, Proceedings of the Royal Society of London, 545

\bibitem[{Thomas \& Stamnes(2002)}]{thomas2002radiative}
Thomas, G.~E., \& Stamnes, K. 2002, Radiative transfer in the atmosphere and
  ocean (Cambridge University Press)

\bibitem[{Toon {et~al.}(1989)Toon, McKay, Ackerman, \&
  Santhanam}]{toon1989rapid}
Toon, O.~B., McKay, C., Ackerman, T., \& Santhanam, K. 1989, Journal of
  Geophysical Research: Atmospheres, 94, 16287

\bibitem[{Van~de Hulst(1963)}]{van1963new}
Van~de Hulst, H. 1963, Rept, Inst. Space Studies, NASA, New York

\bibitem[{van~de Hulst(1980)}]{van1980multiple}
van~de Hulst, H.~C. 1980, Multiple light scattering,  Elsevier

\bibitem[{van Wijngaarden \&
  Happer(2022)}]{https://doi.org/10.48550/arxiv.2205.09713}
van Wijngaarden, W.~A., \& Happer, W. 2022, 2n-Stream Radiative Transfer,
  arXiv, \dodoi{10.48550/ARXIV.2205.09713}

\bibitem[{Virtanen {et~al.}(2020)Virtanen, Gommers, Oliphant, Haberland, Reddy,
  Cournapeau, Burovski, Peterson, Weckesser, Bright, {van der Walt}, Brett,
  Wilson, Millman, Mayorov, Nelson, Jones, Kern, Larson, Carey, Polat, Feng,
  Moore, {VanderPlas}, Laxalde, Perktold, Cimrman, Henriksen, Quintero, Harris,
  Archibald, Ribeiro, Pedregosa, {van Mulbregt}, \& {SciPy 1.0
  Contributors}}]{2020SciPy-NMeth}
Virtanen, P., Gommers, R., Oliphant, T.~E., {et~al.} 2020, SciPy 1.0:
  Fundamental Algorithms for Scientific Computing in Python,
  \dodoi{10.1038/s41592-019-0686-2}

\bibitem[{Walt {et~al.}(2011)Walt, Colbert, \& Varoquaux}]{walt2011numpy}
Walt, S. v.~d., Colbert, S.~C., \& Varoquaux, G. 2011, Computing in Science \&
  Engineering, 13, 22

\bibitem[{Wiscombe(1977)}]{wiscombe1977delta}
Wiscombe, W. 1977, Journal of Atmospheric Sciences, 34, 1408

\bibitem[{Wiscombe \& Joseph(1977)}]{wiscombe1977range}
Wiscombe, W., \& Joseph, J. 1977, Icarus, 32, 362

\bibitem[{Zhang \& Li(2013)}]{zhang2013doubling}
Zhang, F., \& Li, J. 2013, Journal of the atmospheric sciences, 70, 3084

\end{thebibliography}
\bibliographystyle{aasjournal}

\end{document}